\documentstyle[12pt,epsf]{article}
 \newcommand{\insertplot}[5]{\begin{figure}
 \hfill\hbox to 0.05in{\vbox to #5in{\vfill
 \inputplot{#1}{#4}{#5}}\hfill}
 \hfill\vspace{-.1in}
 \caption{#2}\label{#3}
 \end{figure}}
 \newcommand{\inputplot}[3]{
 \special{ps: plotfile #1}
}
\newcommand{\vphi}{\varphi}
\newcommand{\pih}{\frac{\pi}{2}}

\newcommand{\xh}{x_{\rm H}}

\newcommand{\DS}{\displaystyle}
\newcommand{\oh}{\omega_{\rm H}}
\newcommand{\tB}{\tilde{B}}
\newcommand{\tH}{\tilde{H}}
\newcommand{\hPsi}{\hat{\Psi}}

\newcounter{fixy}
\begin{document}
 \newenvironment{fixy}[1]{\setcounter{figure}{#1}}
{\addtocounter{fixy}{1}}
\renewcommand{\thefixy}{\arabic{fixy}}
\renewcommand{\thefigure}{\thefixy\alph{figure}}
\setcounter{fixy}{1}

\title{
Rotating Einstein-Yang-Mills Black Holes}
\vspace{1.5truecm}
\author{
{\bf Burkhard Kleihaus}\\
Department of Mathematical Physics, University College, Dublin,\\
Belfield, Dublin 4, Ireland\\
{\bf Jutta Kunz and Francisco Navarro-L\'erida$^1$}\\
Fachbereich Physik, Universit\"at Oldenburg, Postfach 2503\\
D-26111 Oldenburg, Germany
}

\vspace{1.5truecm}

%\date{March 14, 2002}
\date{\today}

\maketitle
\vspace{1.0truecm}

\begin{abstract}
We construct rotating hairy black holes in SU(2) Einstein-Yang-Mills theory.
These stationary axially symmetric black holes are asymptotically flat.
They possess non-trivial non-Abelian gauge fields
outside their regular event horizon,
and they carry non-Abelian electric charge.
In the limit of vanishing angular momentum,
they emerge from the neutral static spherically symmetric
Einstein-Yang-Mills black holes,
labelled by the node number of the gauge field function.
With increasing angular momentum and mass, 
the non-Abelian electric charge of the solutions increases,
but remains finite.
The asymptotic expansion for these black hole solutions includes
non-integer powers of the radial variable.
\end{abstract}
\vfill
$^1$ Dept. de F\'{\i}sica Te\'orica II, Ciencias F\'{\i}sicas,
Universidad Complutense de Madrid, E-28040 Madrid, Spain\\
%\noindent {Preprint hep-th/0207042} \hfill\break
\vfill\eject

\section{Introduction}

The unique family of stationary asymptotically flat
black holes of Einstein-Maxwell (EM) theory
comprises the rotating Kerr-Newman and Kerr black holes
and the static Reissner-Nordstr\o m and Schwarzschild black holes.
EM black holes are completely determined
by their mass, their charge and their angular momentum,
i.e.~EM black holes have ``no hair'' \cite{nohair1,nohair2}.

The EM ``no-hair'' theorem does not generalize to theories with
non-Abelian gauge fields coupled to gravity \cite{review}.
The generic black hole solutions of SU(2) Einstein-Yang-Mills (EYM) theory
possess non-trivial magnetic fields outside their regular event horizon,
representing non-Abelian ``hair'' \cite{review}.
Besides static spherically symmetric hairy black holes \cite{su2bh},
there are also static hairy black holes, which are not
spherically but only axially symmetric \cite{kkbh,map3,hkk}.
This shows that Israel's theorem neither generalizes to theories
with non-Abelian gauge fields coupled to gravity.

Obviously, also hairy stationary black hole solutions,
representing the non-Abelian generalizations of the Kerr-Newman
black hole solutions, should exist, as conjectured long ago \cite{sud-wald}.
The construction of such hairy rotating black hole solutions, however,
appeared very difficult. First of all, it was not clear,
whether the standard Lewis-Papapetrou parametrization 
of the stationary axially symmetric metric \cite{book}
would be sufficiently general
and whether an ansatz for the gauge fields,
satisfying the Ricci circularity and Frobenius conditions \cite{book,circ},
was available \cite{heus,vs,galt}.
Second, even within the standard metric parametrization 
the black hole solutions would only possess axial symmetry.
Therefore the construction of such solutions would involve the solution 
of a large system of coupled non-linear partial differential equations
for the metric and gauge field functions and thus
represent a numerical challenge.

First progress was achieved with the construction of
perturbative stationary non-Abelian black hole solutions
in SU(2) EYM theory \cite{vs}.
Based on the static hairy black hole solutions, 
these slowly rotating black hole solutions revealed an unexpected
property. The rotating black holes carry non-Abelian electric charge,
whereas their static counterparts are electrically neutral \cite{vs}.
In the static spherically symmetric case 
electrically charged SU(2) black holes are even prohibited 
by the `non-Abelian baldness' theorem \cite{ge}.
Indeed, the non-Abelian electric charge 
of the slowly rotating black hole solutions
turned out to be proportional to their angular momentum,
and thus vanishes in the static limit.
Subsequently, perturbative calculations with more general
boundary conditions predicted
even more exotic stationary hairy black hole solutions
\cite{bhsv}. 

Only recently non-pertubative rotating hairy black hole solutions 
were obtained in SU(2) EYM theory \cite{kkrot}, 
confirming the perturbative calculations \cite{vs}. 
They are obtained within the standard Lewis-Papapetrou parametrization
of the metric, and the ansatz for the gauge fields is consistent
with the circularity and Frobenius conditions.
Representing the first set of non-perturbative rotating 
hairy black hole solutions,
they possess three global charges, a mass, an angular momentum, and a small
non-Abelian electric charge. They do not carry non-Abelian magnetic charge,
although they possess non-trivial magnetic gauge fields
outside their regular event horizon.

Here we present a detailed account of these
rotating hairy black holes, announced in \cite{kkrot}.
We analyze their properties. In particular we discuss their
global charges and their horizon charges \cite{cs}.
We further introduce local charges, to illustrate the contributions
of the gauge fields outside the event horizon.
The black hole solutions depend on two continuous parameters,
the horizon size, and the angular velocity of the horizon.
Like their static spherically symmetric counterparts,
they further depend on the integer node number $k$ of the gauge field functions.
Thus for a given horizon size, when a small angular velocity of the horizon
is imposed, a sequence of rotating black hole solutions,
labelled by the node number $k$,
emerges from the sequence of static black hole solutions.

Furthermore, we present the expansions of the metric and gauge field functions
at the horizon and at infinity for these black holes.
The asymptotic expansion for these black holes has the surprising feature,
that the magnetic gauge field functions approach their asymptotic values
with non-integer powers of the radial coordinate.
In particular, the non-integer powers depend on the non-Abelian
electric charge. In the static limit, the non-Abelian charge vanishes
and the well-known integer power fall-off of the static spherically
symmetric functions is recovered.
%This non-integer fall-off presents an obstacle
%for the identification of a $g$-factor of these rotating hairy black holes.
%Intriguingly, (embedded) Abelian black hole solutions
%possess the $g$-factor of a Dirac particle \cite{carter}.

In section 2 we recall the SU(2) EYM action and the equations of motion.
We present the stationary ansatz for the metric and the gauge field,
and discuss its residual U(1) gauge invariance.
The global properties of the black hole solutions 
and their horizon properties are presented in section 3.
They are obtained from the expansions at infinity and at the horizon.
The expansions also suggest the set of boundary conditions
to be satisfied by the solutions at the horizon and at infinity.
The boundary conditions along the axes, follow
from symmetry and regularity conditions.
For comparison,
we present in section 4 the embedded Kerr-Newman solutions.
In particular, we transform the Kerr-Newman solutions
from Boyer-Lindquist to isotropic coordinates,
and further transform them to the gauge employed for the non-Abelian solutions.
Our numerical results are discussed in section 5.
In section 6 we present our conclusions.
Appendix A demonstrates the circularity condition,
Appendix B and C give details of the expansion at infinity
and at the origin, respectively.

\section{\bf SU(2) EYM action and stationary Ansatz}

We here briefly recall the SU(2) EYM action and the general
set of EYM equations to be satisfied by the stationary
hairy black hole solutions.
We then discuss the ansatz for the metric and the gauge field
functions \cite{kkrot}. The metric chosen is the 
stationary axially symmetric Lewis-Papapetrou metric \cite{book}
in isotropic coordinates. The ansatz for the gauge field
represents a generalization of the previously employed
static axially symmetric parameterization \cite{kkreg,kkbh}
to the stationary case,
satisfying the Ricci circularity and Frobenius conditions \cite{book}.

\subsection{\bf SU(2) EYM equations}

We consider the SU(2) Einstein-Yang-Mills action
\begin{equation}
S=\int \left ( \frac{R}{16\pi G}
  - \frac{1}{2} {\rm Tr} (F_{\mu\nu} F^{\mu\nu})
 \right ) \sqrt{-g} d^4x
\ \label{action} \end{equation}
with curvature scalar $R$,
Newton's constant $G$,
and SU(2) field strength tensor
 \begin{equation}
F_{\mu \nu} =
\partial_\mu A_\nu -\partial_\nu A_\mu + i e \left[A_\mu , A_\nu \right]
 \ , \label{fmn} \end{equation}
where $A_{\mu}$ denotes the gauge fields
\begin{equation}
A_{\mu} = \frac{1}{2} \tau^a A_\mu^a
\   \label{amu} \end{equation}
%$A_\mu = 1/2 \tau^a A_\mu^a$,
and $e$ the Yang-Mills coupling constant.
The gauge fields transform as
\begin{equation}
A_{\mu}' = U A_{\mu} U^\dagger + \frac{i}{e} (\partial_\mu U) U^\dagger
\   \label{gtgen} \end{equation}
under SU(2) gauge transformations $U$.

Variation of the action (\ref{action}) with respect to the metric
$g^{\mu\nu}$ leads to the Einstein equations
\begin{equation}
G_{\mu\nu}= R_{\mu\nu}-\frac{1}{2}g_{\mu\nu}R = 8\pi G T_{\mu\nu}
\  \label{ee} \end{equation}
with stress-energy tensor
\begin{eqnarray}
T_{\mu\nu} &=& g_{\mu\nu}L_M -2 \frac{\partial L_M}{\partial g^{\mu\nu}}
 \nonumber \\
  &=&
      2{\rm Tr}
    ( F_{\mu\alpha} F_{\nu\beta} g^{\alpha\beta}
   -\frac{1}{4} g_{\mu\nu} F_{\alpha\beta} F^{\alpha\beta})
\ , \label{tmunu}
\end{eqnarray}
variation with respect to the gauge field $A_\mu$
leads to the matter field equations,
\begin{eqnarray}
& &\frac{1}{\sqrt{-g}} D_\mu(\sqrt{-g} F^{\mu\nu}) = 0 \ .
\label{feqA} \end{eqnarray}

\subsection{\bf Stationary ansatz for the metric}

To construct rotating axially symmetric EYM black hole solutions,
we employ isotropic coordinates for the metric.
In terms of spherical coordinates $r$, $\theta$ and $\vphi$
the Lewis-Papapetrou metric is parameterized as \cite{kkrot}
\begin{equation}
ds^2 = -fdt^2+\frac{m}{f}dr^2+\frac{m r^2}{f} d\theta^2
       +\frac{l r^2 \sin^2\theta}{f}
          \left(d\vphi+\frac{\omega}{r}dt\right)^2
\ , \label{metric} \end{equation}
where the four metric functions $f$, $m$, $l$ and $\omega$
depend only on the coordinates $r$ and $\theta$.
The $z$-axis represents the symmetry axis.

The metric has Killing vector fields
$\xi = \partial_t$ and $\eta=\partial_\varphi$.
The regularity condition along the $z$-axis
\cite{book},
\begin{equation}
\frac{X,_\mu X^{,\mu}}{4X} \longrightarrow 1 \ , \ \ \
X=\eta^\mu \eta_\mu \
\ ,  \label{regcond} \end{equation}
requires
\begin{equation}
m|_{\theta=0}=l|_{\theta=0}
\ . \label{lm} \end{equation}

The ansatz for the metric, Eq.~(\ref{metric}),
satisfies the Ricci circularity conditions 
\cite{book,circ,nohair2}
\begin{equation}
 \xi^\mu R_{\mu[\alpha}\xi_\beta \eta_{\gamma]}
 = 0 =
 \eta^\mu R_{\mu[\alpha}\xi_\beta \eta_{\gamma]} 
\ ,  \label{cfc1} \end{equation}
where $R_{\mu\nu}$ is the Ricci tensor,
and the Frobenius conditions
\begin{equation}
 \xi_{[\mu} \eta_\nu \eta_{\lambda];\tau}
 = 0 =
  \eta_{[\mu} \xi_\nu \xi_{\lambda];\tau}
\ , \label{cfc2} \end{equation}
implying the corresponding conditions for the stress-energy tensor
for solutions of the EYM equations (see section 2.3.2 and Appendix A).

The event horizon of stationary black hole solutions
resides at a surface of constant radial coordinate, $r=r_{\rm H}$,
and is characterized by the condition $f(r_{\rm H})=0$ \cite{kkrot}.
The Killing vector field
\begin{equation}
\chi = \partial_t - \omega_{\rm H}/r_{\rm H} \partial_{\vphi}
\  \label{chi} \end{equation}
is orthogonal to and null on the  horizon \cite{wald}.

The ergosphere, defined as the region in which $\xi_\mu \xi^\mu$ is positive,
is bounded by the event horizon and by the surface where
\begin{equation}
 -f +\sin^2\theta \frac{l}{f} \omega^2 = 0 \ .
 \label{ergo}
\end{equation}

\subsection{\bf Stationary ansatz for the gauge field}

For the gauge field $A_\mu$ we choose the ansatz \cite{kkrot}
\begin{equation}
A_\mu dx^\mu
  =   \Psi dt +A_\vphi (d\vphi+\frac{\omega}{r} dt)
+\left(\frac{H_1}{r}dr +(1-H_2)d\theta \right)\frac{\tau_\vphi}{2e}
\ , \label{a1} \end{equation}
with
\begin{equation}
\Psi =B_1 \frac{\tau_r}{2e} + B_2 \frac{\tau_\theta}{2e}
\ , \label{a2} \end{equation}
and
\begin{equation}
A_\vphi=   -\sin\theta\left[H_3 \frac{\tau_r}{2e}
            +(1-H_4) \frac{\tau_\theta}{2e}\right] 
\ . \label{a3} \end{equation}
Here the symbols $\tau_r$, $\tau_\theta$ and $\tau_\vphi$
denote the dot products of the Cartesian vector of Pauli matrices,
$\vec \tau = ( \tau_x, \tau_y, \tau_z) $,
with the spherical spatial unit vectors,
$$
\vec e_r = (\sin \theta \cos  \vphi, \sin \theta \sin  \vphi, \cos \theta) \ ,
$$
$$
\vec e_\theta= (\cos \theta \cos \vphi, \cos \theta \sin \vphi,-\sin \theta) \ ,
$$
\begin{equation}
\vec e_\vphi= (-\sin \vphi, \cos \vphi,0) \ .
\end{equation}
The two electric gauge field functions $B_i$
and the four magnetic gauge field functions $H_i$
depend only on the coordinates $r$ and $\theta$.

In the static limit,
the ansatz reduces to the static spherically symmetric SU(2) EYM ansatz,
where
\begin{equation}
l=m \ , \ \ \ \omega=0 \ , \ \ \
H_2=H_4 \ , \ \ \
H_1=H_3=B_1=B_2=0
\ , \label{sss} \end{equation}
and all non-trivial functions depend only on the radial coordinate $r$.

\subsubsection{Residual gauge invariance}

The ansatz (\ref{a1})-(\ref{a3}) is axially symmetric in the sense,
that a rotation around the symmetry axis can be compensated
by a gauge rotation.
The ansatz is form-invariant under Abelian gauge transformations where
\cite{kkreg,kkbh,hkk,kkrot}
\begin{equation}
 U= \exp \left({\frac{i}{2} \tau_\vphi \Gamma(r,\theta)} \right)
\ .\label{gauge} \end{equation}
The functions $H_1$ and $H_2$ transform inhomogeneously
under such gauge transformations,
\begin{eqnarray}
  H_1 & \rightarrow & H_1 -  r \partial_r \Gamma \ , \nonumber \\
  H_2 & \rightarrow & H_2 +   \partial_\theta \Gamma
\ , \label{gt1} \end{eqnarray}
like a 2-dimensional gauge field.
The functions $H_3$ and $H_4$ combine to form a scalar doublet,
$(H_3+{\rm cot} \theta, -H_4)$,
\begin{eqnarray}
H_3+{\rm cot} \theta & \rightarrow & \cos\Gamma ( H_3+{\rm cot}\theta)
                                    +\sin\Gamma (-H_4) \ , \nonumber \\
-H_4    & \rightarrow & \cos\Gamma (-H_4) - \sin\Gamma( H_3+{\rm cot}\theta)
\ . \label{gt2} \end{eqnarray}
Similarly, the functions $B_1$ and $B_2$ transform as
\begin{eqnarray}
B_1 +\cos\theta \frac{\omega}{r} & \rightarrow &
 \cos\Gamma (B_1 +\cos\theta \frac{\omega}{r})
+\sin\Gamma (B_2 -\sin\theta \frac{\omega}{r}) \ , \nonumber \\
B_2 -\sin\theta \frac{\omega}{r} & \rightarrow &
 \cos\Gamma (B_2 -\sin\theta \frac{\omega}{r})
-\sin\Gamma (B_1 +\cos\theta \frac{\omega}{r}) \ .
\label{gt3} \end{eqnarray}
As previously \cite{kkreg,kkbh,kkrot},
we choose the gauge condition
\begin{equation}
r\partial_r H_1-\partial_\theta H_2 =0
\   \label{gc1} \end{equation}
with respect to this residual gauge degree of freedom.

\subsubsection{Stress-energy tensor}

Let us now address the stress-energy tensor $T_{\mu\nu}$
and the consistency of the ansatz for the gauge fields,
Eqs.~(\ref{a1})-(\ref{a3}),
with the Ricci circularity conditions.
The stress-energy is circular, when
\cite{book,circ,nohair2}
\begin{equation}
 \xi^\mu T_{\mu[\alpha}\xi_\beta \eta_{\gamma]}
 = 0 =
 \eta^\mu T_{\mu[\alpha}\xi_\beta \eta_{\gamma]} 
\ .  \label{cfc3} \end{equation}

To verify these circularity conditions for the stress-energy tensor
$T_{\mu\nu}$, Eq.~(\ref{tmunu}),
we expand the field strength tensor in the form
\begin{equation}
F_{\mu\nu} = \sum_{\alpha} F_{\mu\nu}^{(\alpha)} \frac{\tau_\alpha}{2e} \ ,
\ \ \ \ \alpha=r , \ \theta , \ \vphi \ ,
\label{fmunu}
\end{equation}
where its non-vanishing components $F_{\mu\nu}^{(\alpha)}$ are given by
\begin{eqnarray}
F_{r\theta}^{(\vphi)} & = &
-\frac{1}{r} \left[H_{1,\theta}+ r H_{2,r}\right]
\ , \nonumber \\
F_{r\vphi}^{(r)} & = &
-\frac{\sin\theta}{r}\left[rH_{3,r}-H_1 H_4\right]
\ , \nonumber \\
F_{r\vphi}^{(\theta)} & = &
 \frac{\sin\theta}{r}\left[rH_{4,r}+H_1(H_3 +{\rm cot}\theta)\right]
\ , \nonumber \\
F_{\theta\vphi}^{(r)} & = &
-\sin\theta\left[ H_{3,\theta}+H_3{\rm cot}\theta +H_2 H_4 -1 \right]
\ , \nonumber \\
F_{\theta\vphi}^{(\theta)} & = &
 \sin\theta\left[ H_{4,\theta}+{\rm cot}\theta (H_4-H_2)-H_2 H_3 \right]
\ , \nonumber \\
F_{tr}^{(r)} & = &
-\frac{1}{r}\left[rB_{1,r} + H_1 B_2
        -\frac{\omega}{r}\sin\theta(H_1(1-H_4)-H_3+rH_{3,r})
	-\omega_{,r}\sin\theta H_3\right]
\ , \nonumber \\
F_{tr}^{(\theta)} & = &
-\frac{1}{r}\left[rB_{2,r} - H_1 B_1
        +\frac{\omega}{r}\sin\theta(H_1 H_3 +(1-H_4)+rH_{4,r})
	-\omega_{,r}\sin\theta (1-H_4)\right]
\ , \nonumber \\
F_{t\theta}^{(r)} & = &
-\left[B_{1,\theta} - H_2 B_2
     +\frac{\omega}{r}\sin\theta (H_2(1-H_4)-{\rm cot}\theta H_3-H_{3,\theta})
     -\frac{\omega_{,\theta}}{r}\sin\theta H_3\right]
\ , \nonumber \\
F_{t\theta}^{(\theta)} & = &
-\left[B_{2,\theta} + H_2 B_1
     -\frac{\omega}{r}\sin\theta (H_2 H_3+{\rm cot}\theta (1-H_4)
                                     -H_{4,\theta})
     -\frac{\omega_{,\theta}}{r}\sin\theta  (1-H_4)\right]
\ , \nonumber \\
F_{t\vphi}^{(\vphi)} & = &
-\sin\theta\left[B_1 H_4 +B_2(H_3+{\rm cot}\theta)
-\frac{\omega}{r}\sin\theta ({\rm cot}\theta(1-H_4)+H_3)\right]
\ .
\label{f_comp}
\end{eqnarray}
Now it is easily seen, that $T_{tr}=T_{t\theta}=T_{\vphi r}=T_{\vphi \theta}=0$.
Thus the circularity conditions are satisfied
(see Appendix A).

Of particular interest are also the energy density of the matter fields,
$\varepsilon = -T_0^0$, and the angular momentum density $j = T_\vphi^0$.
They are given by
\begin{eqnarray}
-T_0^0 & = &
\frac{1}{2e^2r^4m}
\left\{
\frac{f^2}{m} \left(r F_{r\theta}^{(\vphi)}\right)^2
+\left(\frac{f^2}{l\sin^2\theta} -\omega^2\right)
\left[\left(r F_{r\vphi}^{(r)}\right)^2
     +\left(r F_{r\vphi}^{(\theta)}\right)^2
     +\left( F_{\theta\vphi}^{(r)}\right)^2
     +\left( F_{\theta\vphi}^{(\theta)}\right)^2
     \right]
\right.
\nonumber \\
 & &
\left.
+r^2
\left[\left(r F_{tr}^{(r)}\right)^2
     +\left(r F_{tr}^{(\theta)}\right)^2
     +\left( F_{t\theta}^{(r)}\right)^2
     +\left( F_{t\theta}^{(\theta)}\right)^2
     +\frac{m}{l\sin^2\theta}\left( F_{t\vphi}^{(\vphi)}\right)^2
     \right]
\right\} \ ,
\label{edens}
\end{eqnarray}
and
\begin{eqnarray}
T_\vphi^0 & = &
\frac{1}{r^2 m}
\left\{
r^2\left( F_{r\vphi}^{(r)}  F_{tr}^{(r)} +  F_{r\vphi}^{(\theta)}  F_{tr}^{(\theta)}  \right)
+ F_{\theta\vphi}^{(r)}  F_{t\theta}^{(r)} +  F_{\theta\vphi}^{(\theta)}  F_{t\theta}^{(\theta)}
\right.
\nonumber \\
 & &
\left.
+\frac{\omega}{r}\left[
\left(r F_{r\vphi}^{(r)} \right)^2 + \left(r F_{r\vphi}^{(\theta)} \right)^2
+\left( F_{\theta\vphi}^{(r)}   \right)^2  +\left( F_{\theta\vphi}^{(\theta)}   \right)^2
\right]
\right\} \ ,
\label{jdens}
\end{eqnarray}
respectively.
As seen from Eqs.~(\ref{f_comp}) and  (\ref{edens}),
regularity of the energy density on the $z$-axis requires
\begin{equation}
H_2|_{\theta=0}=H_4|_{\theta=0}
\ . \label{h2h4} \end{equation}

\section{Black Hole properties}

To obtain stationary axially symmetric black hole solutions
which are asymptotically flat, and possess a regular event horizon,
as well as a finite mass, angular momentum and electric charge,
we need to impose the appropriate set of boundary conditions.
Looking for black hole solutions with parity reflection symmetry,
we need to consider the solutions only
for $0 \le \theta \le \pi/2$. 
Thus appropriate boundary conditions must be imposed 
at infinity and at the horizon,
along the $\rho$-axis and along the $z$-axis 
(i.~e.~for $\theta=\pi/2$ and $\theta=0$).

In the following we present these boundary conditions
at infinity, at the horizon, and along the axes.
The choice of boundary conditions is consistent
with the equations of motion, as seen from the expansions
of the metric and gauge field functions
at infinity and at the horizon.
These expansions also allow the extraction of the physical
properties of these solutions, such as their global charges
or their horizon charges. We also introduce local charges
of the black hole solutions.

\subsection{Behaviour at infinity}

\subsubsection{Boundary conditions at infinity}

For notational simplicity we now introduce the dimensionless coordinate $x$,
\begin{equation}
x=\frac{e}{\sqrt{4\pi G}} r
\ , \label{dimless} \end{equation}
and the dimensionless electric gauge field functions
${\bar B}_1$ and ${\bar B}_2$,
\begin{equation}
{\bar B}_1 = \frac{\sqrt{4 \pi G}}{e}  B_1 \ , \ \ \
{\bar B}_2 = \frac{\sqrt{4 \pi G}}{e}  B_2 \ .
\label{barb} \end{equation}

To obtain asymptotically flat solutions, we impose
on the metric functions at infinity ($x=\infty$)
the boundary conditions
\begin{equation}
f|_{x=\infty}= m|_{x=\infty}= l|_{x=\infty}=1 \ , \ \ \
\omega|_{x=\infty}= 0
\ . \label{bc1a} \end{equation}

By requiring the four magnetic gauge field functions $H_i$ to satisfy
\begin{equation}
H_1|_{x=\infty}= H_3|_{x=\infty}= 0 \ , \ \ \
H_2|_{x=\infty}= H_4|_{x=\infty}= \pm 1
\ , \end{equation}
and the two electric gauge field functions $\bar B_i$ to satisfy
\begin{equation}
\bar B_1|_{x=\infty}= \bar B_2|_{x=\infty}= 0
\ , \label{bc1c} \end{equation}
the black hole solutions are magnetically neutral, but they may carry a
non-Abelian electric charge.

The node number $k$ of the gauge field functions
is defined by the number of nodes of
the functions $H_2$ and $H_4$ \cite{kkreg,kkbh}.
For each node number there are two degenerate solutions,
specified by the value of $H_2$ and $H_4$ at infinity, $H_2(\infty)=
H_4(\infty) = \pm 1$.
These solutions are related by the large gauge transformation
$U=\tau_r$, transforming the gauge field functions according to
$$H_1 \rightarrow -H_1 \ , \ \ \
  H_2 \rightarrow -H_2 \ , \ \ \
  H_3 \rightarrow +H_3 \ , \ \ \
  H_4 \rightarrow -H_4 $$
$$
  \bar B_1 \rightarrow +\bar B_1 \ , \ \ \
  \bar B_2 \rightarrow -\bar B_2 + \sin\theta \frac{\omega}{x}
$$
Because of this gauge symmetry
we can choose the gauge field functions
to satisfy $H_2(\infty)=H_4(\infty)=-1$.
Solutions with an odd number of nodes
then have positive values of $H_2$ and $H_4$ at the horizon,
whereas solutions with an even number of nodes
have negative values of $H_2$ and $H_4$ at the horizon.

\subsubsection{Expansion at infinity}

The asymptotic expressions for the metric and gauge field functions are

\begin{equation}
f=1-\frac{2 M}{x} + O \left(\frac{1}{x^2}\right)
\ ,\label{f_asymp_red}
\label{exif} \end{equation}

\begin{equation}
m=1 + \frac{C_1}{x^2} + \frac{Q^2-M^2-2 C_1}{x^2}\sin^2{\theta}
+ o\left( \frac{1}{x^2}\right)
\ , \label{m_asymp_red}
\label{exim} \end{equation}

\begin{equation}
l=1+\frac{C_1}{x^2} + o\left(\frac{1}{x^2}\right)
\ , \label{l_asymp_red}
\label{exil} \end{equation}

\begin{equation}
\omega=-\frac{2 a M}{x^2} + O\left(\frac{1}{x^3}\right)
\ , \label{om_asymp_red}
\label{exio} \end{equation}

\begin{equation}
H_1=\left[\frac{2 C_5}{x^2} + \frac{8 C_4}{\beta-1} x^{-\frac{1}{2} (\beta-1)}
 -\frac{2 C_2 C_3 (\alpha+3)}{ (\alpha+5)Q^2 } x^{-\frac{1}{2}(\alpha+1)}\right]
\sin{\theta} \cos{\theta} + o\left(\frac{1}{x^2}\right)
\ ,\label{H1_asymp_red}
\label{exiH1} \end{equation}

\begin{equation}
H_2=-1+C_3 x^{-\frac{1}{2}(\alpha-1)}
+ o\left(x^{-\frac{1}{2}(\alpha-1)}\right)
\ ,\label{H2_asymp_red}
\label{exiH2} \end{equation}

\begin{equation}
H_3=\left(\frac{C_2}{x}
+ C_3 x^{-\frac{1}{2}(\alpha-1)} \right) \sin{\theta} \cos{\theta}
+ o\left(\frac{1}{x}\right) 
\ ,\label{H3_asymp_red}
\label{exiH3} \end{equation}

\begin{equation}
H_4=-1+C_3 x^{-\frac{1}{2}(\alpha-1)} - \left(\frac{C_2}{x}
+ C_3 x^{-\frac{1}{2}(\alpha-1)} \right) \sin^2{\theta}
+ o\left(\frac{1}{x}\right)
\ , \label{H4_asymp_red}
\label{exiH4} \end{equation}

\begin{equation}
{\bar B}_1 = \frac{Q \cos{\theta}}{x}
+ O\left(\frac{1}{x^2}\right) 
\ , \label{B1_asymp_red}
\label{exiB1} \end{equation}

\begin{equation}
{\bar B}_2 = \frac{Q \sin{\theta}}{x}
+ O\left(\frac{1}{x^2}\right) 
\ , \label{B2_asymp_red} 
\label{exiB2} \end{equation}
where $a$, $M$, $Q$, and $C_1, \dots, C_5$ are dimensionless constants,
and $\alpha$ and $\beta$ determine the non-integer fall-off
of the magnetic gauge field functions $H_i$, with
\begin{equation}
\alpha = \sqrt{9-4 Q^2} \ , \ \ \
\beta = \sqrt{25-4 Q^2} \ .
\label{alpha-beta} \end{equation}
Further details of the asymptotic expansion are presented in Appendix B.

\subsubsection{Global charges}

Let us now obtain the global charges of the EYM solutions
from the asymptotic behaviour of the metric and gauge field functions.

The mass and the angular momentum are obtained
from the metric components $g_{tt}$ and $g_{t\varphi}$, respectively
\cite{islam}.
The asymptotic expansions for the metric functions $f$ and $\omega$,
Eqs.~(\ref{exif}) and (\ref{exio}),
yield for the dimensionless mass $M$ and the dimensionless
angular momentum $J=aM$ the expressions
\begin{equation}
 M= \frac{1}{2}\lim_{x\rightarrow\infty}x^2\partial_x f \ , \ \ \
 J = \frac{1}{2}\lim_{x\rightarrow\infty}x^2\omega
\ . \label{MJ} \end{equation}

The asymptotic behaviour of the gauge fields yields the
non-Abelian electric charge $Q^{\rm YM}$ and magnetic charge $P^{\rm YM}$
of the black hole solutions.
Let us define the gauge-invariant
non-Abelian electric charge $Q^{\rm YM}$ \cite{cs}
\begin{equation}
 Q^{\rm YM} = \frac{1}{4\pi}
 \oint \sqrt{\sum_i{\left( ^*F^i_{\theta\varphi}\right)^2}}
 d\theta d\varphi = \frac{Q}{e}
\ , \label{Qdelta} \end{equation}
where the integral is evaluated at spatial infinity,
and $ ^*F$ represents the dual field strength tensor.
Insertion of the asymptotic expansion of the gauge field functions,
Eqs.~(\ref{exiH1})-(\ref{exiB2}),
into the respective field strength tensor components, Eq.~(\ref{fmunu}),
then yields for the dimensionless non-Abelian electric charge
\begin{equation}
  Q =
\lim_{x\rightarrow\infty}x \left( \cos \theta \bar B_1
                                 +\sin \theta \bar B_2 \right)
\ , \label{Q} \end{equation}
i.~e.~the non-Abelian electric charge can be read off directly
from the asymptotic behaviour of the electric gauge field functions
$\bar B_1$ and $\bar B_2$.

Likewise, the gauge-invariant non-Abelian magnetic charge $P^{\rm YM}$
is obtained from \cite{cs}
\begin{equation}
 P^{\rm YM} = \frac{1}{4\pi}
\oint \sqrt{\sum_i{\left(F^i_{\theta\varphi}\right)^2}} d\theta d\varphi
 = \frac{P}{e}
\ , \label{Pdelta} \end{equation}
where the integral is evaluated at spatial infinity.
Again, insertion of the asymptotic expansion of the gauge field functions
into the respective field strength tensor components
yields the global charge.
For the dimensionless non-Abelian magnetic charge we thus obtain
\begin{equation}
  P = 0
\ . \label{P} \end{equation}
As imposed by the boundary conditions,
the EYM solutions carry no magnetic charge.

\subsection{Boundary conditions along the axes}

The boundary conditions along the $\rho$- and $z$-axis
($\theta=\pi/2$ and $\theta=0$) are determined by the
symmetries.
The metric functions satisfy along the axes
\begin{eqnarray}
& &\partial_\theta f|_{\theta=0} =
   \partial_\theta m|_{\theta=0} =
   \partial_\theta l|_{\theta=0} =
   \partial_\theta \omega|_{\theta=0} = 0 \ ,
\nonumber \\
& &\partial_\theta f|_{\theta=\pih} =
   \partial_\theta m|_{\theta=\pih} =
   \partial_\theta l|_{\theta=\pih} =
   \partial_\theta \omega|_{\theta=\pih} = 0 \ .
\label{bc4a}
\end{eqnarray}
For the gauge field functions symmetry considerations
lead to the boundary conditions
$$
 H_1|_{\theta=0}=H_3|_{\theta=0}=0 \ , \ \ \
   \partial_\theta H_2|_{\theta=0} =
   \partial_\theta H_4|_{\theta=0}  = 0 \ ,
$$
$$
    \bar B_2|_{\theta=0}=0 \ , \ \ \ \partial_\theta \bar B_1|_{\theta=0}=0 \ ,
$$
$$
 H_1|_{\theta=\pih}=H_3|_{\theta=\pih}=0 \ , \ \ \
    \partial_\theta H_2|_{\theta=\pih} =
    \partial_\theta H_4|_{\theta=\pih} = 0 \ ,
$$
\begin{equation}
    \bar B_1|_{\theta=\pih}=0 \ , \ \ \ 
    \partial_\theta \bar B_2|_{\theta=\pih}=0 \
\   \label{bc4c} \end{equation}
along the axes.
In addition, regularity on the $z$-axis requires condition (\ref{lm})
for the metric functions to be satisfied,
and condition (\ref{h2h4}) for the gauge field functions.

\subsection{Behaviour at the horizon}

\subsubsection{Boundary conditions at the horizon}

The event horizon of stationary black hole solutions
resides at a surface of constant radial coordinate, $x=x_{\rm H}$,
and is characterized by the condition $f(x_{\rm H})=0$ \cite{kkrot}.
Requiring the horizon to be regular, we obtain
the boundary conditions at the horizon.

At a regular horizon the metric functions must satisfy
\begin{equation}
f|_{x=x_{\rm H}}=
m|_{x=x_{\rm H}}=
l|_{x=x_{\rm H}}=0
\ , \ \ \ \omega|_{x=x_{\rm H}}=\omega_{\rm H}
\ , \label{bh2a} \end{equation}
where $\omega_{\rm H}$ is constant at the horizon.

To obtain the boundary conditions for the gauge field functions,
we start by requiring
that the electro-static potential $\Psi = (\chi^\mu A_\mu)$,
with Killing vector $\chi$, Eq.~(\ref{chi}),
is constant at the horizon \cite{isorot},
\begin{equation}
\Psi_{\rm H} = (\chi^\mu A_\mu)|_{x=x_{\rm H}} = {\rm const}
\ , \label{esp} \end{equation}
(see also Appendix C).
This yields the conditions
\begin{eqnarray}
\bar B_1 |_{x=x_{\rm H}} & = & -\cos \theta \Psi_0 \ ,
\nonumber \\
\bar B_2 |_{x=x_{\rm H}} & = & \ \ \ \sin \theta \Psi_0 \ .
\label{BC_Q}
\end{eqnarray}
The equations of motion, Eqs.~(\ref{feqA}), then
yield for non-Abelian solutions the conditions
\begin{equation}
\Psi_0 = \omega_{\rm H}/x_{\rm H}
\ , \label{Psi_0} \end{equation}
\begin{equation}
\partial_x \omega |_{x=x_{\rm H}} = \omega_{\rm H}/x_{\rm H}
\ , \label{newcond} \end{equation}
as well as
\begin{equation}
(x\partial_x \bar B_1) |_{x=x_{\rm H}}= 0  \ , \ \
(x\partial_x \bar B_2) |_{x=x_{\rm H}}= 0  \ ,
\label{BC_b}
\end{equation}
and, taking these conditions into account, 
\begin{eqnarray}
(\partial_\theta H_1 + x\partial_x H_2)|_{x=x_{\rm H}} & = & 0  \ ,
\nonumber \\
(x\partial_x H_3 -H_1H_4)|_{x=x_{\rm H}} & = & 0  \ ,
\nonumber \\
(x\partial_x H_4 +H_1 (H_3+{\rm cot}\theta))|_{x=x_{\rm H}} & = & 0  \ ,
\label{BC_h}
\end{eqnarray}
which derive from $F_{r\theta}^{(\vphi)}$,
$F_{r\phi}^{(r)}$, and $F_{r\phi}^{(\theta)}$, respectively.
Thus the regularity conditions at the horizon imply
\begin{equation}
\chi^\mu F_{\mu\nu}|_{x=x_{\rm H}} = 0 \ , \ \ \ \
F_{r\theta}|_{x=x_{\rm H}}= 0 \ , \ \ \ \
F_{r\vphi}|_{x=x_{\rm H}}= 0 \ .
\end{equation}

Furthermore, the gauge condition, Eq.~(\ref{gc1}), implies
\begin{eqnarray}
(x\partial_x H_1 -\partial_\theta H_2)|_{x=x_{\rm H}} & = & 0  \ .
\nonumber 
\label{BC_gauge}
\end{eqnarray}
However, for black hole solutions
the gauge condition (\ref{gc1})
still allows for non-trivial gauge transformations satisfying
\begin{equation}
x^2 \partial^2_x \Gamma
+x \partial_x \Gamma
+  \partial^2_\theta \Gamma = 0
\ . \label{gfree} \end{equation}
To fix the gauge,
we choose the additional condition \cite{kkreg,kkbh,kkrot}
\begin{equation}
(\partial_\theta H_1) |_{x=x_{\rm H}} = 0 ,
\   \label{gfree2} \end{equation}
which implies ${\DS H_1 |_{x=x_{\rm H}} = 0}$, when we take into account
the boundary condition on the axes, ${\DS H_1|_{\theta=0,\pi/2}=0}$,
Eq.~(\ref{bc4c}).

For the numerical solutions we then impose on the gauge field functions
the set of boundary conditions
%\begin{equation}
$$
           H_1 |_{x=x_{\rm H}}= 0 \ , \ \ \
\partial_x H_2 |_{x=x_{\rm H}}= 0 \ , \ \ \
\partial_x H_3 |_{x=x_{\rm H}}= 0 \ , \ \ \
\partial_x H_4 |_{x=x_{\rm H}}= 0 \ ,
%\label{bhbc1} \end{equation}
$$
\begin{equation}
 x_{\rm H} \bar B_1 |_{x=x_{\rm H}}  = -\cos\theta \omega_{\rm H} \ , \ \ \
 x_{\rm H} \bar B_2 |_{x=x_{\rm H}} = \sin\theta \omega_{\rm H}
 \ . \label{bhbc2}
\end{equation}

\subsubsection{Expansion at the horizon}

Expanding the metric and gauge field functions at the horizon
in powers of
\begin{equation}
\delta=\frac{x}{x_{\rm H}}-1
\end{equation}
yields
\begin{equation}
f(\delta,\theta)=\delta^2 f_2 (1 -\delta) + O(\delta^4) \ , \label{f_hor_red}
\end{equation}
\begin{equation}
m(\delta,\theta)=\delta^2 m_2 (1 -3\delta) + O(\delta^4) \ , \label{m_hor_red}
\end{equation}
\begin{equation}
l(\delta,\theta)=\delta^2 l_2 (1 -3\delta) + O(\delta^4) \ , \label{l_hor_red}
\end{equation}
\begin{equation}
\omega(\delta,\theta)=\omega_{\rm H} (1 + \delta) + O(\delta^2) \ ,\label{om_hor_red}
\end{equation}
\begin{equation}
H_1(\delta,\theta)=\delta \left(1 -\frac{1}{2}\delta \right) H_{11}
+ O(\delta^3) \ , \label{H1_hor_red}
\end{equation}
\begin{equation}
H_2(\delta,\theta)=H_{20}+O(\delta^2) \ ,\label{H_2_hor_red}
\end{equation}
\begin{equation}
H_3(\delta,\theta)=H_{30} + O(\delta^2) \ ,\label{H3_hor_red}
\end{equation}
\begin{equation}
H_4(\delta,\theta)=H_{40}+ O(\delta^2) \ ,\label{H4_hor_red}
\end{equation}
\begin{equation}
{\bar B}_1(\delta,\theta)=-\frac{\omega_{\rm H} \cos{\theta}}{x_{\rm H}}
+ O(\delta^2) \ ,\label{B1_hor_red}
\end{equation}
\begin{equation}
{\bar B}_2(\delta,\theta)=\frac{\omega_{\rm H} \sin{\theta}}{x_{\rm H}}
+ O(\delta^2) \ .\label{B2_hor_red}
\end{equation}
The expansion coefficients $f_2$, $m_2$, $l_2$, $H_{11}$,
$H_{20}$, $H_{30}$, and $H_{40}$ are functions of the variable $\theta$.
Among these coefficients the following relations hold,
\begin{equation}
0=\frac{\partial_{\theta}m_2}{m_2}
-2 \frac{\partial_{\theta}f_2}{f_2}
\ , \label{relation_hor_1} \end{equation}
\begin{equation}
H_{11}=\partial_{\theta} H_{20}
\ . \label{relation_hor_2} \end{equation}
Further details of the expansion at the horizon
are given in Appendix C.

\subsubsection{Horizon properties}

Let us now obtain the horizon properties of the EYM solutions
from the expansion at the horizon of the metric and gauge field functions.

The first quantity of interest is the area of the horizon. The dimensionless
area $A$ is given by
\begin{equation}
A = 2 \pi \int_0^\pi  d\theta \sin \theta
\frac{\sqrt{l_2 m_2}}{f_2} x_{\rm H}^2
\ . \label{area} \end{equation}
The area $A$ of the black hole horizon
defines the area parameter $x_\Delta$ via
\begin{equation}
A = 4 \pi x_\Delta^2 \ .
\label{xDelta} \end{equation}
The entropy $S$ of the black hole then corresponds to
\begin{equation}
S = \frac{A}{4} \ .
\label{entro} \end{equation}

To obtain a measure for the deformation of the horizon we
compare the dimensionless circumference of the horizon along the equator, $L_e$,
with the dimensionless circumference of the horizon along the poles, $L_p$,
\begin{equation}
L_e = \int_0^{2 \pi} { d \vphi \left.
 \sqrt{ \frac{l}{f}} x \sin\theta
 \right|_{x=x_{\rm H}, \theta=\pi/2} } \ , \ \ \
L_p = 2 \int_0^{ \pi} { d \theta \left.
 \sqrt{ \frac{m  }{f}} x
 \right|_{x=x_{\rm H}, \vphi={\rm const.}} }
\ , \label{lelp} \end{equation}
and consider, in particular, their ratio $L_e/L_p$.

The surface gravity of the black hole solutions is obtained from \cite{wald}
\begin{equation}
\kappa^2_{\rm sg} = -1/4 (D_\mu \chi_\nu)(D^\mu \chi^\nu)
\ , \label{sgwald} \end{equation}
with Killing vector
$\chi = \xi -(\omega_{\rm H}/x_{\rm H}) \eta$.
Inserting the expansion in $\delta = (x/x_{\rm H}-1)$
at the horizon, Eqs.~(\ref{f_hor_red})-(\ref{om_hor_red}),
yields for the dimensionless surface gravity
\begin{equation}
\kappa_{\rm sg} = \frac{f_2(\theta)}{x_{\rm H} \sqrt{m_2(\theta)}}
\ . \label{temp} \end{equation}
As seen from Eq.~(\ref{relation_hor_1}),
$\kappa_{\rm sg}$ is indeed constant on the horizon,
as required by the zeroth law of black hole mechanics.
The dimensionless temperature $T$ 
of the black hole is proportional to the surface gravity,
\begin{equation}
T = \frac{\kappa_{\rm sg}}{2 \pi} 
\ . \label{tempt} \end{equation}

Let us now consider the Yang-Mills horizon charges.
By evaluating the integrals, Eqs.~(\ref{Qdelta}) and (\ref{Pdelta}),
at the horizon, we obtain the horizon electric charge $Q_\Delta$
and the horizon magnetic charge $P_\Delta$ \cite{cs},  respectively,
\begin{equation}
 Q_\Delta = \frac{e}{4\pi}
 \oint \sqrt{\sum_i{\left(^*F^i_{\theta\varphi}\right)^2}}
 d\theta d\varphi
\ , \label{Qdelta1} \end{equation}
\begin{equation}
 P_\Delta = \frac{e}{4\pi}
\oint \sqrt{\sum_i{\left(F^i_{\theta\varphi}\right)^2}} d\theta d\varphi
\ , \label{Pdelta1} \end{equation}
where $Q_\Delta$ and $P_\Delta$ again represent the dimensionless quantities.
%For the non-Abelian solutions, $Q_\Delta$ and $P_\Delta$ differ from
%$Q^{\rm YM}$ and $P^{\rm YM}$, respectively. 
%$Q_\Delta \ne Q$ and $P_\Delta \ne 0$.

\subsection{Local charges}

To define the local mass let us first consider 
the general definition of  the total mass
(see e.~g.~\cite{wald})
\begin{equation}
{\cal M} = 2 \int_\Sigma \left( T_{\mu\nu} -\frac{1}{2} T g_{\mu\nu}\right)
                        n^\mu \xi^\nu dV
 -\frac{1}{8\pi G} \int_{\rm H} \frac{1}{2}\varepsilon_{\mu\nu\rho\sigma} 
\nabla^\rho \xi^\sigma dx^\mu dx^\nu 
\ , \label{Mtot} \end{equation}
where
$\Sigma$ denotes an asymptotically flat hypersurface 
bounded by the horizon ${\rm H}$,
$dV$ is the natural volume element on $\Sigma$,
$n^\mu = (1, 0, 0, -\omega/r)/\sqrt{f}$ 
is normal to $\Sigma$ with $n_\mu n^\mu = -1$,
and $\xi^\nu$ denotes the time like Killing vector field.

Note, that $T=0$ for the action, Eq.~(\ref{action}).
Straightforward calculation yields
$$
T_{\mu\nu}   n^\mu \xi^\nu dV   = - T_0^0 \sqrt{-g} dr d\theta d\vphi
 = -\frac{1}{ 8 \pi G} R_0^0 \sqrt{-g} dr d\theta d\vphi  \  ,
$$
where in the last step the Einstein equations have been used.

Integration over $\Sigma$ yields
$$
 -\frac{1}{ 8 \pi G} \int_\Sigma  R_0^0 \sqrt{-g} dr d\theta d\vphi      =
 \frac{2\pi}{4\pi G} \int_0^{\frac{ \pi}{2} }
 \left. \left[ \frac{\sqrt{l}}{f}  r^2  \sin\theta
 \left( \frac{\partial f}{\partial r} -
 \frac{l}{f} \sin^2\theta \omega
 \left(\frac{\partial \omega}{\partial r} - \frac{\omega}{r}  \right)
 \right)
 \right]  \right|_{r_{\rm H}}^{\infty}  d\theta
 \ .
$$
The calculation of the boundary term in Eq.~(\ref{Mtot}) 
yields the same expression, but evaluated at $r= r_{\rm H} $.

Consequently ,
\begin{eqnarray}
 {\cal M}  & = & \lim_{r\to\infty}
 \frac{2\pi }{4\pi G} \int_0^{\frac{\pi}{2}}
 \left. \left[\frac{\sqrt{l}}{f}  r^2  \sin\theta
 \left( \frac{\partial f}{\partial r} -
 \frac{l}{f} \sin^2\theta \omega
 \left(\frac{\partial \omega}{\partial r} - \frac{\omega}{r}   \right)
 \right)
 \right]  \right|_{r} d\theta
 \nonumber\\
 & = &
 \lim_{x\to\infty}\frac{\sqrt{4\pi G}}{e}\frac{1}{G} \frac{1}{2}
 \int_0^{\frac{\pi}{2}}
 \left. \left[ \frac{\sqrt{l}}{f}  x^2  \sin\theta
 \left( \frac{\partial f}{\partial x} -
 \frac{l}{f} \sin^2\theta \omega
 \left(\frac{\partial \omega}{\partial x} - \frac{\omega}{x}    \right)
 \right)
 \right]  \right|_{x} d\theta
 \nonumber\\
 & = &
 \lim_{x\to\infty}\frac{\sqrt{4\pi G}}{e}\frac{1}{G}  M(x)  \ ,
 \nonumber
 \end{eqnarray}
where we changed to the dimensionless coordinate $x$, Eq.~(\ref{dimless}),
%$x = r  e/\sqrt{4\pi G} $ 
and defined the dimensionless local mass $M(x)$,
\begin{eqnarray}
M(x)&\!\!\!=\!\!\!&\frac{1}{2}
\int_0^{\frac{\pi}{2}}  \frac{\sqrt{l}}{f}  x^2   \sin \theta
\left( \frac{\partial f}{\partial x}
-\frac{l}{f} \sin^2 \theta \omega
 \left(\frac{\partial \omega}{\partial x} - \frac{\omega}{x}    \right)
\right) d\theta \ .
\label{localM} \end{eqnarray}

In a similar way we obtain the dimensionless local angular momentum $J(x)$,
\begin{eqnarray}
J(x)&\!\!\!=\!\!\!&\frac{1}{4}x^2
\int_0^{\frac{\pi}{2}} \sin^3\theta \frac{l^{3/2}}{f^2}
\left(x \frac{\partial\omega}{\partial x} -\omega\right) d\theta \ .
\label{localJ} \end{eqnarray}
from the expression for the total angular momentum,
$$
 {\cal J} = -\left[
 \frac{1}{8 \pi G}    \int_\Sigma R_{\mu\nu}      n^\mu \eta^\nu dV
 -\frac{1}{16\pi G} \int_{\rm H} \frac{1}{2}
 \varepsilon_{\mu\nu\rho\sigma} \nabla^\rho \eta^\sigma
 dx^\mu dx^\nu \right]  \ .
$$

Defining the horizon mass of the black hole
$M_\Delta = M(x_{\rm H})$ and its horizon angular momentum
$J_\Delta = J(x_{\rm H})$,
we obtain the relation
\begin{equation}
M_\Delta = 2  TS - 2 \frac{\omega_{\rm H}}{x_{\rm H}} J_\Delta
\ . \label{relmassd} \end{equation}

We further define the dimensionless local non-Abelian electric 
and magnetic charges, $Q(x)$, and $P(x)$,
\begin{equation}
 Q(x) = \frac{e}{4\pi}
 \oint \sqrt{\sum_i{\left( ^*F^i_{\theta\varphi}\right)^2}}
 d\theta d\varphi
\ ,  \end{equation}
and
\begin{equation}
 P(x)= \frac{e}{4\pi}
\oint \sqrt{\sum_i{\left(F^i_{\theta\varphi}\right)^2}} d\theta d\varphi
\ ,  \end{equation}
respectively,
where the integrals are evaluated on surfaces with fixed radial coordinate $x$.
Explicitly,
\begin{eqnarray}
Q(x)&\!\!\!=\!\!\!&\int_0^{\frac{\pi}{2}} \sin\theta
\frac{\sqrt{l}}{f} \Bigg\{ \Bigg[-x\sin^2\theta
\frac{\partial \omega}{\partial x}H_3
+ \sin\theta \cos\theta \omega (1-H_4)
- \nonumber\\&&x \sin\theta \cos\theta
\frac{\partial \omega}{\partial x} (1-H_4)+\sin^2\theta \omega H_3
-x H_1 (\cos\theta {\bar B}_1 -\sin\theta {\bar B}_2) -\nonumber\\&&
\omega H_1+x^2\left(\sin\theta \frac{\partial {\bar B}_1}{\partial x}
+ \cos\theta \frac{\partial {\bar B}_2}{\partial x}\right)\Bigg]^2
+\Bigg[ -x \sin\theta \cos\theta \frac{\partial \omega}{\partial x}H_3
-\nonumber\\&&\sin^2\theta \omega (1-H_4)
+ x \sin^2\theta \frac{\partial \omega}{\partial x} (1-H_4)
+ \sin\theta \cos\theta \omega H_3+ \nonumber \\ &&
x H_1 (\sin\theta {\bar B}_1 + \cos\theta {\bar B}_2)
+ x^2\left(\cos\theta \frac{\partial {\bar B}_1}{\partial x}
- \sin\theta \frac{\partial {\bar B}_2}{\partial x}\right)
\Bigg]^2\Bigg\}^{1/2} d\theta \ ,
\label{localQ} \end{eqnarray}
\begin{eqnarray}
P(x)&\!\!\!=\!\!\!&\int_0^{\frac{\pi}{2}}\Bigg\{ \Bigg[
-\sin^2\theta + \sin\theta \cos\theta H_3 (1+H_2)
+ \cos^2\theta (H_2-H_4)
+\nonumber\\&& \sin^2\theta \frac{\partial H_3}{\partial \theta}
-\sin\theta \cos\theta \frac{\partial H_4}{\partial \theta}
+ \sin^2\theta H_2 H_4 \Bigg]^2
+\Bigg[ -\sin\theta \cos\theta-\nonumber\\&&\sin^2\theta H_3 (1+H_2)
+H_3-\sin\theta \cos\theta (H_2-H_4)
+\sin\theta \cos\theta \frac{\partial H_3}{\partial \theta}
+\nonumber\\&&\sin^2\theta \frac{\partial H_4}{\partial \theta}
+\sin\theta \cos\theta H_2 H_4 \Bigg]^2 \Bigg\}^{1/2} d\theta \ .
\label{localP} \end{eqnarray}

\section{\bf Embedded Kerr-Newman Black Holes}

Here we briefly recall Kerr-Newman black holes,
embedded in SU(2) EYM theory \cite{perry}.
For better comparison with the non-Abelian rotating black holes,
we first consider the coordinate transformation 
from Boyer-Lindquist coordinates to isotropic coordinates.
Then we perform an SU(2) gauge transformation to
the gauge employed for the non-Abelian black holes.

\subsection{\bf Boyer-Lindquist coordinates}

In Boyer-Lindquist coordinates the metric of Kerr-Newman black holes
is given by
\begin{equation}
ds^2 = -\frac{\Delta}{\rho^2} \left(dt + a \sin^2 \theta d \varphi \right)^2
  + \frac{\sin^2\theta}{\rho^2} \left( a dt + \rho_0^2 d \varphi \right)^2
  + \frac{\rho^2}{\Delta} d \tilde x^2 + \rho^2 d\theta^2
\ , \label{KN0} \end{equation}
where
\begin{equation}
\rho^2 = \tilde x^2+a^2\cos^2\theta \ , \ \ \
\rho^2_0 = \tilde x^2+a^2 \ , \ \ \  \Delta = \tilde x^2-2M \tilde x+a^2+C^2
\ , \label{KN1} \end{equation}
and $M$ denotes the black hole mass, $a$ the angular momentum
per unit mass, $a=J/M$, and $C$ the ``total charge''
(see Eq.~(\ref{tq})).

The gauge field of embedded Kerr-Newman solutions is given by \cite{perry}
\begin{equation}
A_\mu^a dx^\mu =
\frac{ Q^a \tilde x}{\rho^2} \left(
 dt + a \sin^2 \theta d \varphi \right)
 + \frac{ P^a \cos \theta }{\rho^2} \left(
 a dt + \rho_0^2 d \varphi \right)
\ , \label{KN2} \end{equation}
where $Q^a$ and $P^a$ are constant vectors in the Lie algebra,
considered the Yang-Mills analogue of the electric and
magnetic charge, respectively \cite{perry,foot0}.
They define the ``total charge'' $C$,
\begin{equation}
C^2 = Q^aQ^a + P^a P^a
\ , \label{tq} \end{equation}
and $Q^a$ is proportional to $P^a$ \cite{perry}.

The condition $\Delta(\tilde x_{\rm H})=0$ yields
the regular event horizon of the Kerr-Newman solutions,
\begin{equation}
\tilde x_{\rm H} = M + \sqrt{M^2 - (a^2+C^2)}
\ , \label{KN3} \end{equation}
extremal Kerr-Newman solutions satisfy
$\tilde x_{\rm H} = M = \sqrt{a^2+C^2}$.

\subsection{\bf Isotropic coordinates}

Let us now consider the Kerr-Newman solution
in isotropic coordinates.
For Kerr-Newman solutions
the isotropic radial coordinate $x$ is related to the Boyer-Lindquist
radial coordinate $\tilde x$ by
\begin{equation}
x(\tilde x)= \frac{1}{2}\left[\sqrt{\tilde x^2-2M \tilde x+(a^2+C^2)}
 +(\tilde x-M)\right]
\ , \label{KN4} \end{equation}
and the metric functions $f(x,\theta)$, $m(x,\theta)$,
$l(x,\theta)$ and $\omega(x,\theta)$ (see Eq.~(\ref{metric}))
are given by
\begin{equation}
f(x,\theta)= \frac{\rho^2 \Delta}{\rho_0^4 - a^2 \sin^2 \theta \Delta}
\ , \label{KN6} \end{equation}
\begin{equation}
\frac{m(x,\theta)}{x} = \frac{\tilde x^2 + a^2 \cos^2 \theta}{x^2}
\ , \label{KN7} \end{equation}
\begin{equation}
\frac{x^2 l(x,\theta)}{f(x,\theta)} =
 \rho_0^2 +a^2\sin^2\theta
              \frac{2M \tilde x-C^2}{\rho^2}
\ , \label{KN8} \end{equation}
and
\begin{equation}
\frac{\omega(x,\theta)}{x} =
 \frac{a(2M \tilde x-C^2)}{\rho_0^4 - a^2 \sin^2 \theta \Delta}
\ , \label{KN9} \end{equation}
where
\begin{equation}
  \tilde x(x) = x + M + \frac{M^2 - (a^2+C^2)}{4 x}
\ . \label{KN14} \end{equation}

In isotropic coordinates the event horizon resides at
\begin{equation}
x_{\rm H} = 1/2 \sqrt{M^2 - (a^2+C^2)}
\ , \label{KN5} \end{equation}
thus extremal Kerr-Newman solutions satisfy $x_{\rm H} = 0$.
At the event horizon, the metric function $f$ is identically zero,
the metric function ratios $(m/f)$ and $(l/f)$ are finite,
and the metric function $\omega$ is constant,
\begin{equation}
\omega(x_{\rm H})=\frac{a  x_{\rm H}}{4M x_{\rm H}+2M^2-C^2} = \omega_{\rm H}
\ . \end{equation}
Consequently the Kerr-Newman solutions satisfy
\begin{equation}
\frac{\omega_{\rm H}}{x_{\rm H}}
=\frac{\sqrt{M^2-4 x_{\rm H}^2-C^2}} {2M(M+2x_{\rm H})-C^2}
=\frac{a}{(M+2 x_{\rm H})^2+a^2}
\ . \label{KNrel} \end{equation}

\subsection{\bf Gauge transformation}

Let us now consider a Kerr-Newman solution with
Lie algebra vectors $Q^a$ and $P^a$ pointing in the $z$-direction,
%i.e.~$Q^a = \delta^{a,z} Q$, $P^a = \delta^{a,z} P$.
\begin{equation}
A_\mu dx^\mu =
\frac{ Q \tilde x}{\rho^2} \left(
 dt + a \sin^2 \theta d \varphi \right)
 \frac{\tau_z}{2}
 + \frac{ P \cos \theta }{\rho^2} \left(
 a dt + \rho_0^2 d \varphi \right)
 \frac{\tau_z}{2}
\ . \label{KN15} \end{equation}
To obtain the solution in the gauge employed for the non-Abelian ansatz
(\ref{a1})-(\ref{a3}),
we perform a gauge transformation, where
\begin{equation}
U(\theta,\varphi) = 
                 e^{-i \frac{\varphi}{2} \tau_z}
                 e^{-i \frac{\theta}{2} \tau_y}
\ . \label{gttp}\end{equation}
The gauge transformed vector field reads
\begin{equation}
{A'}_\mu dx^\mu =
 \left( \frac{ Q \tilde x}{\rho^2}
 + \frac{ P \cos \theta }{\rho^2} a \right)
 \frac{\tau_r}{2} d t
 + A_\varphi d \varphi
 +\frac{\tau_\varphi}{2} d \theta
\ \label{KN10}\end{equation}
with
\begin{equation}
 A_\varphi =
 \left(
 \frac{ Q \tilde x}{\rho^2} a \sin^2 \theta
 + \frac{ P \cos \theta }{\rho^2} \rho_0^2
 + \cos{\theta} \right) \frac{\tau_r}{2}
 -\sin{\theta} \frac{\tau_\theta}{2}
\ . \label{KN11} \end{equation}

By comparing this embedded Abelian expression for the gauge fields
with the non-Abelian ansatz, Eqs.~(\ref{a1})-(\ref{a3}),
we obtain the embedded Abelian gauge field functions
$H_i$ and $\bar B_i$,
\begin{equation}
%H_1=H_2=H_4=0 \ , \ \ \
%-\sin\theta H_3 =  \cos \theta \left( P \frac{\rho_0^2}{\rho^2} + 1 \right)
 H_1=H_2=H_4=0 \ , \ \ \
 -\sin\theta H_3 =  
 \frac{ Q \tilde x}{\rho^2} a \sin^2 \theta 
 + \frac{ P \cos \theta }{\rho^2} \rho_0^2
 + \cos{\theta}
\ , \label{KN12} \end{equation}
\begin{equation}
 \bar B_1  = \sin{\theta}\, \frac{\omega}{x}\, H_3 +
\frac{Q \tilde x + P \cos{\theta} a}{\rho^2} \ , \ \ \
%\cos \theta \left( P \frac{a}{\rho^2}
%- P \frac{\rho_0^2}{\rho^2} \frac{\omega}{x}
%- \frac{\omega}{x}
%\right)
%\ , \ \ \
 \bar B_2 = \sin{\theta}\, \frac{\omega}{x}
\ . \label{KN13} \end{equation}

The asymptotic expansion for the Abelian gauge field and metric functions 
is given in Appendix B, the expansion at the horizon in Appendix C.

\section{Numerical Results} 

The rotating hairy black hole solutions of SU(2) EYM theory emerge from the
static hairy black hole solutions, when a small value of the
angular velocity of the horizon is imposed
via the boundary conditions.
We therefore briefly recall the properties of the
static hairy black hole solutions of SU(2) EYM theory \cite{su2bh}.

The static hairy black hole solutions of SU(2) EYM theory
carry non-trivial magnetic gauge fields outside their
regular event horizon, but carry no non-Abelian magnetic charge.
Since their electric fields vanish identically,
their only global charge is their mass.
These black holes can be characterized by their mass and by two integers,
the azimuthal winding number $n$ of their gauge fields \cite{foot1}
and the node number $k$ of their 
magnetic gauge field functions \cite{su2bh,kkbh,kksw}.
For a given winding number $n$, the solutions form sequences
labelled by the node number $k$. With increasing $k$,
the non-Abelian black hole solutions approach a limiting solution,
corresponding to an embedded Abelian solution with magnetic charge $n$
\cite{foot2}.

In contrast to the static hairy black hole solutions,
their rotating generalizations carry electric gauge fields.
Notably, the rotating hairy black hole solutions possess
a non-Abelian electric charge \cite{vs,kkrot},
whereas their static counterparts are neutral.
Thus their global charges are their mass, their angular momentum
and their non-Abelian electric charge.
Their non-Abelian magnetic charge is identically zero.
Like their static counterparts, the rotating hairy black hole solutions
form sequences depending on the winding number $n$ and the node number
$k$ of the gauge fields.

Here we consider the sequence of rotating hairy black hole solutions
with winding number $n=1$ and node number $k$.
In the static limit these black hole solutions approach 
spherically symmetric SU(2) EYM black hole solutions \cite{su2bh}.
Rotating black hole solutions with higher winding number $n$
approach in the static limit
SU(2) EYM black hole solutions, which possess only axial symmetry \cite{kkbh}.
Such black hole solutions will be considered elsewhere.

For a given node number $k$, the rotating hairy black hole solutions
depend on the isotropic horizon radius $x_{\rm H}$
%(or the area parameter $x_\Delta$),
and on the value of the metric function $\omega$ at the horizon,
$\omega_{\rm H}$,
via the boundary conditions.
The ratio $\omega_{\rm H}/x_{\rm H}$ represents the rotational velocity
of the horizon.

When the black hole solutions are constructed,
their global charges are obtained from the asymptotic fall-off
of the metric and gauge field functions of the solutions.
Likewise, their horizon properties are obtained from
the expansion at the horizon.

To construct the rotating hairy black hole solutions,
we solve the set of ten coupled non-linear
elliptic partial differential equations numerically \cite{schoen},
subject to the above boundary conditions.
We employ compactified dimensionless coordinates,
mapping spatial infinity to the finite value $\bar x=1$,
where
\begin{equation}
\bar x = 1-\frac{x_{\rm H}}{x}
\ . \label{barx2} \end{equation}
We furthermore introduce the functions 
$\bar{f}(\bar x,\theta)$,
$\bar{l}(\bar x,\theta)$,
$\bar{m}(\bar x,\theta)$, and
$g(\bar x,\theta)$
\cite{kkreg,kkbh,hkk},
\begin{equation}
\bar{f}(\bar x,\theta)=\frac{f(\bar x,\theta)}{\bar x^2} \ , \ \ \
\bar{l}(\bar x,\theta)=\frac{l(\bar x,\theta)}{\bar x^2} \ , \ \ \
\bar{m}(\bar x,\theta)=\frac{m(\bar x,\theta)}{\bar x^2} \ , \ \ \
g(\bar x,\theta)=\frac{m(\bar x,\theta)}{l(\bar x,\theta)} \ .
\label{barflm} \end{equation}
The numerical calculations, based on the Newton-Raphson method,
are performed with help of the program FIDISOL \cite{schoen}.
The equations are discretized on a non-equidistant grid in $\bar x$ and  $\theta$.
Typical grids used have sizes $100 \times 20$, covering the integration region
$0\leq \bar x\leq 1$ and $0\leq\theta\leq\pi/2$.
(See \cite{kkreg,kkbh,hkk} and \cite{schoen}
for further details on the numerical procedure.)

\subsection{Global charges}

To construct a rotating black hole solution with one node
and horizon radius $x_{\rm H}$,
we start from the static spherically symmetric
SU(2) EYM black hole solution with one node and
the same horizon radius,
which represents the limiting solution in the static limit
$\omega_{\rm H} \rightarrow 0$.
By imposing a small but finite value of $\omega_{\rm H}$
via the boundary conditions,
a rotating black hole solution is obtained,
which possesses non-trivial functions $\omega$, $H_1$, $H_3$, 
$\bar B_1$, $\bar B_2$.
The rotating black hole solution is only axially symmetric.
It possesses a non-vanishing electric gauge field,
giving rise to a non-Abelian electric charge \cite{vs,kkrot}.

When increasing $\omega_{\rm H}$ from zero, while keeping $x_{\rm H}$ fixed,
a branch of black hole solutions forms, the lower branch.
The lower branch extends up to a maximal value $\omega^{\rm max}_{\rm H}$,
where a second branch,
the upper branch, bends backwards towards $\omega_{\rm H}=0$.
Along both branches the mass $M$, the angular momentum $J$, and the
non-Abelian electric charge $Q$ continuously increase \cite{kkrot}.
This is seen in Figs.~1a-c,
where the mass $M$,
the angular momentum per unit mass $a=J/M$
and the non-Abelian electric charge $Q$
are shown as functions of the parameter $\omega_{\rm H}$
for three values of the isotropic horizon radius,
$x_{\rm H}=0.1$, 0.5 and 1.
%(The calculations along the upper branch have been continued
%to smaller values of $\omega_{\rm H}$
%with respect to those of Ref.~\cite{kkrot}).

Whereas both mass $M$ and angular momentum per unit mass $a$
of the non-Abelian solutions increase strongly
along the upper branch, diverging with $\omega_{\rm H}^{-1}$ in the
limit $\omega_{\rm H} \rightarrow 0$,
the non-Abelian electric charge $Q$ remains small.
It approaches apparently a finite limiting value,
$Q_{\rm lim} \approx 0.124$
(as discussed in section 5.3),
independent of the isotropic horizon radius $x_{\rm H}$.

For comparison, we exhibit in Figs.~1a-b also
the mass and the angular momentum of embedded Abelian solutions
with the same horizon radii: those of the Kerr solutions 
and those of the Kerr-Newman solutions with $Q=0$ and $P=-1$.
(Mass and angular momentum of embedded Kerr-Newman solutions,
possessing the same charge $Q$ as the non-Abelian solutions but with $P=0$
are graphically indistinguishable from the mass and angular momentum
of the Kerr solutions, shown.
Likewise, mass and angular momentum of embedded Kerr-Newman solutions,
possessing the same charge $Q$ as the non-Abelian solutions and $P=-1$
are graphically indistinguishable from the mass and angular momentum
of the Kerr-Newman solutions, shown.)

Interestingly, both mass and angular momentum of the non-Abelian solutions,
which carry a small electric charge $Q$ and no magnetic charge,
are close to mass and angular momentum of the embedded
Kerr-Newman solutions with magnetic charge $P=-1$.
This is seen in particular
for solutions with small horizon radii $x_{\rm H}$,
where large deviations of these global properties from 
those of the corresponding Kerr solutions arise.
For non-Abelian black hole solutions with large horizon radii,
these global charges are also close to those of the Kerr solutions.
Here a magnetic charge of magnitude $|P|=1$
is less important for the global properties of the solutions.
Thus also Kerr and Kerr-Newman solutions differ less for large horizon radii.

As a point in case, let us inspect the
maximal value of $\omega_{\rm H}$ reached, $\omega_{\rm H}^{\rm max}$.
The maximal value $\omega_{\rm H}^{\rm max}$ 
of the non-Abelian black hole solutions
depends on the horizon radius $x_{\rm H}$,
and increases monotonically with $x_{\rm H}$.
For the solutions shown in Figs.~1a-c,
$\omega^{\rm max}_{\rm H}( x_{\rm H} =0.1) \approx 0.0288$,
$\omega^{\rm max}_{\rm H}( x_{\rm H} =0.5) \approx 0.0649$, and
$\omega^{\rm max}_{\rm H}( x_{\rm H} =1.0) \approx 0.0719$.
For Kerr-Newman solutions with $Q=0$ and $P=-1$
very similar values for $\omega^{\rm max}_{\rm H}$ are obtained,
$\omega^{\rm max}_{\rm H}( x_{\rm H} =0.1) \approx 0.0278$,
$\omega^{\rm max}_{\rm H}( x_{\rm H} =0.5) \approx 0.0646$, and
$\omega^{\rm max}_{\rm H}( x_{\rm H} =1.0) \approx 0.0718$.
In contrast, for the Kerr solutions the maximal value of $\omega_{\rm H}$,
$$\omega_{\rm H}^{\rm max,K}
 = \frac{1}{2(\sqrt{5}+1)}\sqrt{\frac{\sqrt{5}-1}{\sqrt{5}+3}} 
\approx 0.07507 \ , $$
is independent of the horizon radius $x_{\rm H}$.
We conjecture that, in the limit $x_{\rm H} \to \infty$,
the maximal value $\omega^{\rm max}_{\rm H}$
of the non-Abelian solutions 
tends to the maximal value $\omega_{\rm H}^{\rm max,K}$
of the Kerr solutions.

So far we have limited the discussion to non-Abelian solutions
with one node. When we consider black hole solutions with higher
node numbers, we observe similar features.
The rotating hairy black hole solution with $k$ nodes
emerges from the corresponding static black hole solution, 
when a finite value of $\omega_{\rm H}$
is imposed via the boundary conditions.
When $\omega_{\rm H}$ increases,
the lower branch of black hole solutions forms.
It extends up to a maximal value $\omega^{\rm max}_{\rm H}$,
where the upper branch forms and bends backwards towards $\omega_{\rm H}=0$.
Along both branches mass, angular momentum, and
non-Abelian electric charge continuously increase.

For node number $k=3$, the mass and angular momentum of the black hole solutions
differ only little from the mass and angular momentum of the $k=1$ 
black hole solutions.
In general, the values are still closer to those of the
Kerr-Newman solutions with $Q=0$ and $P=-1$. 
Also, the values of $\omega^{\rm max}_{\rm H}$ are very close
to those of the Kerr-Newman solutions with $Q=0$ and $P=-1$. 
Along the lower branch,
the non-Abelian electric charge of the $k=3$ solutions is much smaller than
the electric charge of the $k=1$ solutions, 
typically it is on the order of $10^{-4}$.
Along the upper branch the non-Abelian electric charge increases
relatively strongly. However, because of numerical inaccuracies
encountered along the upper branch for the higher node solutions,
we cannot infer from the numerical calculations,
that the charge tends to a limiting value $Q_{\rm lim}$.

\subsection{Energy density}

Let us now discuss the rotating black hole solutions
in more detail and begin with the energy density of the black hole
solutions, $\varepsilon = - T^0_0$.

Due to the rotation, the energy density of the matter fields
is angle-dependent, and, in particular, not constant at the horizon.
The maximum of the energy density resides on the $\rho$-axis at the horizon,
as seen in the following two representative examples.

In Figs.~2a-d we exhibit the energy density of the matter fields
of the black hole solution with horizon radius $x_{\rm H}=1$
and $\omega_{\rm H}=0.04$ on the lower branch.
Fig.~2a shows a 3-dimensional plot of the energy density
as a function of the coordinates $\rho = x \sin \theta$ and
$z= x \cos \theta$ together with a contour plot,
and Figs.~2b-d show surfaces of constant energy density.
The surfaces of constant energy density appear ellipsoidal,
being flatter at the poles than in the equatorial plane.
For the largest values of the energy density, as shown in Fig.~2d, the horizon
is seen in the pole region.

Analogously,
we exhibit in Figs.~3a-d the energy density of the matter fields
of the black hole solution with horizon radius $x_{\rm H}=1$
and $\omega_{\rm H}=0.04$ on the upper branch.
The energy density is much stronger deformed now, showing a
large peak on the $\rho$-axis. Consequently,
the surfaces of constant energy density appear torus-shaped,
with the horizon seen in the center of the torus.
Further away form the black hole horizon, however,
the surfaces of constant energy density appear again ellipsoidal,

We do not observe a strong dependence
of the energy density on the node number $k$.
On the lower branch the $k=3$ densities are similar
to the $k=1$ densities. On the upper branch,
the numerical determination of the $k=3$ densities becomes, however, 
unreliable.

\subsection{Gauge field and metric functions}

We now turn to the metric and gauge field functions of the 
non-Abelian black hole solutions.
In Figs.~4a-j we show the functions for the solutions
with one node on the upper and lower branch 
for horizon radius $x_{\rm H}=1$ and $\omega_{\rm H}=0.04$.
Also shown are the functions of the three node solution
on the lower branch for the same set of parameters.
We first discuss the magnetic gauge field functions, shown in Figs.~4a-4d,
then the electric gauge field functions, shown in Figs.~4e-4f,
and finally the metric functions, shown in Figs.~4g-4j.
Since the global mass and angular momentum of the
non-Abelian solutions are rather close to those of the
Kerr-Newman solutions with $Q=0$ and $P=-1$,
we compare with the corresponding functions of these
Kerr-Newman solutions.

The magnetic gauge field function $H_1$, shown in Fig.~4a,
is always very small.
For the $k=1$ solutions,
its magnitude is two orders of magnitude smaller on the lower branch 
than on the upper branch.
For the $k=3$ lower branch solution it is still smaller
by two orders of magnitude than for the lower branch $k=1$ solution.
$H_1$ shows a distinct angle dependence, which is similar for 
the three black holes exhibited. 
The Kerr-Newman solutions have vanishing $H_1$.
This is consistent with 
the observed smallness of $H_1$ for the non-Abelian solutions.

The magnetic gauge field function $H_2$ is almost spherically symmetric 
for the $k=1$ solutions on both branches as well as for the $k=3$
solution on the lower branch. It is exhibited in Fig.~4b.
Whereas $H_2$ decreases monotonically for the $k=1$ solutions
from a finite positive value at the horizon 
to its boundary value of $-1$ at infinity,
$H_2$ oscillates around zero (possessing three nodes)
for the $k=3$ solution in an extended interior region,
before it decreases to its boundary value $-1$ for $x \rightarrow \infty$.
Thus $H_2$ precisely keeps its features of the static non-Abelian
solutions, where with increasing node number $k$,
it becomes close to zero in an increasing interior region.
In particular, in the limit $k \rightarrow \infty$,
a limiting solution with $H_2 =0$ is reached \cite{foot2}.

The magnetic gauge field function $H_3$, shown in Fig.~4c,
is very similar for the $k=1$ and $k=3$ solutions on the lower branch,
and about an order of magnitude smaller on the lower branch
than on the upper branch.
Comparison of the function $H_3$ of the non-Abelian solutions
with the function $H_3$, Eq.~(\ref{KN12}),
of the corresponding Kerr-Newman solutions 
with $Q=0$ and $P=-1$ reveals very good agreement. 
Thus $H_3$, like $H_1$, corresponds approximately to the Abelian function.

The magnetic gauge field function $H_4$,
shown in Fig.~4d, is rather similar to the function $H_2$
on the lower branch, both for the $k=1$ and the $k=3$ solutions.
Its angle dependence on the lower branch is only small.
On the upper branch, however, a distinct angle dependence appears.
$H_4$, like $H_2$, retains the non-Abelian character of the solutions.

Turning to the electric gauge field functions $\bar{B}_1$ and $\bar{B}_2$
we observe, that the angle dependence is largely determined by
the boundary conditions at the horizon.
To eliminate this trivial angle dependence,
we consider the new functions $\hat B_1$ and $\hat B_2$, defined via
$${\hat B}_1 = -\cos\theta \bar{B}_1 + \sin\theta \bar{B}_2 \ , \ \ \
  {\hat B}_2 =  \sin\theta \bar{B}_1 + \cos\theta \bar{B}_2 \ . $$
As seen in Fig.~4e,
the function ${\hat B}_1$ then shows only a relatively small deviation 
from spherical symmetry for all three black hole solutions considered.
The function $\hat{B}_2$ is considerably smaller 
in amplitude than the function $\hat{B}_1$. 
Its angle dependence
is similar for the $k=1$ and $k=3$ solutions on the lower branch,
but different on the upper branch.

Let us again compare with the electric gauge field functions of the
corresponding Kerr-Newman solutions, Eq.~(\ref{KN13}),
with $Q=0$ and $P=-1$.
We again find good agreement between these Abelian and the
non-Abelian functions.
Thus $\hat B_1$ and $\hat B_2$ also correspond in good approximation
to the Abelian functions.

Thus we conclude, that the non-Abelian gauge field functions,
which are non-vanishing in the static limit, $H_2$ and $H_4$,
retain their non-Abelian features.
Whereas all gauge field functions which vanish in the static limit,
$H_1$, $H_3$, $B_1$ and $B_2$,
are very close to the corresponding functions of the Kerr-Newman
solution with $Q=0$ and $P=-1$.
  
Let us now turn to the metric functions.
In order to exhibit the behaviour of the metric functions $f$, $m$ and $l$
at the horizon more clearly we show the functions
$\bar{f}(x,\theta)$,
$\bar{l}(x,\theta)$, and
$g(x,\theta)$, Eq.~(\ref{barflm}).

The function $\bar f$  is shown in Fig.~4g.
It is almost spherically symmetric on the lower branch 
and deviates from spherical symmetry only slightly on the upper branch. 
On the upper branch the magnitude of $\bar f$ and its slope at the
horizon are considerably smaller than on the lower branch.
There is very little dependence of the metric function $\bar f$
on the node number. This is also true for the other metric functions.

In Fig.~4h we show the ratio $g=m/l$.
Note, that for a spherically symmetric metric the function $g$ 
is identical to one, 
thus the deviation of this function from one 
indicates the deviation from spherical symmetry.
We observe that on the lower branch $g$ is close to one for all $x$ 
except in a region near the horizon, where it
deviates slightly from one.
On the upper branch, in contrast, 
the deviation from one is large and $g$ approaches its asymptotic value
only for larger values of $x$.

The metric function $\bar l$ is shown in Fig.~4i.
For solutions with $k=1$ and $k=3$ nodes, 
as well as for solutions on the lower and upper branch,
the functions $\bar l$ are almost identical.
Also, their deviation from spherical symmetry is extremely small.

We exhibit the function $\omega$ in Fig.~4j.
The shape of the function $\omega$ is similar 
for solutions on the lower and upper branch.
However, on the upper branch the maximum is larger than on the lower branch. 
The angle dependence is small, 
except near the maximum of the function on the upper branch.

In general, the deviation from spherical symmetry is small on the lower branch.
We also observe little dependence of the metric functions on the node number,
although for solutions with smaller horizon radius
the node number dependence of the metric functions increases slightly.
All metric functions are rather close to the corresponding functions
of the Kerr-Newman solutions with $Q=0$ and $P=-1$.

We conclude, that the functions of the non-Abelian solutions
are rather close to those of the embedded Kerr-Newman solutions
with $Q=0$ and $P=-1$,
except for those gauge field functions, 
which do not vanish in the static limit. 
It is surprising, that already the one node solutions follows so
closely these Abelian solutions, because their functions $H_2$ and $H_4$
are still very different from zero, the value assumed by the
Abelian solutions, Eq.~(\ref{KN12}).
With increasing node number, though, one expects the solutions
to get close to the Abelian solution, since
$H_2$ and $H_4$ are close to zero in an increasing interval.

Let us now address the limiting non-Abelian solutions, obtained
in the limit $\omega_{\rm H} \rightarrow 0$.
Along the lower branch the limiting solution
corresponds to a static spherically symmetric non-Abelian black hole solution.
In contrast, along the upper branch mass and angular momentum
of the solution diverges, thus the limiting solution is singular.
Comparison with the Abelian solutions shows, 
that in the limit $\omega_{\rm H} \rightarrow 0$,
electric and magnetic charge become negligible
when $Q \ll 1$ and $|P|=1$.
Consequently, the ratio $M/a$ tends to one for fixed $x_{\rm H}$.

We gain some understanding of the limiting solution by noting,
that it can be obtained numerically, as a non-Abelian gauge field solution
in an extremal background metric.
Scaling the radial coordinate and angular momentum per unit mass by the mass,
and taking the limit $M \rightarrow \infty$,
the metric decouples from the gauge field, i.~e.~$G_{\mu\nu} = 0$.
Taking the limiting extremal Kerr solution as background metric,
one can solve the gauge field equations numerically.
The corresponding solution is then independent of the isotropic
horizon radius $x_{\rm H}$.
The limiting value of the electric charge obtained 
in this way is $Q_{\rm lim}=0.124$.

\subsection{Local charges}

In Figs.~5a-d we present the local charges
$M(x)$, $J(x)$, $Q(x)$ and $P(x)$, 
Eqs.~(\ref{localM})-(\ref{localJ}) and
Eqs.~(\ref{localQ})-(\ref{localP}), 
for the black hole solutions with one node on the lower and
upper branch, for horizon radius $x_{\rm H}=1$ and $\omega_{\rm H}=0.04$.
We also show these charges for the black hole solution
with three nodes on the lower branch for the same set of parameters.

In Fig.~5a the local mass $M(x)$ is shown.
We observe, that the mass at infinity, representing the global mass,
is only slightly larger than the mass at the horizon.
For the solutions shown, we find on the lower branch the global masses
$M_{k=1}\approx 2.357$ and
$M_{k=3}\approx 2.371$,
and on the upper branch we find
$M_{k=1}\approx 10.25$.
Thus the fields outside the horizon contribute little to the mass.

In Fig.~5b the local angular momentum $J(x)$ is shown.
Again, the global angular momentum, read off at infinity,
is only slightly larger than the angular momentum at the horizon.
For the solutions shown, the global angular momentum on the lower branch is
$J_{k=1}\approx 1.852$ and
$J_{k=3}\approx 1.872$,
and on the upper branch it is
$J_{k=1}\approx 102.5$.

The local electric charge $Q(x)$ is shown in Fig.~5c.
For solutions on both branches $Q(x)$ 
decreases significantly with increasing $x$, 
giving rise to a large difference between the horizon electric charge
and the global electric charge.
In general $Q(x)$ is smaller on the lower branch
than on the upper branch,
where it assumes values between $\approx 0.9$ at the horizon and
$0.09$ at infinity for the solutions shown.
The decrease with $x$ indicates the presence of
a charge density outside the horizon, cancelling part of the horizon charge.

We show the local magnetic charge $P(x)$ in Fig.~5d.
In contrast to the local electric charge, 
the local magnetic charge is not monotonic but
possesses a maximum of magnitude one,
for the solutions on both branches. 
At infinity $P(x)$ vanishes, in accordance with the boundary conditions,
corresponding to a vanishing global magnetic charge.
With increasing node number, the local magnetic charge remains
close to the value
one in an increasing region, as in the static case \cite{review}.
This again indicates, that in the limit $k \rightarrow \infty$,
the rotating non-Abelian black hole solutions
tend to a limiting solution with magnetic charge $|P|=1$,
like their static counterparts \cite{foot2}.

\subsection{Horizon properties}

We now turn to the horizon properties of the rotating non-Abelian
black holes.
In Figs.~6a-d we demonstrate the dependence of the horizon mass,
the horizon angular momentum,
and the horizon electric and magnetic charges
on $\omega_{\rm H}$ for fixed isotropic horizon radius $x_{\rm H}$,
for black hole solutions with one node.
The corresponding horizon size, the deformation of the horizon and
the surface gravity are shown in Figs.~7a-c.

The horizon mass $M_\Delta$ follows closely the global mass $M$,
as seen in Fig.~6a, where the horizon mass is shown for
black hole solutions with horizon radius $x_{\rm H}=0.1$ and
$x_{\rm H}=1$. 
Similarly, the horizon angular momentum $J_\Delta$ closely follows
the global angular momentum $J$.
The horizon angular momentum per unit mass, $a_\Delta=J_\Delta/M_\Delta$,
is shown in Fig.~6b.
Both horizon mass and horizon angular momentum are close
to the corresponding Kerr-Newman values for solutions
with $Q=0$ and $P=-1$.

The horizon electric charge $Q_\Delta$, shown in Fig.~6c,
increases monotonically along both branches, like the global
non-Abelian electric charge $Q$.
It is, however, about an order of magnitude larger 
than the global non-Abelian electric charge $Q$.
The horizon magnetic charge $P_\Delta$ is shown in Fig.~6d.
Whereas the black holes carry no global non-Abelian magnetic charge,
their horizon magnetic charge is on the order of one.
Starting from finite values on the lower branch,
which correspond to the horizon magnetic charges of the
static solutions, $P_\Delta$ decreases monotonically along the lower branch.
It then reaches a minimum along the upper branch and increases again.
The location of the minimum strongly depends on $x_{\rm H}$.

The horizon size as quantified by the area parameter $x_\Delta$
is shown in Fig.~7a for black holes with isotropic horizon radii
$x_{\rm H}=0.1$ and $x_{\rm H}=1$
as a function of $\omega_{\rm H}$.
The horizon area grows monotonically along both branches,
and diverges along the upper branch.
Comparison with the corresponding Abelian black hole solutions 
again shows,
that the non-Abelian horizon size follows closely the Kerr-Newman 
horizon size for $Q=0$ and $P=-1$, whereas only for large 
$x_{\rm H}$ it is also close to the Kerr horizon size.

The deformation of the horizon is revealed by measuring
the circumference of the horizon along the equator, $L_e$,
and the circumference of the horizon along the poles, $L_p$,
Eqs.~(\ref{lelp}).
The deformation of the horizon,
quantified by $L_e/L_p$, Eq.~(\ref{lelp}),
is shown in Fig.~7b.
The ratio $L_e/L_p$ grows monotonically along both branches.
As $\omega_{\rm H}$ tends to zero on the upper branch, the ratio
tends to the value  $L_e/L_p \approx 1.645$,
corresponding to the value of the limiting extremal Kerr solution,
for $M \to \infty$.
On the lower branch the ratio $L_e/L_p$ assumes the value one
in the limit $\omega_{\rm H} \to 0$,
corresponding to the value of
a static spherically symmetric hairy black hole.
Also shown are the corresponding embedded Abelian solutions,
approaching the same limiting values of the deformation.

The surface gravity $\kappa_{\rm sg}$ is exhibited in Fig.~7c,
along with the surface gravity of the Kerr and the Kerr-Newman
solutions for $Q=0$ and $P=-1$.
The surface gravity of the non-Abelian black holes
decreases monotonically along both branches,
starting from the value of the corresponding 
static non-Abelian black hole solution 
on the lower branch in the limit $\omega_{\rm H} \rightarrow 0$.
On the upper branch the surface gravity tends to zero
in the limit $\omega_{\rm H} \rightarrow 0$, the value assumed by
extremal black hole solutions.
The surface gravity of Kerr-Newman and non-Abelian
solutions agrees well for black hole solutions with large horizon radii, 
for black hole solutions with small horizon radii, however,
a difference is observed on the lower branch.
Here, in the limit $\omega_{\rm H} \rightarrow 0$,
the rotating non-Abelian black hole solutions approach the corresponding static
non-Abelian black hole, and the Kerr-Newman solutions
approach the corresponding 
Reissner-Nordstr\o m black hole with $Q=0$ and $P=-1$.
For these static solutions 
the difference in surface gravity in known to decrease 
with increasing horizon size \cite{review}.

\boldmath
\subsection{Fixed $\omega_{\rm H}$}
\unboldmath

Having considered the rotating hairy black hole solutions
for fixed horizon radius $x_{\rm H}$ as a function of $\omega_{\rm H}$,
let us now keep $\omega_{\rm H}$ fixed and vary the horizon radius.

In Fig.~8 we show the mass $M$ of the non-Abelian black hole solutions
as a function of the isotropic horizon radius $x_{\rm H}$
for $\omega_{\rm H}=0.01$, $\omega_{\rm H}=0.02$, and $\omega_{\rm H}=0.05$.
For a given value of $\omega_{\rm H}$ there is
a minimal value of the horizon radius $x_{\rm H}$.
In particular, the limit $x_{\rm H} \rightarrow 0$
is only reached for $\omega_{\rm H} \rightarrow 0$.
Thus we do not obtain globally regular rotating solutions in
the limit $x_{\rm H} \rightarrow 0$ \cite{vs}.

For comparison, Fig.~8 also presents
the mass $M$ of the corresponding Kerr solutions and Kerr-Newman
solutions with $Q=0$ and $P=-1$.
For a fixed value of $\omega_{\rm H}$, the mass of
the Kerr solutions forms two straight lines, extending from the origin.
The non-Abelian solutions tend toward these lines
for large values of the horizon radius.
The mass of the Kerr-Newman solutions again is close to 
the mass of the non-Abelian solutions. In particular, we observe,
that the minimal values of the horizon radius only differ slightly
for the non-Abelian and Kerr-Newman solutions.

\boldmath
\subsection{Fixed $\omega_{\rm H}/x_{\rm H}$}
\unboldmath

Let us finally consider variation of the parameters
$x_{\rm H}$ and $\omega_{\rm H}$, while keeping their ratio 
$\omega_{\rm H}/x_{\rm H}$ fixed.

The global charges of the black hole solutions with one node
are shown in Fig.~9
for fixed ratio $\omega_{\rm H}/x_{\rm H}=0.04$.
The mass $M$ and the angular momentum per unit mass $a$
increase monotonically along both branches, 
reaching finite limiting values on the upper branch for
$\omega_{\rm H} \rightarrow 0$.
Except in a small region close to $\omega_{\rm H}=0$ on the lower branch
the non-Abelian electric charge $Q$ also increases monotonically.
The non-Abelian electric charge also approaches a finite
limiting value on the upper branch, 
apparently close to the limiting value 
$Q_{\rm lim} \approx 0.124$, observed previously.

To gain a better understanding of the limiting behaviour
on the upper branch, we consider also the horizon properties
for these black hole solutions.
In Fig.~10 we show their horizon size $x_\Delta$, their deformation
$L_e/L_p$ and their surface gravity $\kappa_{\rm sg}$.
On the upper branch 
the horizon size, the deformation and the surface gravity
of the non-Abelian black hole solutions are very close
to those of the Kerr-Newman solution with $Q=0$ and $P=-1$.
In particular,
the horizon size and deformation remain finite
on the upper branch in the limit $\omega_{\rm H} \rightarrow 0$,
and the surface gravity tends to zero.
This indicates, that an extremal black hole is
approached in this limit.
Since the limiting black hole solution retains
its non-Abelian character in the gauge field functions,
it should correspond to a rotating hairy extremal black hole.

On the lower branch in the limit $\omega_{\rm H} \rightarrow 0$,
the corresponding Bartnik-McKinnon solution is approached
by the non-Abelian solutions, whereas the Kerr-Newman solutions
approach a Reissner-Nordstr\o m solution.
This limiting behaviour is suggested by a detailed inspection
of the metric and gauge field functions.
In Fig.~10 this limiting behaviour is reflected in the fact, 
that the horizon size
of the non-Abelian solutions tends to zero, whereas the horizon
size of the Kerr-Newman solutions tends to a finite value.
Furthermore, the surface gravity of the non-Abelian solutions
apparently diverges in the limit, in agreement with
the static non-Abelian results \cite{review},
whereas the surface gravity of the Kerr-Newman solutions
tends to zero, the value of an extremal Reissner-Nordstr\o m solution.

\section{Conclusions}

We have given a detailed account of a new class of black hole solutions
in SU(2) EYM theory, which represent the first examples of non-perturbative
stationary non-Abelian black hole solutions \cite{kkbh}.
These black hole solutions carry mass, angular momentum
and a non-Abelian electric charge.
Although they do not carry a non-Abelian magnetic charge,
they still possess non-trivial magnetic gauge fields
outside their regular event horizon.
They therefore represent rotating hairy black hole solutions.

The global charges of the rotating hairy black hole solutions are
not independent. For a given mass and angular momentum,
as well as node number of the solution,
a unique electric charge is obtained.
Whereas mass and angular momentum are unbounded,
we observe, that the electric charge remains very small. 

The event horizon of the static axially symmetric black hole solutions
resides at a surface of constant isotropic radial coordinate, $x=x_{\rm H}$.
The boundary conditions at the horizon ensure regularity of the horizon.
The horizon mass and horizon angular momentum are only slightly smaller
than the global mass and the global angular momentum of the
black hole solutions. 
The horizon electric charge is, however, significantly larger than the
global electric charge, and the solutions possess
horizon magnetic charge of order one, 
whereas their global magnetic charge vanishes.

The rotating hairy black hole solutions emerge from the static
hairy black hole solutions in the limit of vanishing angular momentum.
Since the static spherically symmetric black hole solutions form a sequence
labelled by the node number $k$ of the gauge field function, we obtain
the corresponding sequence of rotating black hole solutions,
by starting from the static black hole solutions
and imposing a small angular velocity of the horizon
via the boundary conditions.

The rotating non-Abelian black hole solutions are rather close
to the Kerr-Newman solutions with $Q=0$ and $P=-1$.
In particular, the metric and gauge field functions of the
non-Abelian solutions are very close to those of the Abelian solutions,
except for those gauge field function, which do not vanish
in the static limit. These retain the non-Abelian character of the solutions.
However, with increasing node number, also these functions
tend to their Abelian counterparts.

The asymptotic expansion performed for the metric and
gauge field functions, contains non-integer powers for the
magnetic gauge field functions. 
In particular, 
the non-integer exponents depend on the non-Abelian electric charge.
The expansion then imposes constraints on the possible values 
of the electric charge.
Since in the static limit the non-Abelian electric charge vanishes,
the well known power law decay of the static gauge field functions
is recovered.
%The non-integer fall-off of the gauge field functions
%does not admit the standard extraction of the gyromagnetic ratio.
%Possessing the Dirac value for the embedded Kerr-Newman solutions,
%the value of the gyromagnetic ratio 
%for the rotating hairy black hole solutions thus remains open.

The expansions of the metric and gauge field functions at the horizon
show, that the rotating hairy black hole solutions satisfy 
the zeroth law of black hole mechanics \cite{wald}.
Relations obtained recently within the isolated horizon framework
\cite{cs,isorot} concerning various horizon properties,
such as the horizon mass and the horizon charges,
will be considered elsewhere.
 
The hairy black hole solutions constructed here non-perturbatively,
were first considered perturbatively \cite{vs}.
To compare with these perturbative calculations,
where linear rotational excitations of the static EYM black holes 
were studied, we consider the slowly rotating non-Abelian solutions 
in the limit $\omega_{\rm H} \rightarrow 0$.
In the perturbative calculations $Q \propto J$ \cite{vs},
and the ratio $Q/J$ depends only on the horizon radius.
The non-perturbative calculations show good agreement
with the non-perturbative results for the slowly rotating solutions
with large values of the horizon radius.

Besides these non-Abelian stationary black hole solutions
with finite angular momentum $J$ and finite electric charge $Q$,
perturbative studies \cite{bhsv} have predicted two more types
of stationary non-Abelian black hole solutions.
These correspond to rotating black hole solutions 
which are uncharged ($J>0$, $Q = 0$),
and non-static charged black hole solutions, which
have vanishing angular momentum ($J=0$, $Q\ne 0$).
Both types satisfy a different set of boundary conditions at infinity.

This different set of boundary conditions at infinity
should also be observed by rotating regular non-Abelian solutions \cite{bhsv}.
The numerical construction of such non-perturbative regular solutions
has been attempted recently \cite{radu2},
and arguments have been put forward, that such solutions
should not exist.
Our attempts to obtain numerically the non-perturbative
counterparts of the predicted further types of
black hole solutions have met with the same difficulties
for the same reasons \cite{next}.

Besides the rotating black hole solutions considered here, 
there might be rapidly rotating branches of non-Abelian black holes solutions
in EYM theory, not connected to the static solutions.
There should also be
rotating hairy black hole solutions in other non-Abelian theories
such as Einstein-Yang-Mills-Higgs theory.
For instance, we expect rotating black hole solutions with
magnetic charge and with magnetic dipole hair \cite{hkk,map3},
Furthermore, the recent conjecture,
that ``any dyon solution with nonzero angular momentum
necessarily contains an event horizon''
\cite{radu2} awaits investigation.

\section{Appendix A: Ricci circularity conditions}

Let us demonstrate that the ansatz for the metric (\ref{metric})
satisfies the Ricci circularity conditions, Eq.~(\ref{cfc1}).

The Killing vectors
$\xi = \partial_t $ and $\eta = \partial_\vphi $
have components
\begin{eqnarray}
\xi^\mu = (1,0,0,0) \ , \ \ \ & &
\xi_\mu = \xi^\nu g_{\mu\nu} = (g_{tt}, 0,0, g_{t\vphi}) \ ,
\nonumber\\
\eta^\mu = (0,0,0,1) \ , \ \ \ & &
\eta_\mu = \eta^\nu g_{\mu\nu} = (g_{\vphi t}, 0,0, g_{\vphi\vphi}) \ .
%\nonumber
\end{eqnarray}
Consider the Ricci circularity condition
$\xi^\mu R_{\mu[\alpha}\xi_\beta \eta_{\gamma]}=0$, Eq.~(\ref{cfc1}),
where the square bracket denotes antisymmetrization.
Hence, $\alpha\neq\beta\neq\gamma$.
Thus $\xi^\mu R_{\mu[\alpha}\xi_\beta \eta_{\gamma]}\neq 0$
is only possible if either $\beta=0,\gamma=3$ or  $\beta=3,\gamma=0$.
In both cases $\alpha= 1$ or $\alpha=2$.

For $\alpha=1$,
\begin{eqnarray}
\xi^\mu R_{\mu[1}\xi_0 \eta_{3]} & =  & R_{0[1}\xi_0 \eta_{3]}
\nonumber\\
& = & \frac{1}{6}(R_{01}\xi_0 \eta_3 +R_{00}\xi_3 \eta_1 +R_{03}\xi_1 \eta_0
                -R_{03}\xi_0 \eta_1 -R_{00}\xi_1 \eta_3-R_{01}\xi_3 \eta_0)
\nonumber\\
& = &\frac{1}{3}R_{01}(\xi_0 \eta_3-\xi_3 \eta_0)
\nonumber\\
& = &\frac{1}{3}R_{01}(g_{tt}g_{\vphi\vphi}-g_{\vphi t}^2)
\nonumber
\ , \end{eqnarray}
and similarly, for $\alpha=2$,
$$
\xi^\mu R_{\mu[2}\xi_0 \eta_{3]}=
\frac{1}{3}R_{02}(g_{tt}g_{\vphi\vphi}-g_{\vphi t}^2) \ ,
$$
since all other components vanish due to $\xi_1=0,\xi_2=0$
and $\eta_1=0,\eta_2=0$.

On the other hand, the ansatz for the metric Eq.~(\ref{metric}) yields
$$
R_{01} = R_{tr} =0 \ , \ \ \ R_{02} = R_{t\theta} =0 \ .
$$
Consequently, $\xi^\mu R_{\mu[\alpha}\xi_\beta \eta_{\gamma]}=0$.

Note that $R_{tr} =0$, $R_{t\theta} =0$
implies $G_{tr} = R_{tr}-1/2 g_{tr}R = 0$
and $G_{t\theta} = R_{t\theta}-1/2 g_{t\theta}R = 0$, respectively.
Therefore $T_{tr} =0$ and $T_{t\theta}=0$
is a necessary condition for the solutions of the Einstein equations. In a similar fashion it can be shown that the Ricci circularity condition
$\eta^\mu R_{\mu[\alpha}\xi_\beta \eta_{\gamma]}=0$, Eq.~(\ref{cfc1}), is satisfied by the metric (\ref{metric}) and  implies $T_{\vphi r} =0$ and $T_{\vphi \theta}=0$. 
However,
for the Ansatz (\ref{a1}) for the gauge field
these conditions are fullfilled identically.
Hence the solutions satisfy the Ricci circularity conditions.

\section{Appendix B: Asymptotic expansion}

\subsection{General asymptotic expansion}

The asymptotic expansion of the fields may be obtained
from the field equations and the corresponding boundary conditions.
However, the process is rather involved.
Indeed, the most natural assumption for the asymptotic $r$-dependence
of the functions, i.e.~polynomial,
seems to be in contradiction with the presence 
of a non-vanishing electric charge.
In view of that and inspired by perturbative results \cite{vs},
we allowed for the presence of logarithms in the expansions.
We then observed that, when keeping only a finite number of logarithmic terms,
the charge was forced to vanish.
By permitting an infinite number of such terms, however,
it became possible to obtain a consistent expansion
in the presence of an electric charge.
This feature then suggested to include non-integer powers of $r$,
with the exponents depending on the electric charge.

Numerically, we realized that there should be a splitting
of some terms in the $1/r$ Taylor series expansion
of the static spherically symmetric case.
However, the exponents of the new terms turned out to be very close
to integer numbers, even though the behavior of the functions
could not be described just by means of a $1/r$ Taylor series.
The procedure of how to compute this expansion was then clear.

First of all, we introduced a formal parameter $\epsilon$
in order to characterize terms of the order of $1/r$,
without assuming a Taylor series in $1/r$ for the functions.
Then we expanded all the functions in this formal parameter,
the coefficients of such expansions depending on $r$ and $\theta$.
Finally, we introduced those $\epsilon$ series
into the system of field equations,
taking into account the explicit dependence on $r$
in the equations by including $\epsilon$ appropriately.
The last step was collecting coefficients in $\epsilon$
for each equation of the system,
and solving the equations so formed, 
order by order in this formal parameter, 
keeping in mind the boundary conditions
and the fact that the coefficients of the $\epsilon$ series
had to behave consistently with the corresponding power of $\epsilon$.

Proceeding in this way,
the expression for this asymptotic expansion is found to be:
\begin{eqnarray}
f&=&1-\frac{2 M}{x} + \frac{2 M^2 + Q^2}{x^2}
+  o \left(\frac{1}{x^2}\right)\ , \nonumber\\
m&=&1 + \frac{C_1}{x^2} + \frac{Q^2-M^2-2 C_1}{x^2}\sin^2{\theta}
+ o\left( \frac{1}{x^2}\right)\ , \nonumber\\
l&=&1+\frac{C_1}{x^2} + o\left(\frac{1}{x^2}\right)\ , \nonumber \\
\omega&=&-\frac{2 a M}{x^2} + \frac{6 a  M^2 + C_2 Q}{x^3}
+ o\left(\frac{1}{x^3}\right)\ , \nonumber\\
H_1&=&\left[ \frac{2 C_5}{x^2}
+ \frac{8 C_4}{\beta-1} x^{-\frac{1}{2} (\beta-1)}
-\frac{2 C_2 C_3 (\alpha+3)}{(\alpha+5)Q^2}
 x^{-\frac{1}{2}(\alpha+1)}\right] \sin{\theta} \cos{\theta}
+ o\left(\frac{1}{x^2}\right)\ ,\nonumber\\
H_2&=&-1+C_3 x^{-\frac{1}{2}(\alpha-1)} + \bigg[\frac{C_5}{x^2}
+ C_4 x^{-\frac{1}{2}(\beta-1)}
+ \frac{C_3 (\alpha^2+2 \alpha -11)}{2 (\alpha+1)} \nonumber\\
&&\bigg( M- \frac{2 C_2 (\alpha+3)}{ (\alpha+5)Q^2} \bigg)
 x^{-\frac{1}{2} (\alpha+1)} \bigg]
+ \bigg[-\frac{2 C_5}{x^2} - 2 C_4 x^{-\frac{1}{2}(\beta-1)} \nonumber\\
&&+ \frac{C_2 C_3 (\alpha+1) (\alpha+3)}{2 (\alpha+5) Q^2}
 x^{-\frac{1}{2}(\alpha+1)}\bigg] \sin^2{\theta}
+o\bigg(\frac{1}{x^2}\bigg)\ ,\nonumber\\
H_3&=&\bigg(\frac{C_2}{x} + C_3 x^{-\frac{1}{2}(\alpha-1)} \bigg)
 \sin{\theta} \cos{\theta}
+ \bigg[\frac{C_3 (\alpha^2+2 \alpha-11)}{2 (\alpha+1)} \nonumber\\
&&\bigg(M -\frac{2 C_2 (\alpha+3)}{ (\alpha+5)Q^2}\bigg)
 x^{-\frac{1}{2}(\alpha+1)}+C_4 x^{-\frac{1}{2}(\beta-1)}
+\frac{3 C^2_3}{(\alpha+1)(\alpha-2)}x^{-(\alpha-1)} \nonumber\\
&&+\frac{2 C_5 + C_2 M - 3 a M Q}{2 x^2}\bigg]
 \sin{\theta} \cos{\theta} + o\left(\frac{1}{x^2}\right)\ , \nonumber \\
H_4&=&-1+C_3 x^{-\frac{1}{2}(\alpha-1)} - \bigg(\frac{C_2}{x}
+ C_3 x^{-\frac{1}{2}(\alpha-1)} \bigg) \sin^2{\theta}
+\bigg[\frac{C_3 (\alpha^2+2 \alpha -11)}{2 (\alpha+1)} \nonumber\\
&&\bigg(M-\frac{2 C_2 (\alpha+3)}{(\alpha+5)Q^2}\bigg)x^{-\frac{1}{2}(\alpha+1)}
+C_4 x^{-\frac{1}{2} (\beta-1)} + \frac{C_5}{x^2}\bigg] \nonumber\\
&&- \bigg[\frac{C_3 (\alpha^2+2 \alpha -11)}{2 (\alpha+1)}
\bigg(M-\frac{2 C_2 (\alpha+3)}{(\alpha+5)Q^2}\bigg)x^{-\frac{1}{2}(\alpha+1)}
+C_4 x^{-\frac{1}{2} (\beta-1)} \nonumber \\
&&+ \frac{3 C^2_3}{(\alpha+1)(\alpha-2)} x^{-(\alpha-1)}
+ \frac{2 C_5 + C_2 M - 3 a M Q}{2 x^2}\bigg]\sin^2{\theta}
+ o\left(\frac{1}{x^2}\right)\ , \nonumber\\
{\bar B}_1 &=& \frac{Q \cos{\theta}}{x} -\frac{M Q \cos{\theta}}{x^2}
+ \bigg[\frac{2 C_4 (\beta-5)}{ (\beta-1)Q}x^{-\frac{1}{2} (\beta+1)}
-\frac{4 C_2 C_3 Q}{(\alpha-3) (\alpha+5)} x^{-\frac{1}{2}(\alpha+3)} 
\nonumber\\
&&+\frac{C^2_3 (\alpha^2-\alpha-8) Q}
{\alpha (\alpha-1) (\alpha+2) (\alpha-3)} x^{-\alpha}
+\frac{2 Q^3-C_1 Q+4 M^2 Q + 4 C_5 Q -2 C_6}{6 x^3}\bigg]
 \cos{\theta} \nonumber\\
&&+ \bigg[\frac{C_6}{x^3} + \frac{C^2_3 Q}{(\alpha+2)(\alpha-3)}x^{-\alpha}
+ \frac{8 C_4 Q}{(\beta-1)(\beta+5)} x^{-\frac{1}{2}(\beta+1)}\nonumber \\
&& + \frac{4 C_2 C_3 Q}{(\alpha-3)(\alpha+5)}
 x^{-\frac{1}{2}(\alpha+3)} \bigg] \cos^3{\theta}
+ o\left(\frac{1}{x^3}\right)\ , \nonumber \\
{\bar B}_2 &=& \frac{Q \sin{\theta}}{x} - \frac{M Q \sin{\theta}}{x^2}
+\bigg[\frac{2 Q^3-C_1 Q + 4 M^2Q -2 C_5 Q -2 C_6 -24 a M}{6 x^3} \nonumber\\
&&+\frac{C^2_3  (\alpha^2-\alpha- 8)Q}{\alpha (\alpha-1)(\alpha+2)(\alpha-3)}
x^{-\alpha}\bigg] \sin{\theta} + \bigg[\frac{C_6}{x^3}
+\frac{C^2_3 Q}{(\alpha+2)(\alpha-3)}x^{-\alpha}      \nonumber\\
&& +\frac{4 C_2 C_3 Q}{(\alpha-3)(\alpha+5)} x^{-\frac{1}{2}(\alpha+3)}
-\frac{2 C_4 (\beta-5)}{(\beta-1)Q} x^{-\frac{1}{2}(\beta+1)}\bigg]
\sin{\theta} \cos^2{\theta} + o\left(\frac{1}{x^3}\right) 
\ \ , \label{asymp_general}
\end{eqnarray}
with the notation of section 3.1.

\subsection{Asymptotic expansion: relation with static case}

Here we show how the previous asymptotic expansions reduce
to the ones corresponding to the static case in the limit of vanishing charge.
Moreover, as these are non-perturbative expansions,
they must include the perturbative expansions reported in \cite{vs}.
This is indeed the case.
In order to perform perturbative expansions in terms of the electric charge,
we must recall that, for a fixed value of the horizon radius,
all the parameters in these expansions are functions of $Q$.
Due to the presence of $Q$ in some denominators,
the dependence on $Q$ of these coefficients has to be singular,
in such a way that the resulting series in $Q$ turns out to be regular.
The behavior of the constants as functions of $Q$ reads:
\begin{eqnarray}
M&=&M_0+Q^2 K_0 (Q)\ ,\nonumber\\
C_1&=&-\frac{1}{2} M^2_0+Q^2 K_1 (Q)\ ,\nonumber\\
C_2&=&-b +Q^2 K_2(Q)\ ,\nonumber\\
C_3&=&b+Q^2 K_3(Q)\ ,\nonumber\\
C_4&=&-\frac{5}{4}\frac{b^2}{Q^2}+K_4(Q)\ ,\nonumber\\
C_5&=&\frac{1}{2} \frac{b^2}{Q^2} + \bigg[\frac{7}{48}b^2
+\frac{3}{4}b(K_2(Q)-K_3(Q))-\frac{1}{5} K_4(Q)\bigg] + Q K_5(Q)\ ,\nonumber\\
C_6&=&-\frac{1}{5}\frac{b^2}{Q^2} + Q \bigg[\frac{1}{16}b^2
+\frac{3}{4}b K_2(Q)-\frac{3}{20}b K_3(Q)-\frac{1}{5} K_4(Q)\bigg]
+Q^2 K_6(Q)\ ,\nonumber\\
a&=&Q K_7(Q)\ ,
\end{eqnarray}
where $M_0$ is the dimensionless mass
of the static spherically symmetric solution,
$b$ is the parameter of the gauge field for such a limiting solution,
and $K_1, \dots, K_7$ are regular functions of $Q$.
Introducing these relations in (\ref{asymp_general})
and expanding the result in $Q$,
we recover the perturbative expansions given in  \cite{vs}:
\begin{eqnarray}
f&=&1 - \frac{2 M'}{x} + \frac{2 {M'}^2}{x^2} + O(Q^2)\ ,\nonumber\\
m&=&1-\frac{{M'}^2}{2 x^2} + O(Q^2)\ ,\nonumber\\
l&=&1-\frac{{M'}^2}{2 x^2} + O(Q^2)\ ,\nonumber\\
\omega&=&-\frac{2 J' M'}{x^2} + \frac{6J' {M'}^2- a' Q'}{x^3}\ ,\nonumber\\
H_1&=&O(Q^2)\ ,\nonumber\\
H_2&=&-1+\frac{a'}{x}+\frac{2 a' M'-3{a'}^2}{4 x^2} + O(Q^2)\ ,\nonumber\\
H_3&=&O(Q^2)\ ,\nonumber\\
H_4&=&-1+\frac{a'}{x}+\frac{2 a' M'-3{a'}^2}{4 x^2} + O(Q^2)\ ,\nonumber\\
{\bar B}_1&=&\frac{Q' \cos{\theta}}{x}-\frac{M' Q' \cos{\theta}}{x^2}
+ \frac{1}{60 x^3}\cos{\theta}(40 {a'}^2 Q'+225{M'}^2Q'+120 {c_4}'\nonumber\\
&&-8 {a'}^2 Q' \ln{x}) + O(Q^2)\ ,\nonumber\\
{\bar B}_2&=&\frac{Q' \sin{\theta}}{x} -\frac{M' Q' \sin{\theta}}{x^2}
-\frac{1}{60 x^3}\sin{\theta} (45 {M'}^2 Q' +240 J' M' + 60 {c_4}'\nonumber\\
&& -4 {a'}^2 Q' \ln{x}) + O(Q^2)\ ,
\end{eqnarray}
where primes denote the notation of Volkov and Straumann,
which is related to our notation by
\begin{eqnarray}
M'&=&M_0\ ,\nonumber\\
a'&=&b\ ,\nonumber\\
Q'&=&Q\ ,\nonumber\\
J'&=&Q K_7(0)\ ,\nonumber\\
{c_4}'&=& \frac{Q}{200} \left[100 b (K_2(0)-K_3(0))
- 31 b^2-300 M^2_0-80 K_4(0)\right]\ .
\end{eqnarray}

\subsection{Asymptotic expansion: embedded Kerr-Newman solutions}

For comparison, we present the asymptotic expansion also for
the embedded Kerr-Newman solutions for our choice of coordinates
and gauge. The expansion reads
\begin{eqnarray}
f&=&1-\frac{2M}{x} + \frac{2 M^2 + Q^2+ P^2}{x^2} + O\left(\frac{1}{x^3}\right)
\ ,\\
l&=&1 +\frac{a^2-M^2 +Q^2+P^2}{2 x^2} + O\left(\frac{1}{x^3}\right)
\ ,\\
m&=&1+\frac{a^2-M^2 +Q^2+P^2}{2 x^2}-\frac{a^2 \sin^2 \theta}{x^2}
+ O\left(\frac{1}{x^3}\right)
\ ,\\
\omega&=& \frac{2 a M}{x^2} -\frac{a (Q^2+P^2+6 M^2)}{x^3}
+ O\left(\frac{1}{x^4}\right)
\ ,\\
H_1&=&0
\ ,\\
H_2&=&0
\ ,\\
H_3&=&-(1+P) \cot\theta -\frac{a Q \sin\theta}{x}
-\frac{a \sin\theta (a P \cos\theta -MQ)}{x^2}
+ O\left(\frac{1}{x^3}\right)
\ ,\nonumber\\
H_4&=&0
\ , \nonumber\\
\bar{B}_1&=& \frac{Q}{x} +\frac{a P \cos\theta - MQ}{x^2}
+ O\left(\frac{1}{x^3}\right)
\ ,\\
\bar{B}_2&=&\frac{2 a M \sin\theta}{x^3} +O\left(\frac{1}{x^4}\right)
\ ,
\end{eqnarray}
where $a$ is the angular momentum per unit mass,
$M$ is the mass, $Q$ is the electric charge, ($Q^2 = Q^a Q^a$),
and $P$ is the magnetic charge, ($P^2 = P^a P^a$).

\section{Appendix C: Expansion at the horizon}

We here first motivate our choice of boundary conditions at the
horizon. Then we give the full expansion of the metric and
gauge field functions at the horizon and relate this
general expansion with the static case.
 \subsection{Boundary conditions at the horizon}

Let us begin by noting that the ansatz of the gauge field has the property
\cite{galt,radu2}
\begin{equation}
\partial_\vphi A_\mu = D_\mu u \ ,
\label{axsym}
\end{equation}
with $u = \tau_z/2$.
The components $F_{\mu\vphi}$ can be expressed as
\begin{equation}
F_{\mu\vphi}= D_\mu W \ ,
\label{Fmuphi}
\end{equation}
with
\begin{equation}
W = A_\vphi - u
\ . \label{W} \end{equation}
$W$ transforms as a scalar doublet under gauge transformations
$U= \exp(i \Gamma \tau_\phi/2)$, Eq.~(\ref{gauge}).
Using the definition of $W$, Eq.~(\ref{W}), we find for the component $A_t$
of the gauge field
\begin{equation}
A_t  =  \Psi  + \frac{\omega}{r} \frac{\tau_z}{2} + \frac{\omega}{r} W
     =  \hPsi + \frac{\omega}{r} W \ ,
\label{A_t}
\end{equation}
with
\begin{equation}
\hPsi =\Psi  + \frac{\omega}{r} \frac{\tau_z}{2}
\ . \label{hPsi} \end{equation}
Thus $\hPsi$ also transforms as a scalar doublet under gauge
transformations.

To discuss the behaviour of the solutions at the horizon it is convenient
to rewrite $\Psi$ and $A_\vphi$  as
\begin{equation}
\Psi = -\frac{\tB_1}{\xh} \frac{\tau_z}{2}
        +\frac{\tB_2}{\xh} \frac{\tau_\rho}{2} \ , \ \ \ \
A_\vphi = -\sin\theta\left[ \tH_4 \frac{\tau_z}{2}
                          - \tH_3 \frac{\tau_\rho}{2}\right]
\ . \end{equation}
This yields for the scalar doublets,
$W$ and $\hPsi$, Eqs.~(\ref{W}), (\ref{hPsi}),
\begin{eqnarray}
W & = & \sin\theta \tH_3 \frac{\tau_\rho}{2}
        -(\sin\theta \tH_4+1)\frac{\tau_z}{2} \ ,
\\
\hPsi & = & -(\frac{\tB_1}{\xh}-\frac{\omega}{x})\frac{\tau_z}{2}
            +\frac{\tB_2}{\xh} \frac{\tau_\rho}{2} \ .
\end{eqnarray}

We assume (see below)
that near the horizon the metric functions can be expanded as
\begin{equation}
f= f_2 \delta^2 + O(\delta^3) \ , \ \ \
m= m_2 \delta^2 + O(\delta^3) \ , \ \ \
l= l_2 \delta^2 + O(\delta^3) \ , \ \ \
\omega = \oh + \omega_1 \delta +O(\delta^2) \ ,
\end{equation}
where
$\delta = (x-\xh)/\xh$ and $\oh$ is a constant.
For the gauge field functions we assume an expansion in the form
\begin{eqnarray}
H_1 & = & H_{11} \delta +O(\delta^2) \ ,
\nonumber \\
H_2 & = & H_{20} + H_{21}\delta +O(\delta^2) \ ,
\nonumber \\
\tH_3 & = & \tH_{30} + \tH_{31}\delta +O(\delta^2) \ ,
\nonumber \\
\tH_4 & = & \tH_{40} + \tH_{41}\delta +O(\delta^2) \ ,
\nonumber \\
\tB_1 & = & \tB_{10} + \tB_{11}\delta +O(\delta^2) \ ,
\nonumber \\
\tB_2 & = & \tB_{20} + \tB_{21}\delta +O(\delta^2) \ ,
\end{eqnarray}
where $H_1|_{x=x_{\rm H}}=0$ fixes the gauge freedom.

Let us write the field equations as
\begin{equation}
E^\mu = \frac{1}{\sqrt{-g}} D_\nu(\sqrt{-g}F^{\nu\mu}) = 0 \ , \ \ \ \
E_{\mu\nu} = G_{\mu\nu} - 8 \pi G T_{\mu\nu} = 0 \ .
\end{equation}
The expansion of $E^t$ at the horizon yields
\begin{equation}
\tB_{11} = \tH_{40}(\oh-\omega_1) \sin \theta \ , \ \ \ \
\tB_{21} = \tH_{30}(\oh-\omega_1) \sin \theta \ .
\end{equation}
With this result the expansion of $E^\vphi$ leads to the conditions
\begin{eqnarray}
\left[(\tB_{10}-\oh) \tH_{30}\sin\theta
        - \tB_{20} (\tH_{40}\sin\theta +1)\right]\tB_{20}
& = & 0 \ ,
\nonumber\\
\left[(\tB_{10}-\oh) \tH_{30}\sin\theta
        - \tB_{20} (\tH_{40}\sin\theta +1)\right](\tB_{10}-\oh)
& = & 0 \ .
\end{eqnarray}
In terms of $W$ and $\hPsi$ these conditions are equivalent to
\begin{equation}
\left. \left[\hPsi,[\hPsi,W]\right]\right|_{x=\xh} = 0
\Longleftrightarrow \left. [\hPsi,W]\right|_{x=\xh} = 0
\Longleftrightarrow \left. F_{t\vphi}\right|_{x=\xh} = 0
\ .
\label{bcWH}
\end{equation}

We now assume that the electro-static potential $\Psi$
is constant at the horizon, Eq.~(\ref{esp}),
i.~e. $\tB_{10}={\rm const}$ and $\tB_{20}=0$.

To discuss the boundary conditions (\ref{bcWH}),
let us first assume $\hPsi|_{x=\xh} \neq 0$ and
$W|_{x=\xh}=\lambda\hPsi|_{x=\xh}$, for some function
$\lambda(\theta)$. In this case the expansion yields for
the gauge potential
\begin{equation}
A_\mu dx^\mu =
\left\{ -\left[\frac{\tB_{10}}{\xh}+O(\delta^2)\right]dt
        +\left[\frac{\lambda}{\xh}(\tB_{10}-\oh)
                 +1+O(\delta^2)\right](d\vphi+\frac{\omega}{x}dt)
\right\}\frac{\tau_z}{2}
+O(\delta^4) \ .
\end{equation}
The functions $H_1$, $1-H_2$, $\tH_3$ and $\tB_2$ vanish at least
up to order $O(\delta^4)$,
indicating that we find only embedded Abelian solutions
for $\hPsi|_{x=\xh} \neq 0$.

Assuming $\hPsi|_{x=\xh} = 0$, however,
does not imply restrictions on $H_{20}$, $\tH_{30}$ and
$\tH_{40}$. In this case the expansion yields
$H_{21}=\tH_{31}=\tH_{41}=0$ and $\omega_1=\omega_{\rm H}$.
The last condition implies $\tB_{11}=\tB_{21}=0$.
In this case non-Abelian solutions are possible.

\subsection{General expansion at the horizon}

Here we present the expansion of the functions
of the stationary axially symmetric black hole solutions
at the horizon $x_{\rm H}$ in powers of $\delta$.
These expansions can be obtained from the regularity conditions
imposed on the Einstein equations and the matter field equations:

\begin{eqnarray}
f(\delta,\theta)&=&\delta^2 f_2 \bigg\{1 -\delta
+  \frac{\delta^2}{24}\bigg[\frac{1}{x_{\rm H}^2}\frac{f_2}{l_2}
\bigg[24 \cot{\theta}[-(-H_{30,\theta} +1 -{H_{20}}^2) H_{30} \nonumber\\
&&- H_{20} H_{40,\theta} + H_{40} H_{40, \theta}]
+ 12 \bigg(H^2_{20}(H^2_{30}+H^2_{40}-1)+\frac{(H_{20}-H_{40})^2
+H^2_{30}}{\sin^2{\theta}} \nonumber\\
&&-(H^2_{30}+H^2_{40})+2 H_{20} (-H_{30} H_{40,\theta}
+H_{40} H_{30,\theta})+1 -2 H_{30,\theta} +H^2_{30,\theta}\nonumber\\
&&+H^2_{40,\theta} \bigg) \bigg]
-2 \cot{\theta} \bigg(3\frac{f_{2,\theta}}{f_2}
- 2 \frac{l_{2,\theta}}{l_2}\bigg)
- \bigg(3 \frac{f_{2,\theta}}{f_2}\frac{l_{2,\theta}}{l_2}
+ 6\frac{f_{2,\theta \theta}}{f_2}
+ \left(\frac{l_{2,\theta}}{l_2}\right)^2  \nonumber \\
&& -2 \frac{l_{2,\theta\theta}}{l_2}
-18-6\left(\frac{f_{2,\theta}}{f_2}\right)^2 \bigg)
- \frac{24}{f_2} \bigg[ 4 \sin{\theta} \frac{\omega_2}{x_{\rm H}}
 \bigg( H_{30}B_{12} + (1-H_{40}) B_{22}\bigg)   \nonumber \\
&&-2(B^2_{12} + B^2_{22})
-\sin^2{\theta}\frac{\omega^2_2}{x^2_{\rm H} f_2} \bigg( x^2_{\rm H} l_2
+ 2 f_2 \left(H^2_{30}+(1-H_{40})^2\right) \bigg) \bigg] \bigg] \bigg\}
+O(\delta^5)\ , \nonumber\\
m(\delta,\theta)&=&\delta^2 m_2 \bigg\{1-3 \delta
+ \frac{\delta^2}{24} \bigg[150 -4 \frac{l_{2,\theta \theta}}{l_2}
+ 2 \bigg(\frac{l_{2,\theta}}{l_2}\bigg)^2
+ 3 \frac{l_{2,\theta}}{l_2}\frac{m_{2,\theta}}{m_2}
-6 \frac{m_{2,\theta \theta}}{m_2}  \nonumber\\
&&+6 \bigg( \frac{m_{2,\theta }}{m_2} \bigg)^2
-6 \bigg( \frac{f_{2, \theta}}{f_2} \bigg)^2
+ 2 \cot{\theta} \bigg( 3  \frac{m_{2, \theta}}{m_2}
-4  \frac{l_{2,\theta}}{l_2}\bigg)
+ 24 \sin^2{\theta}\frac{l_2 \omega^2_2}{f^2_2} \bigg] \bigg\}
+ O(\delta^5)\ , \nonumber \\
l(\delta,\theta)&=&\delta^2 l_2 \bigg\{1 -3 \delta
+ \frac{\delta^2}{12} \bigg[ \bigg( \frac{l_{2,\theta}}{l_2}\bigg)^2
- 2 \frac{l_{2,\theta \theta}}{l_2}
+75- 4 \cot{\theta}\frac{l_{2,\theta}}{l_2}\bigg]\bigg\}
+ O(\delta^5)\ ,\nonumber\\
\omega(\delta,\theta)&=&\omega_{\rm H} (1 + \delta)
+ \delta^2\omega_2 + O(\delta^4)\ ,\nonumber\\
H_1(\delta,\theta)&=&\delta \left(1 -\frac{1}{2}\delta \right) H_{11}
+ O(\delta^3)\ , \nonumber\\
H_2(\delta,\theta)&=&H_{20}
+\frac{\delta^2}{4} \bigg[\frac{m_2}{l_2}\bigg(H_{20} (H^2_{30}
+H^2_{40}-1)-H_{30}H_{40,\theta}+H_{40} H_{30,\theta}
+\frac{H_{20}-H_{40}}{\sin^2{\theta}} \nonumber\\
&&-\cot{\theta} (-2 H_{20} H_{30} + H_{40,\theta})\bigg)-(H_{11,\theta}
+ H_{20,\theta \theta})\bigg] + O(\delta^3)\ , \nonumber \\
H_3(\delta,\theta)&=&H_{30}
-\frac{\delta^2}{8} \bigg[-\bigg(2\frac{f_{2,\theta}}{f_2}
-\frac{l_{2,\theta}}{l_2}\bigg) (1-H_{40}H_{20}-H_{30,\theta}
-\cot{\theta} H_{30}) \nonumber\\
&& -2\cot{\theta} H_{20}(H_{20}-H_{40})+2H_{30,\theta \theta}
+ 4 H_{20} H_{40,\theta} -2\bigg(\frac{H_{30}}{\sin^2{\theta}}
-2 \cot{\theta} H_{30,\theta}\bigg)\nonumber\\
&&-2 H_{30}H^2_{20} - 2 H_{40}(2H_{11}-H_{20,\theta})
-8\sin{\theta}\frac{l_2 \omega_2}{f^2_2}(x_{\rm H} B_{12}
-\sin{\theta} \omega_2 H_{30}) \bigg] + O(\delta^3)\ , \nonumber\\
H_4(\delta,\theta)&=&H_{40}
-\frac{\delta^2}{8}\bigg[ \bigg(2\frac{f_{2,	\theta}}{f_2}
- \frac{l_{2,\theta}}{l_2}\bigg) [H_{40,\theta}-H_{20} H_{30}
-\cot{\theta} (H_{20}-H_{40})]\nonumber\\
&&+H_{20}(-4 H_{30,\theta} +2) + 2 [H_{30}(2 H_{11}-H_{20,\theta})
+H_{40,\theta \theta}-H_{40} H^2_{20}]\nonumber\\
&&+ 2 \frac{H_{20}-H_{40}}{\sin^2{\theta}} - 2 \cot{\theta}(-2 H_{11}
-H_{40,\theta} + H_{20} H_{30} + H_{20,\theta})\nonumber\\
&&+8 \sin{\theta}\frac{l_2 \omega_2}{f^2_2}[ x_{\rm H} B_{22}
-\sin{\theta}\omega_2 (1-H_{40})] \bigg] + O(\delta^3)\ ,\nonumber\\
{\bar B}_1(\delta,\theta)&=&-\frac{\omega_{\rm H} \cos{\theta}}{x_{\rm H}}
+\delta^2 (1-\delta) B_{12} + O(\delta^4)\ ,\nonumber\\
{\bar B}_2(\delta,\theta)&=&\frac{\omega_{\rm H} \sin{\theta}}{x_{\rm H}}
+\delta^2 (1-\delta) B_{22} + O(\delta^4)\ . \label{hor_general}
\end{eqnarray}
$B_{12}$ and $B_{22}$ are functions of $\theta$.
Relations (\ref{relation_hor_1}) and (\ref{relation_hor_2}) also hold.

\subsection{Expansion: relation with static case}

The general expansion at the horizon
includes the expansion for the static case.
The static limit corresponds to setting
$\omega_{\rm H}=\omega_2=B_{12}=B_{22}=H_{11}=H_{31}=0$,
$m_2=l_2$, and $H_{40}=H_{20}$;
in addition, $f_2$, $l_2$, and $H_{20}$ become constant.
The expansion reads
\begin{eqnarray}
f(\delta)&=&\delta^2 f_2 \left\{1-\delta
+ \frac{\delta^2}{4}\left[\frac{2}{x^2_{\rm H}}\frac{f_2}{l_2}(H^2_{20}-1)^2
+3\right]\right\}+O(\delta^5)\ ,\nonumber\\
l(\delta)&=&\delta^2 l_2 \left(1-3\delta+\frac{25}{4}\delta^2\right)
+ O(\delta^5)\ ,\nonumber\\
H_2(\delta)&=&H_{20}\left[1+\frac{\delta^2}{4}(H^2_{20}-1)\right]
+ O(\delta^3)\ ,\label{hor_static}
\end{eqnarray}
recovering the known expressions
for the behavior at the horizon of the
static spherically symmetric black hole solutions.

\subsection{Expansion: embedded Kerr-Newman solutions}

For comparison, we present the  expansion at the horizon also for
the embedded Kerr-Newman solutions for our choice of coordinates
and gauge. The expansion reads

\begin{eqnarray}
f&=&\frac{4 x_{\rm H}^2 [(M+2 x_{\rm H})^2+a^2 \cos^2\theta]}{[2 M (M+2 x_{\rm H}) -(Q^2 +P^2)] ^2} \delta^2(1-\delta)
+  O(\delta^4)\ ,\\
m&=&\frac{4 [(M+2 x_{\rm H})^2+a^2 \cos^2\theta]^2}{[2 M (M+2 x_{\rm H}) -(Q^2 +P^2)] ^2} \delta^2(1-3\delta)
+  O(\delta^4)\ ,\\
l&=&4 \delta^2(1-3\delta) +  O(\delta^4)\ ,\\
\omega&=&\frac{a x_{\rm H}}{2 M (M+2 x_{\rm H}) -(Q^2 +P^2)}(1+\delta) \nonumber\\
&&- \frac{2 a x_{\rm H}^2}{[2 M (M+2 x_{\rm H}) -(Q^2 +P^2)] ^3} \{(M+2 x_{\rm H})[2 (M+2 x_{\rm H}) (M+x_{\rm H}) \nonumber \\
&&-(Q^2+P^2)] + 2 a^2 x_{\rm H} \cos^2\theta\}\delta^2 + O(\delta^3)\ ,\\
H_1&=&0\ ,\\
H_2&=&0\ ,\\
H_3&=&-\frac{1}{(M+2 x_{\rm H})^2+a^2\cos^2\theta} \Big\{\cot\theta \Big[(M+2 x_{\rm H})^2
+ P [2 M(M+2 x_{\rm H}) \nonumber\\
&&-(Q^2+P^2)] +a^2\cos^2\theta\Big]+ aQ(M+2 x_{\rm H}) \sin\theta\Big\} \nonumber\\
&&-\frac{a \sin\theta}{2 [(M+2 x_{\rm H})^2+a^2\cos^2\theta]^2} [2 a^2 Q x_{\rm H} \cos^2\theta
- 4 a P x_{\rm H} (M+2 x_{\rm H}) \cos\theta \nonumber\\
&&- 2 Qx_{\rm H} (M +2 x_{\rm H})^2] \delta^2 + O(\delta^3)\ ,\\
H_4&=&0\ ,\\
\bar{B}_1&=&\frac{Q (M+2x_{\rm H}) -a \cos\theta}{2 M (M+2 x_{\rm H}) -(Q^2 +P^2)}\nonumber\\
&&-\frac{1}{2[2 M (M+2 x_{\rm H}) -(Q^2 +P^2)]^3}
\Big\{2 Q x_{\rm H} \Big[ -[M (M+2x_{\rm H})-(Q^2+P^2)]^2\nonumber\\
&&+ (M+2 x_{\rm H})^2[(M+2x_{\rm H})^2 + 4 x_{\rm H}^2]\Big]-4a x_{\rm H} \Big[(M+2 x_{\rm H})[4 M P x_{\rm H} \nonumber\\
&&-(Q^2+P^2)]-2 P x_{\rm H} (Q^2+P^2)+ 2 (M+2 x_{\rm H})^2(M+x_{\rm H})\Big] \cos\theta\nonumber\\
&&+8 a^2 Q x_{\rm H}^2(M +2 x_{\rm H}) \cos^2\theta - 8 a^3 x_{\rm H}^2 \cos^3\theta \Big\}\delta^2 + O(\delta^3)\ ,\\
\bar{B}_2&=&\frac{a \sin\theta}{2 M (M+2 x_{\rm H}) -(Q^2 +P^2)}\nonumber\\
&& -\frac{2 a x_{\rm H} \sin\theta}{[2 M (M+2 x_{\rm H}) -(Q^2 +P^2)]^3}
\{(M+4 x_{\rm H})[2M (M + 2x_{\rm H}) -(Q^2+P^2)]\nonumber\\
&& -2 a^2 x_{\rm H} \sin^2\theta \} \delta^2 + O(\delta^3)\ ,
\end{eqnarray}
where
\begin{eqnarray}
x_{\rm H}&=&\frac{1}{2} \sqrt{M^2-(a^2+Q^2+P^2)}\ ,\\
\delta&=&\frac{x}{x_{\rm H}} -1\ .
\end{eqnarray}

\newpage

\setcounter{fixy}{1}

   \begin{fixy} {0}
\begin{figure}\centering
{\large Fig. 1a} \vspace{0.0cm}\\
\epsfysize=8cm
\mbox{\epsffile{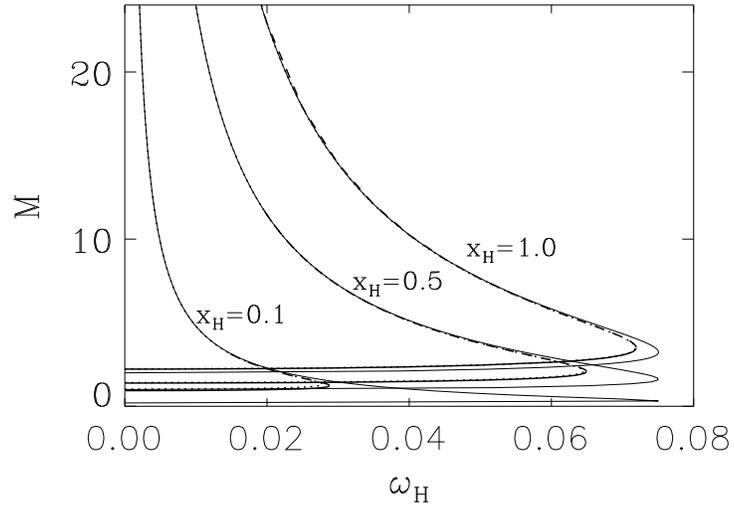}}
\caption{
The dimensionless mass $M$ is shown as a function of $\omega_{\rm H}$ for
node number $k=1$ and $x_{\rm H} = 1.0 $, $0.5$ and $0.1$.
For the same values of parameters
the dimensionless mass of the
Kerr solution (thin solid) and the
Kerr-Newman solution (dotted) for $Q=0$ and $|P|=1$ are also shown.
}
\end{figure}
\begin{figure}\centering
{\large Fig. 1b} \vspace{0.0cm}\\
\epsfysize=8cm
\mbox{\epsffile{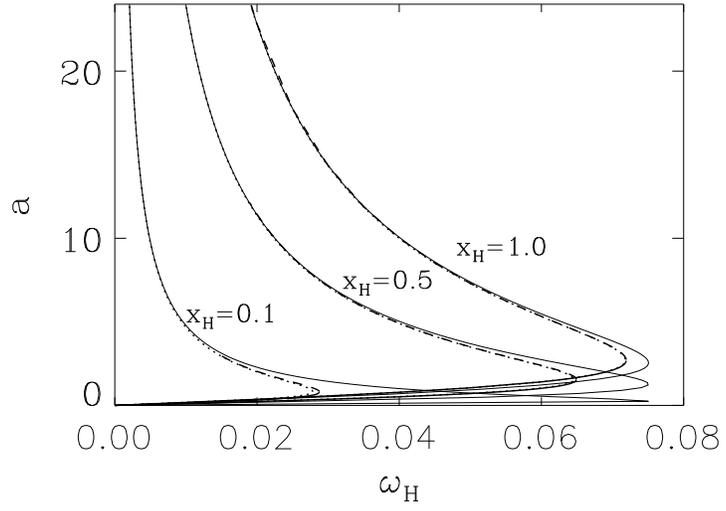}}
\caption{
Same as Fig.~1a for  the ratio  $a=J/M$ .
}
\end{figure}
 \begin{figure}\centering
{\large Fig. 1c} \vspace{0.0cm}\\
 \epsfysize=8cm
\mbox{\epsffile{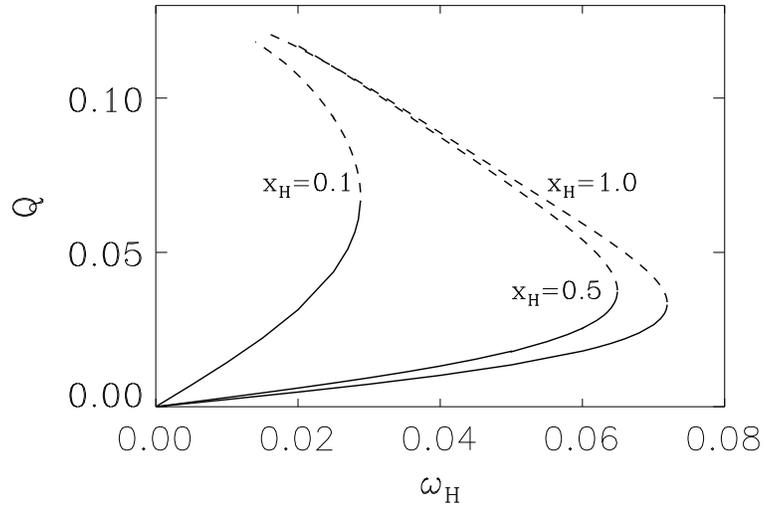}}
\caption{
The electric charge  $Q$ is shown as a function of $\omega_{\rm H}$ for
node number $k=1$ and $x_{\rm H} = 1.0 $, $0.5$ and $0.1$.
}
\end{figure}
     \end{fixy}

     \begin{fixy}{-1}
\begin{figure}\centering
{\large Fig.2} \vspace{1.cm}\\
a\mbox{\epsfysize=10cm  \epsffile{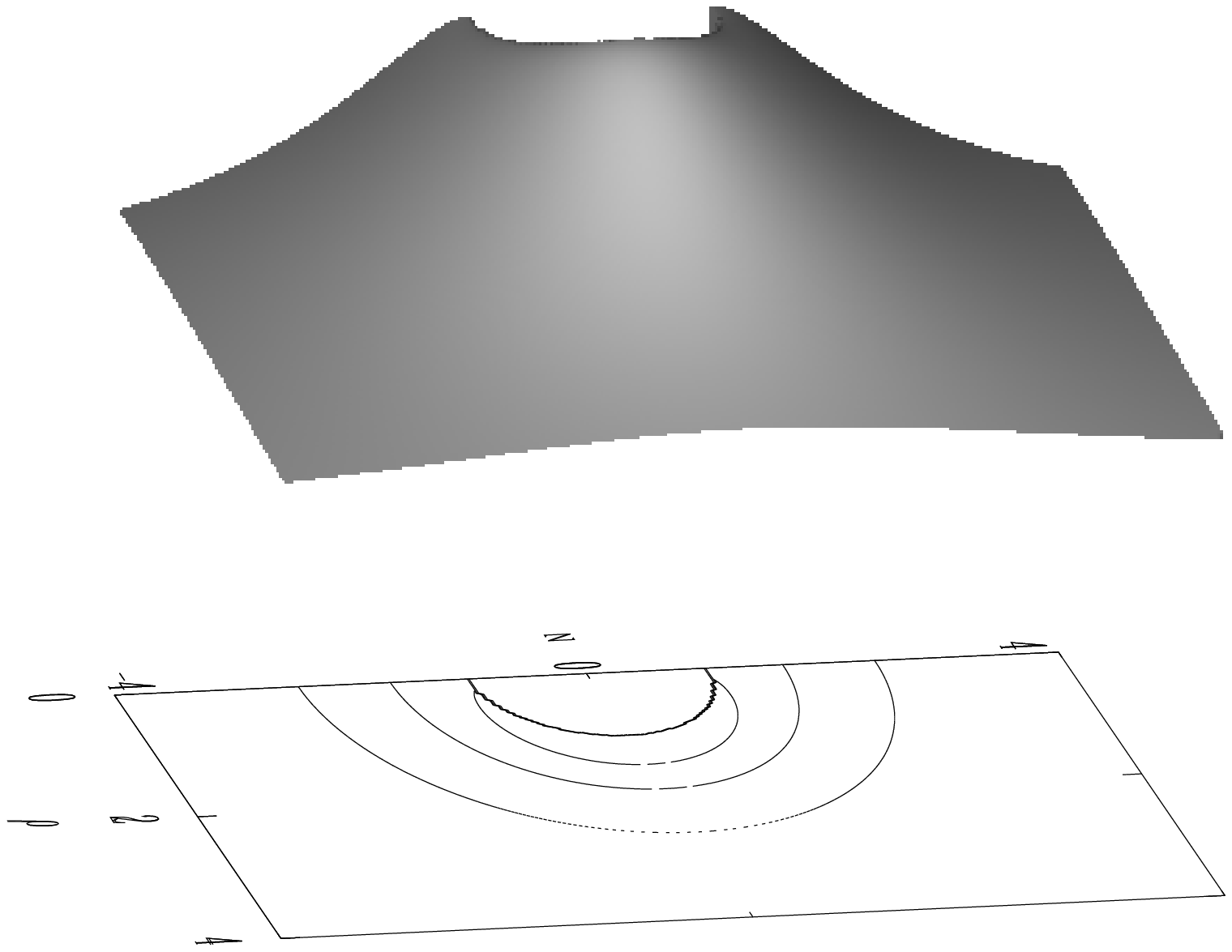}  }
\vspace{1.cm}\\
\begin{tabular}{ccc}
b & c & d \\
\mbox{\hspace*{5.mm}\epsfysize=4.cm\epsffile{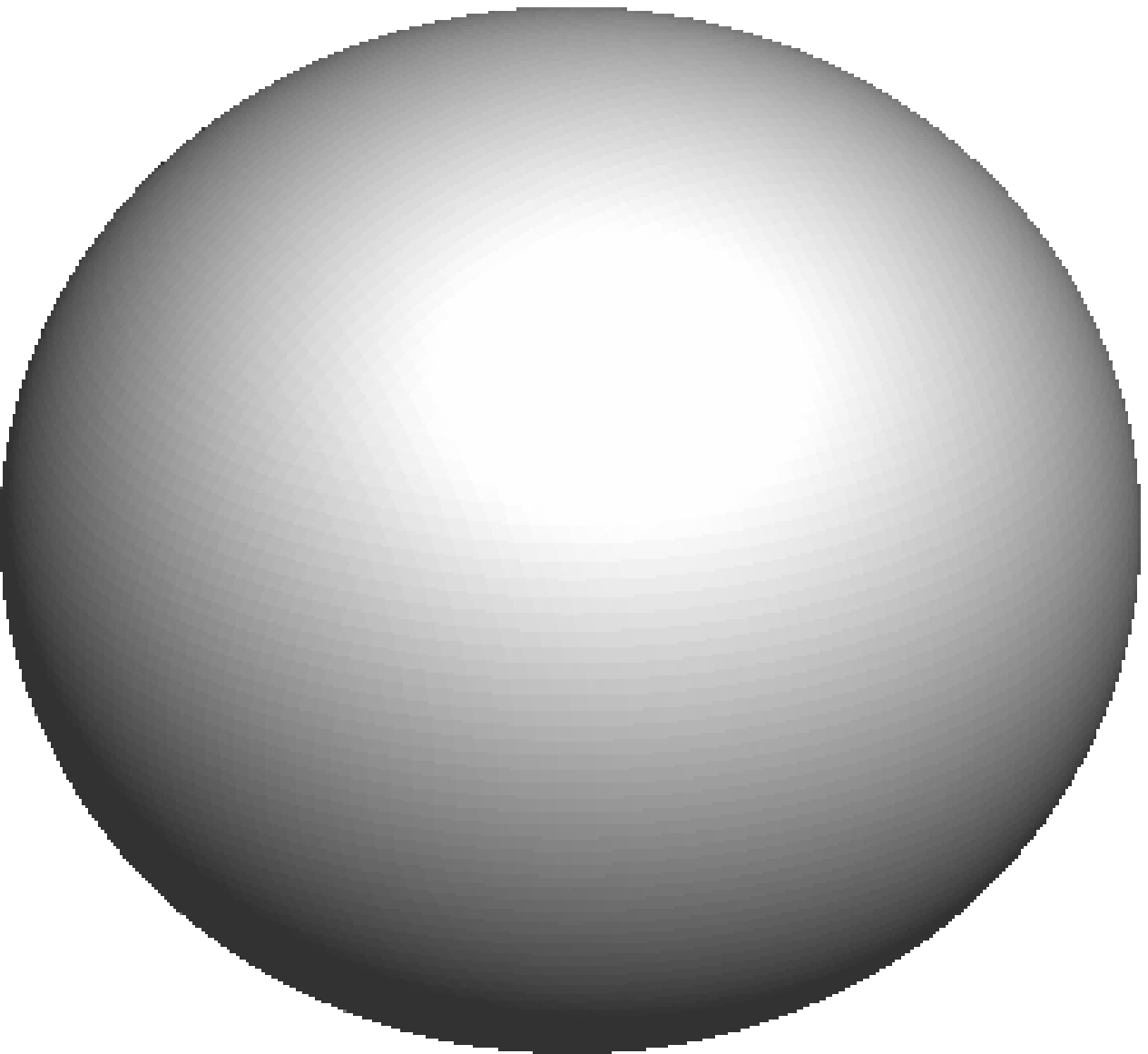}\hspace*{5.mm}}
&
\raisebox{4.mm}{\hspace*{5.mm}\epsfysize=3.2cm\epsffile{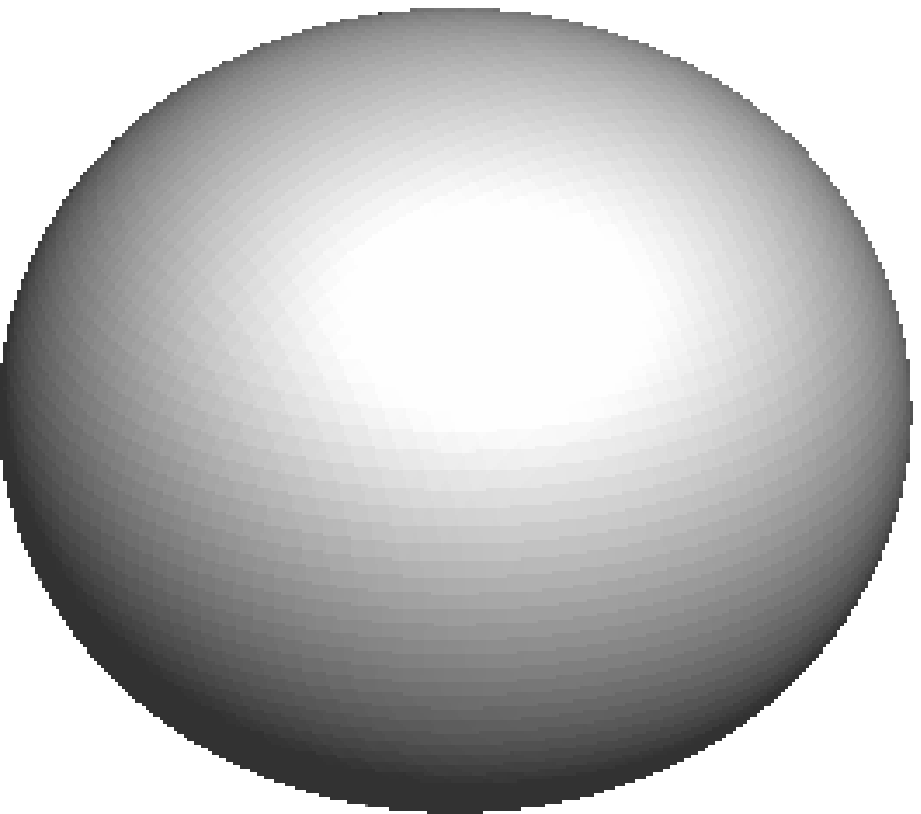}\hspace*{5.mm}}
&
\raisebox{6.mm}{\hspace*{5.mm}\epsfysize=2.4cm\epsffile{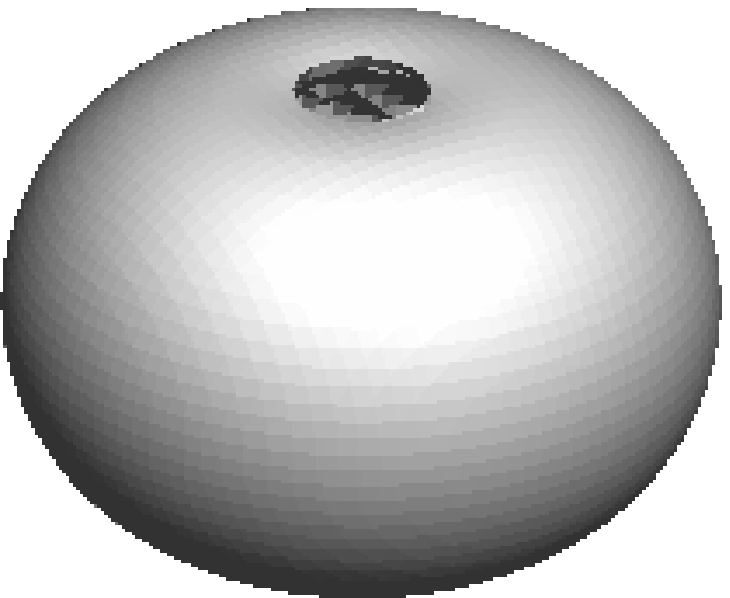}\hspace*{5.mm}}
\\
$\varepsilon = 0.0006$ &
$\varepsilon = 0.0009$ &
$\varepsilon = 0.0011$
\end{tabular}
\caption{
The energy density of the matter fields $\varepsilon=-T_0^0$ is shown as a
funtion of the coordiantes $\rho = x \sin\theta$, $z=x\cos\theta$
for $k=1$, $x_{\rm H}=1.0$, $\omega_{\rm H} = 0.04$ on the lower branch.
}
\end{figure}
      \end{fixy}

      \begin{fixy} {-1}
\begin{figure}\centering
{\large Fig. 3} \vspace{1.cm}\\
\mbox{\epsfysize=10cm  \epsffile{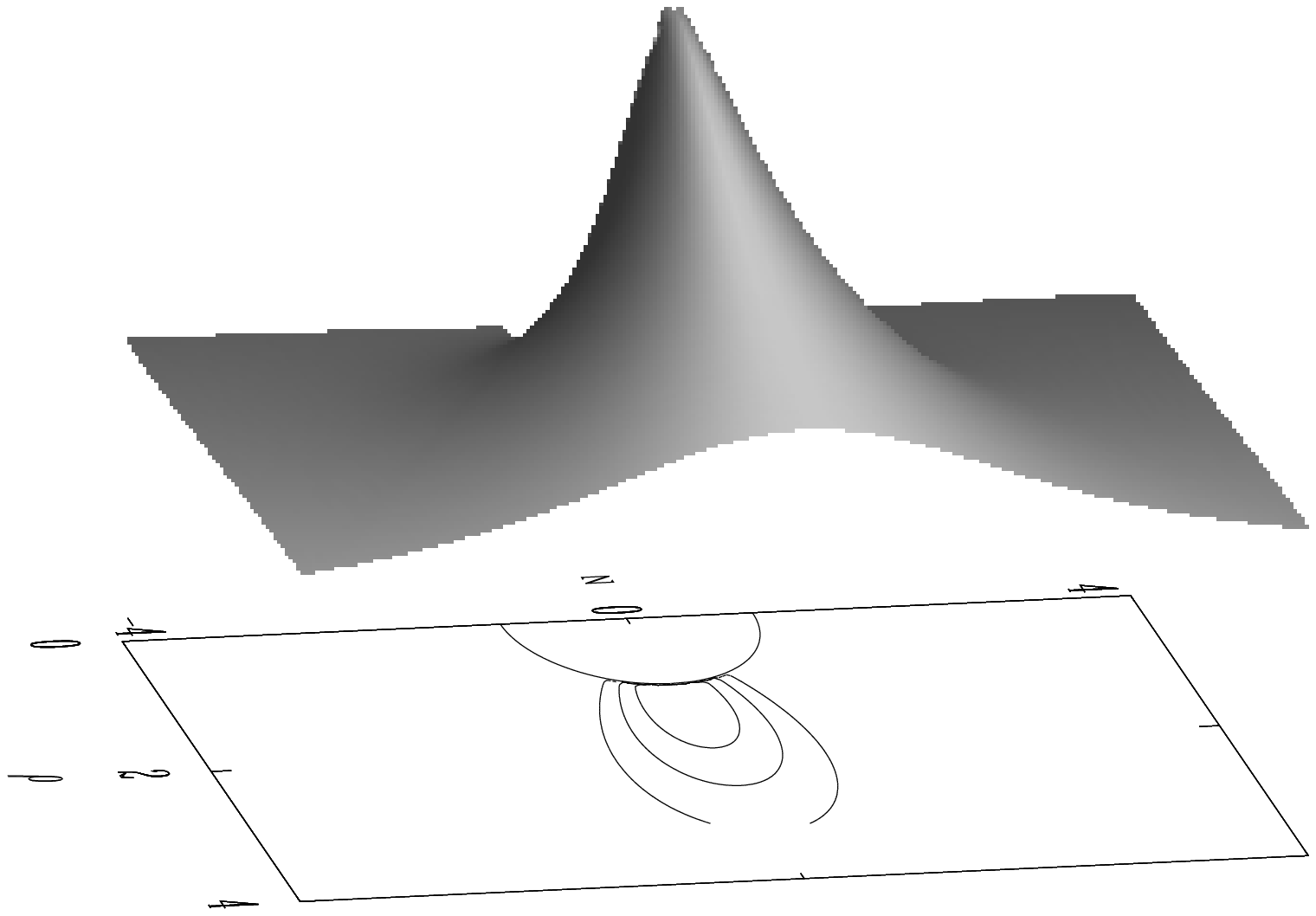}  }
\vspace{1.cm}\\
\begin{tabular}{ccc}
b & c & d \\
\mbox{\hspace*{5.mm}\epsfysize=3.cm\epsffile{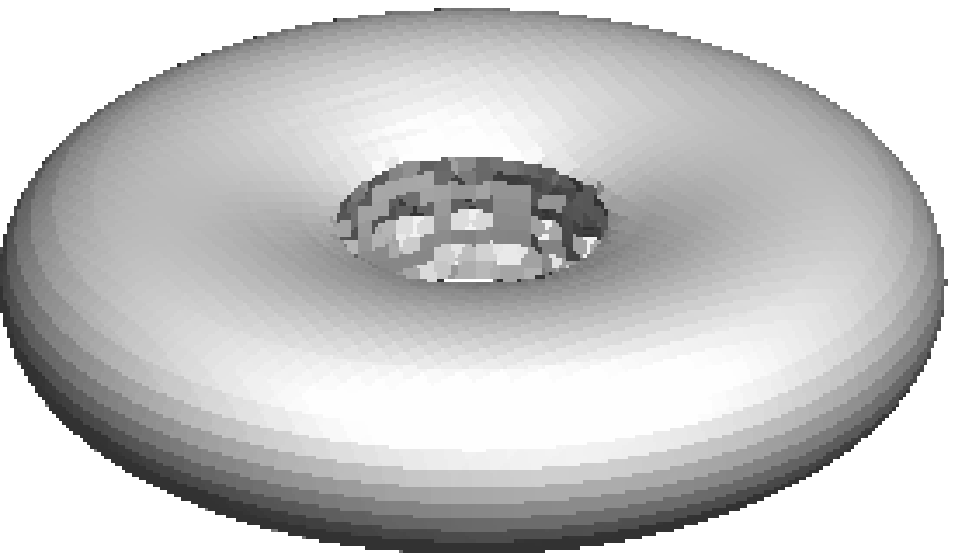}\hspace*{5.mm}}
&
\raisebox{4.mm}{\hspace*{5.mm}\epsfysize=2.1cm\epsffile{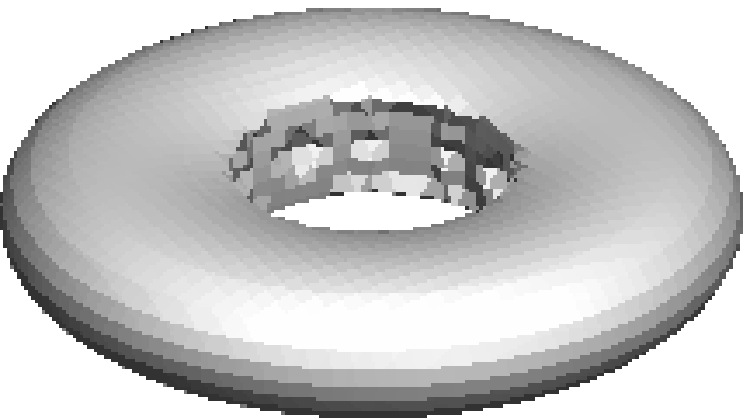}\hspace*{5.mm}}
&
\raisebox{6.mm}{\hspace*{5.mm}\epsfysize=1.6cm\epsffile{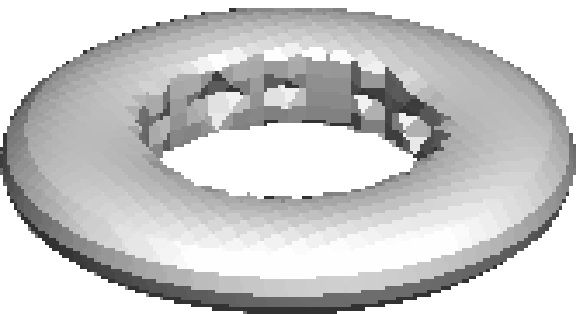}\hspace*{5.mm}}
\\
$\varepsilon = 0.00004$ &
$\varepsilon = 0.00005$ &
$\varepsilon = 0.00009$
\end{tabular}
\caption{
The same as Fig.~2 on the upper branch.
}
\end{figure}
       \end{fixy}

\clearpage

   \begin{fixy} {0}
\begin{figure}\centering
{\large Fig. 4a} \vspace{0.0cm}\\
\epsfysize=8cm
\mbox{\epsffile{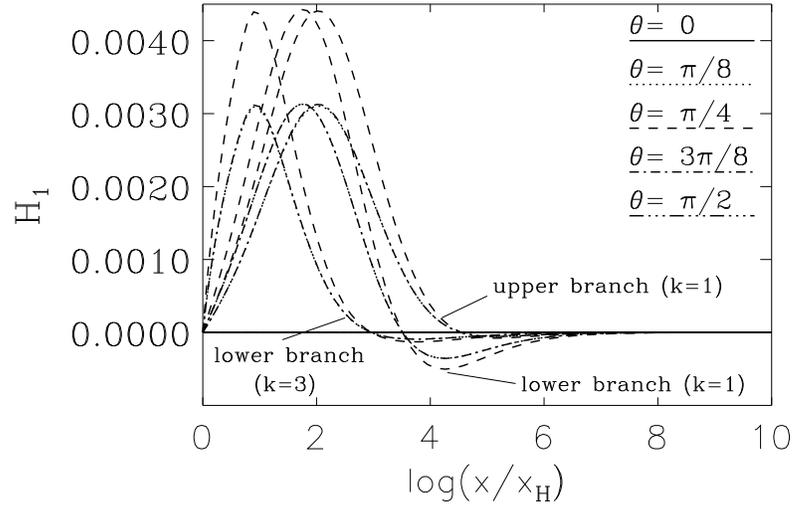}}
\caption{
The function $H_1$  is shown
for $x_{\rm H}=1.0$, $\omega_{\rm H}=0.04$ and $k=1$ on the lower and upper branch,
and for  $k=3$ on the lower  branch.
}
\end{figure}

\begin{figure}\centering
{\large Fig. 4b} \vspace{0.0cm}\\
\epsfysize=8cm
\mbox{\epsffile{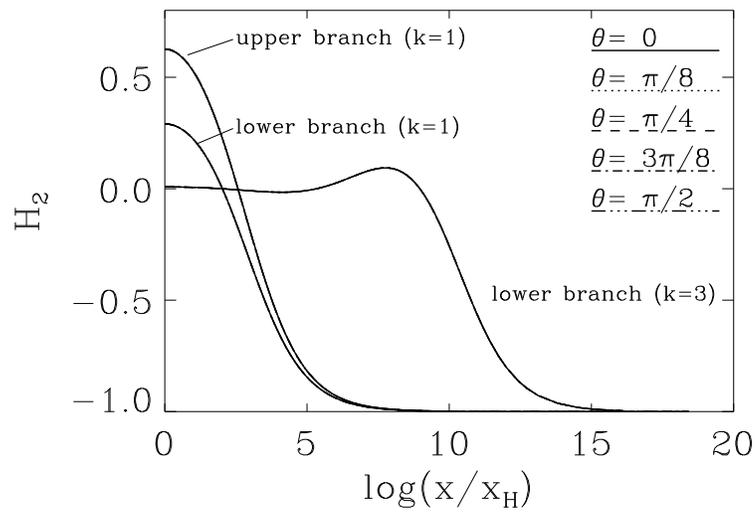}}
\caption{
Same as Fig.~4a for the function $H_2$.
}
\end{figure}

\begin{figure}\centering
{\large Fig. 4c} \vspace{0.0cm}\\
\epsfysize=8cm
\mbox{\epsffile{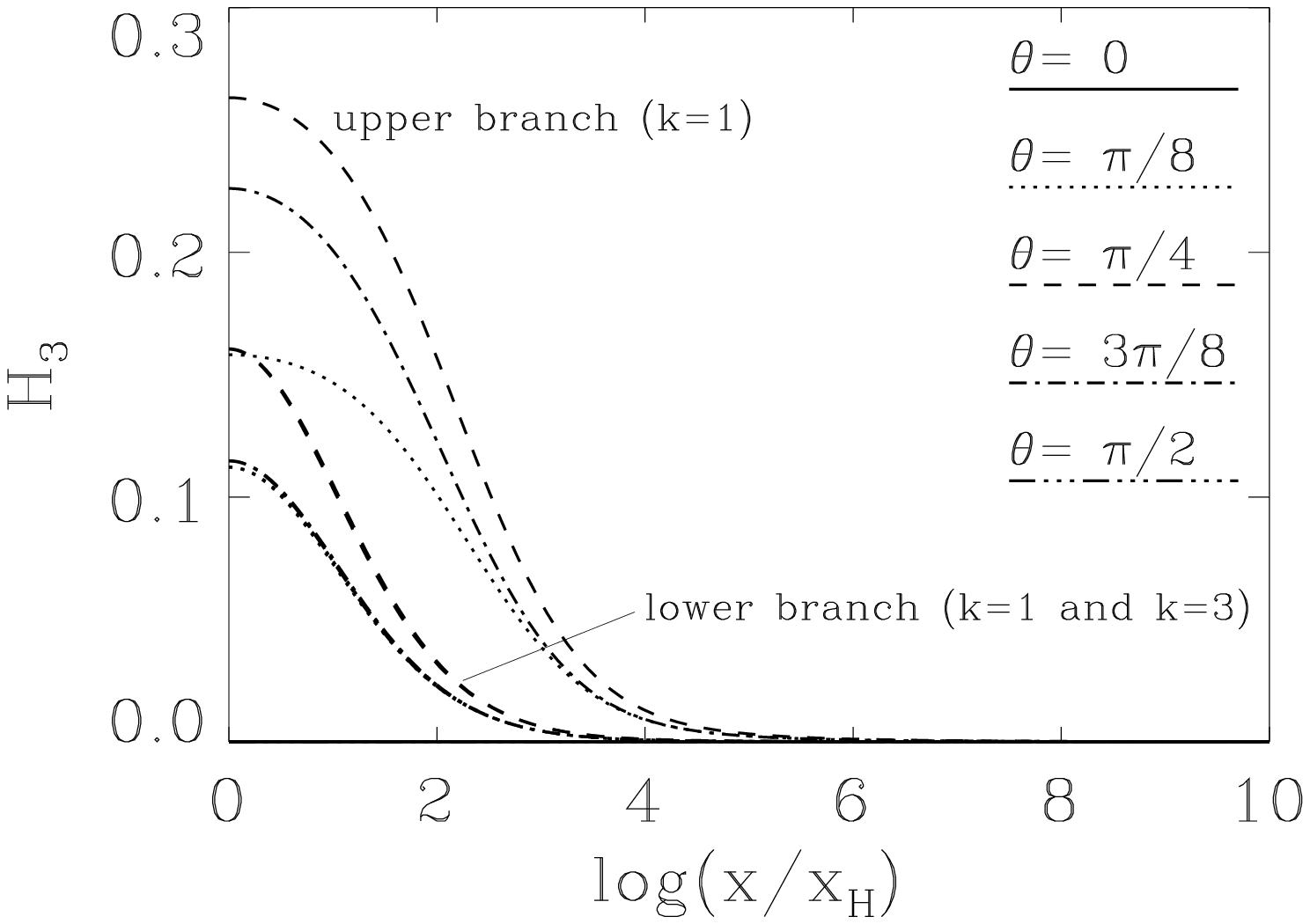}}
\caption{
Same as Fig.~4a for the function $H_3$.
The curves for $k=1$ and $k=3$ on the lower branch coincide.
}
\end{figure}

\begin{figure}\centering
{\large Fig. 4d} \vspace{0.0cm}\\
\epsfysize=8cm
\mbox{\epsffile{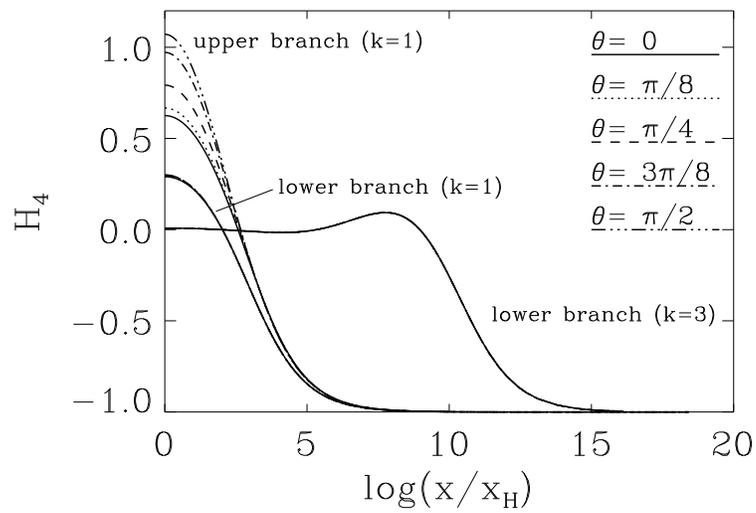}}
\caption{
Same as Fig.~5a for the function $H_4$.
}
\end{figure}

\begin{figure}\centering
{\large Fig. 4e} \vspace{0.0cm}\\
\epsfysize=8cm
\mbox{\epsffile{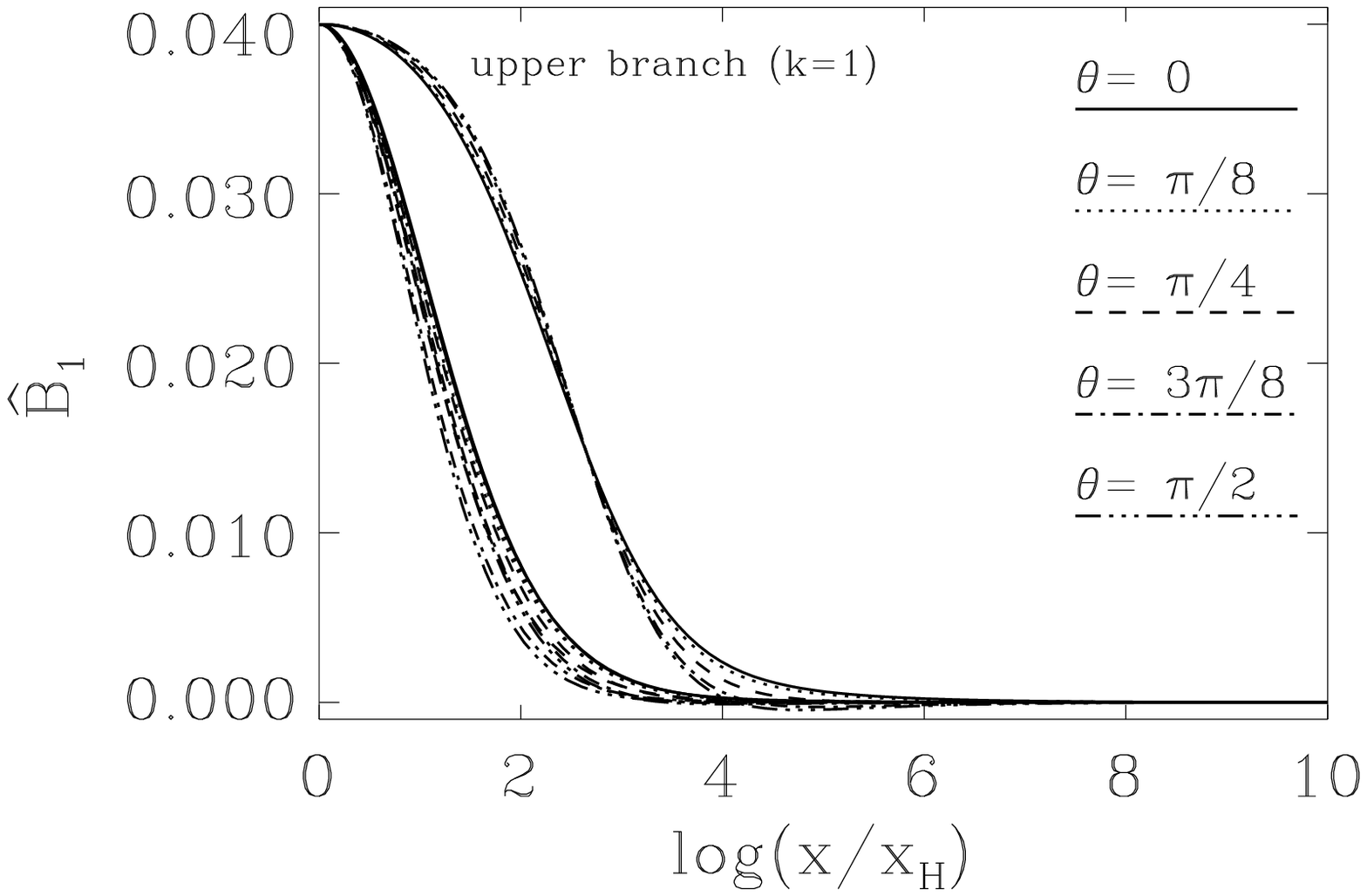}}
\caption{
Same as Fig.~4a for the function $\hat{B}_1$.
}
\end{figure}

\begin{figure}\centering
{\large Fig. 4f} \vspace{0.0cm}\\
\epsfysize=8cm
\mbox{\epsffile{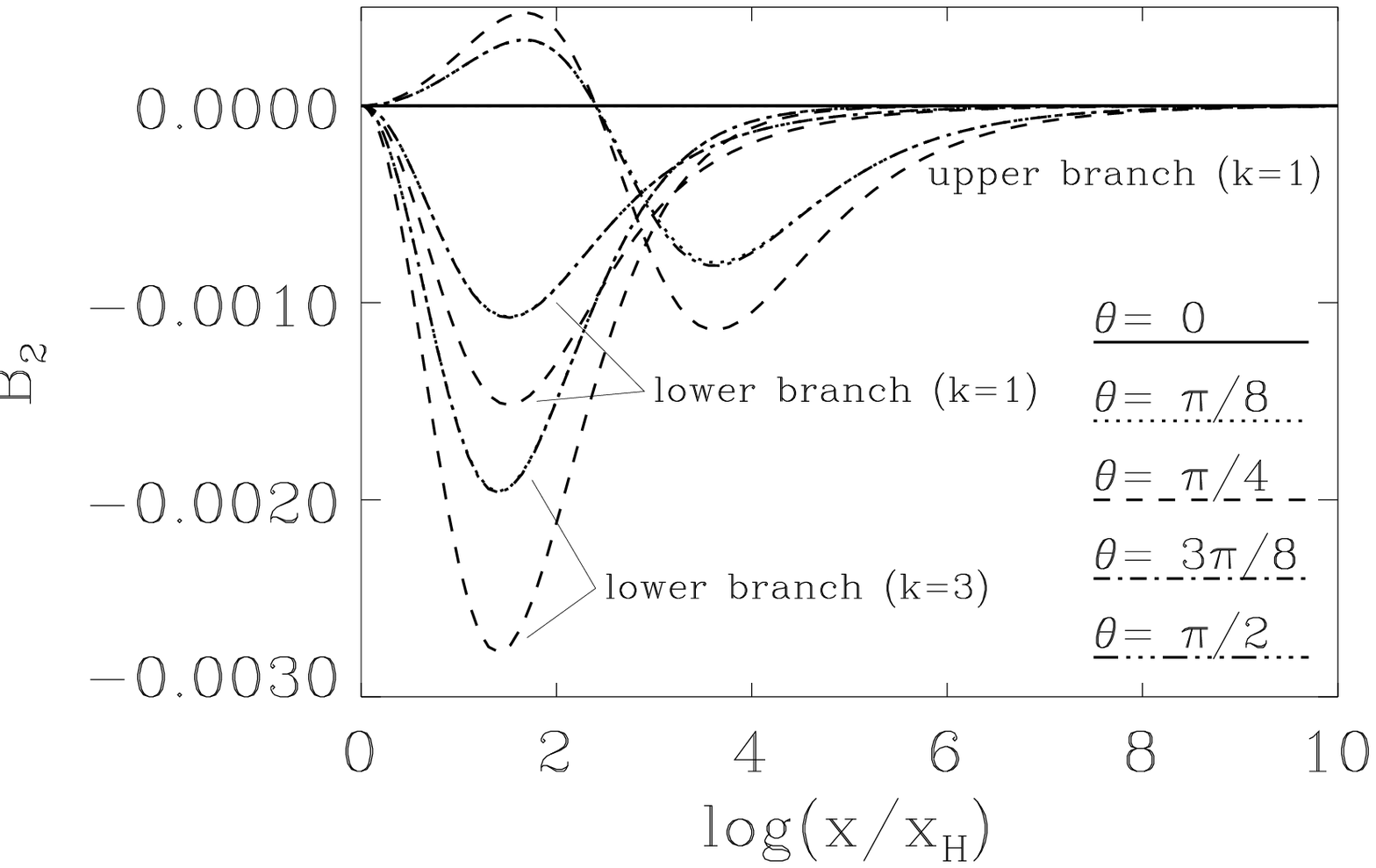}}
\caption{
Same as Fig.~4a for the function $\hat{B}_2$.
}
\end{figure}

\begin{figure}\centering
{\large Fig. 4g} \vspace{0.0cm}\\
\epsfysize=8cm
\mbox{\epsffile{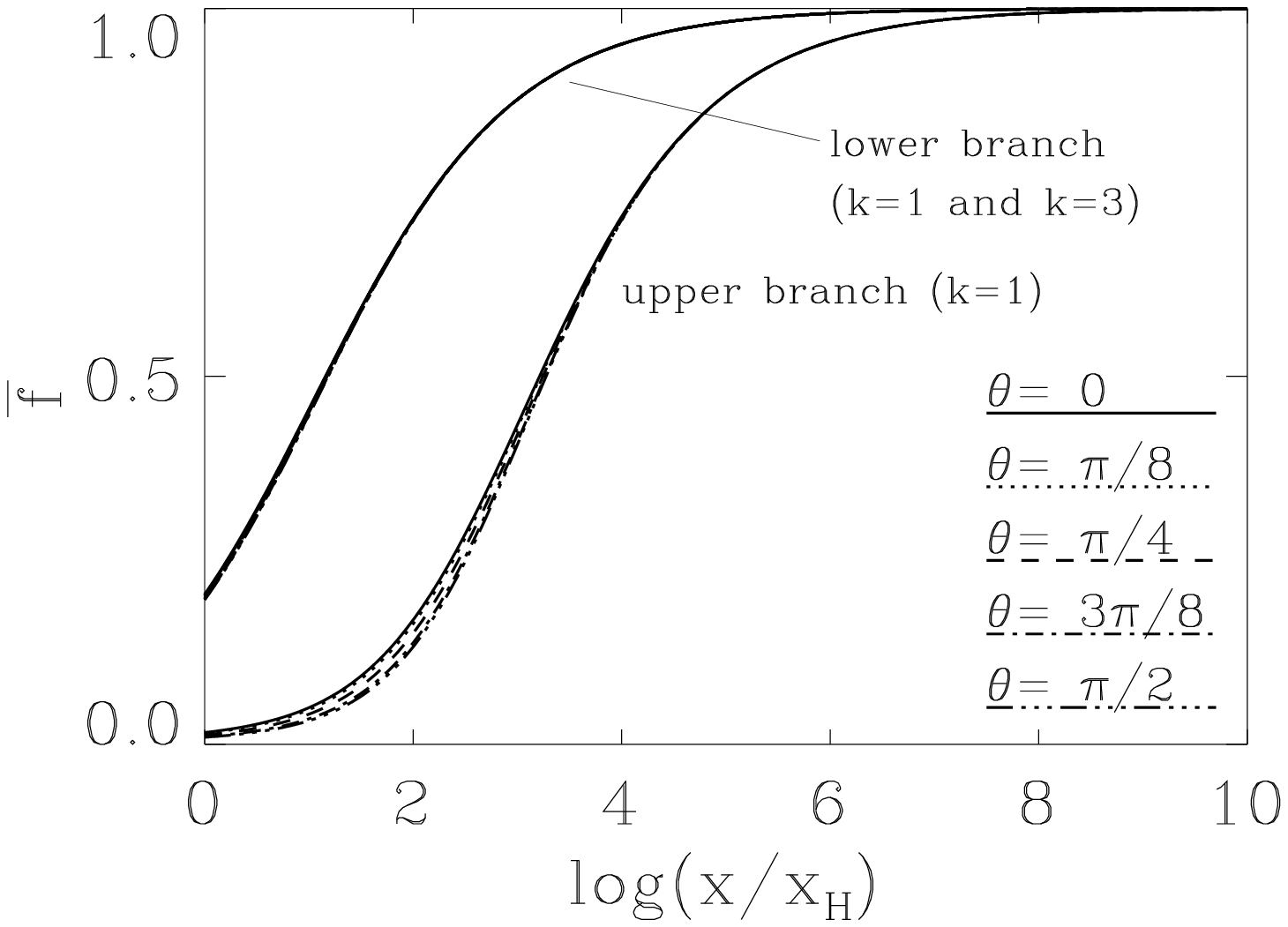}}
\caption{
Same as Fig.~4a for the function $\bar{f}$.
The curves for $k=1$ and $k=3$ on the lower branch coincide.
}
\end{figure}

 \begin{figure}\centering
{\large Fig. 4h} \vspace{0.0cm}\\
 \epsfysize=8cm
\mbox{\epsffile{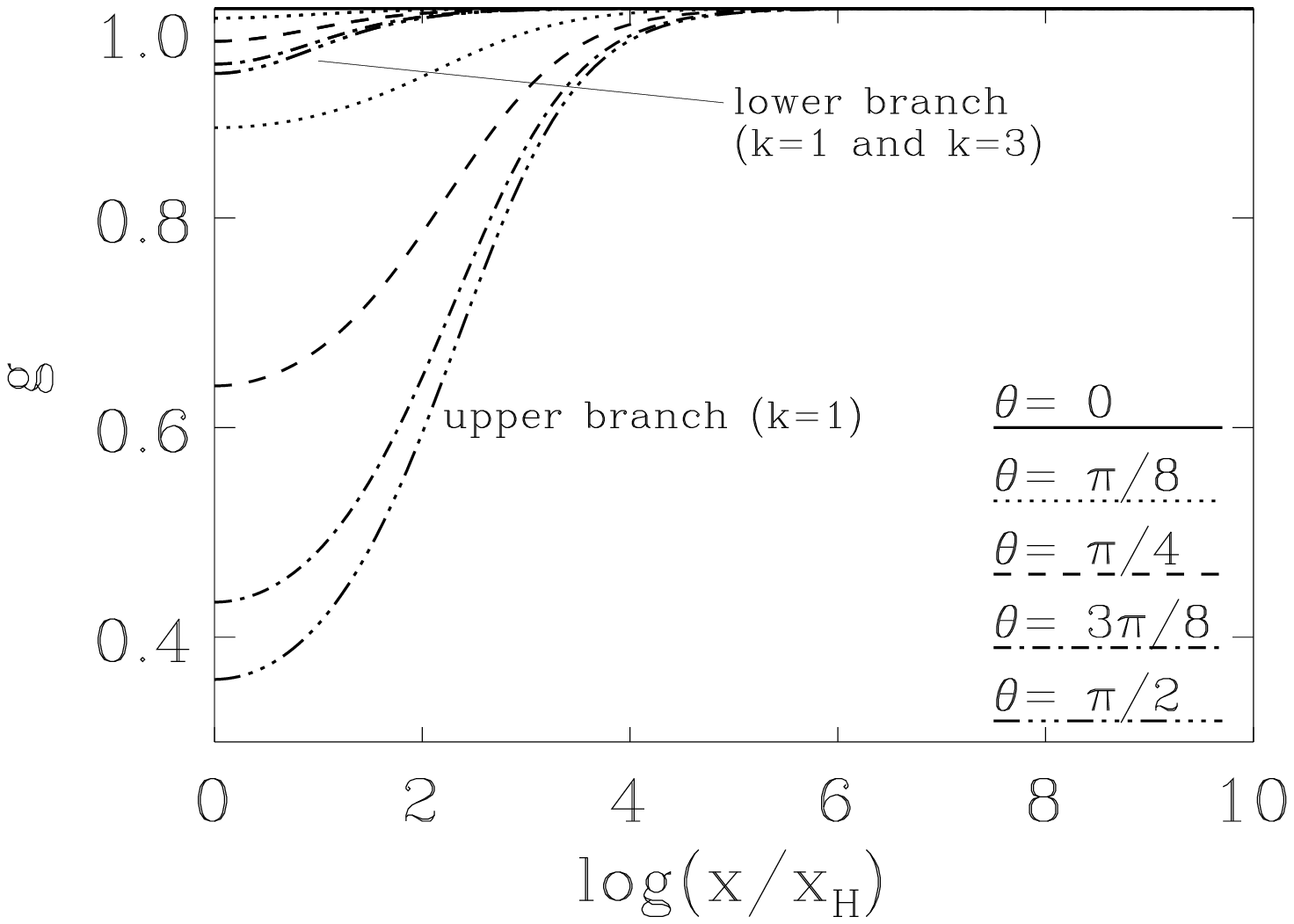}}
\caption{
Same as Fig.~4a for the function $g=m/l$.
The curves for $k=1$ and $k=3$ on the lower branch coincide.
}
\end{figure}

\begin{figure}\centering
{\large Fig. 4i} \vspace{0.0cm}\\
\epsfysize=8cm
\mbox{\epsffile{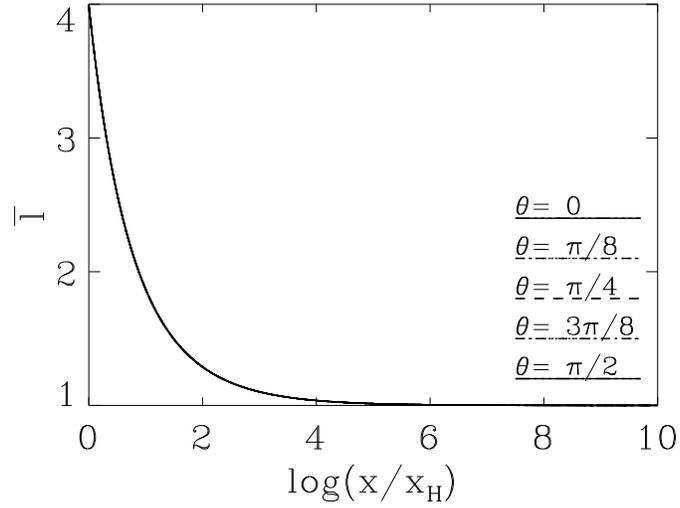}}
\caption{
Same as Fig.~4a for the function $\bar{l}$.
The curves for $k=1$ and $k=3$ on the lower branch
and $k=1$ on the upper branch coincide.
}
\end{figure}

\begin{figure}\centering
{\large Fig. 4j} \vspace{0.0cm}\\
\epsfysize=8cm
\mbox{\epsffile{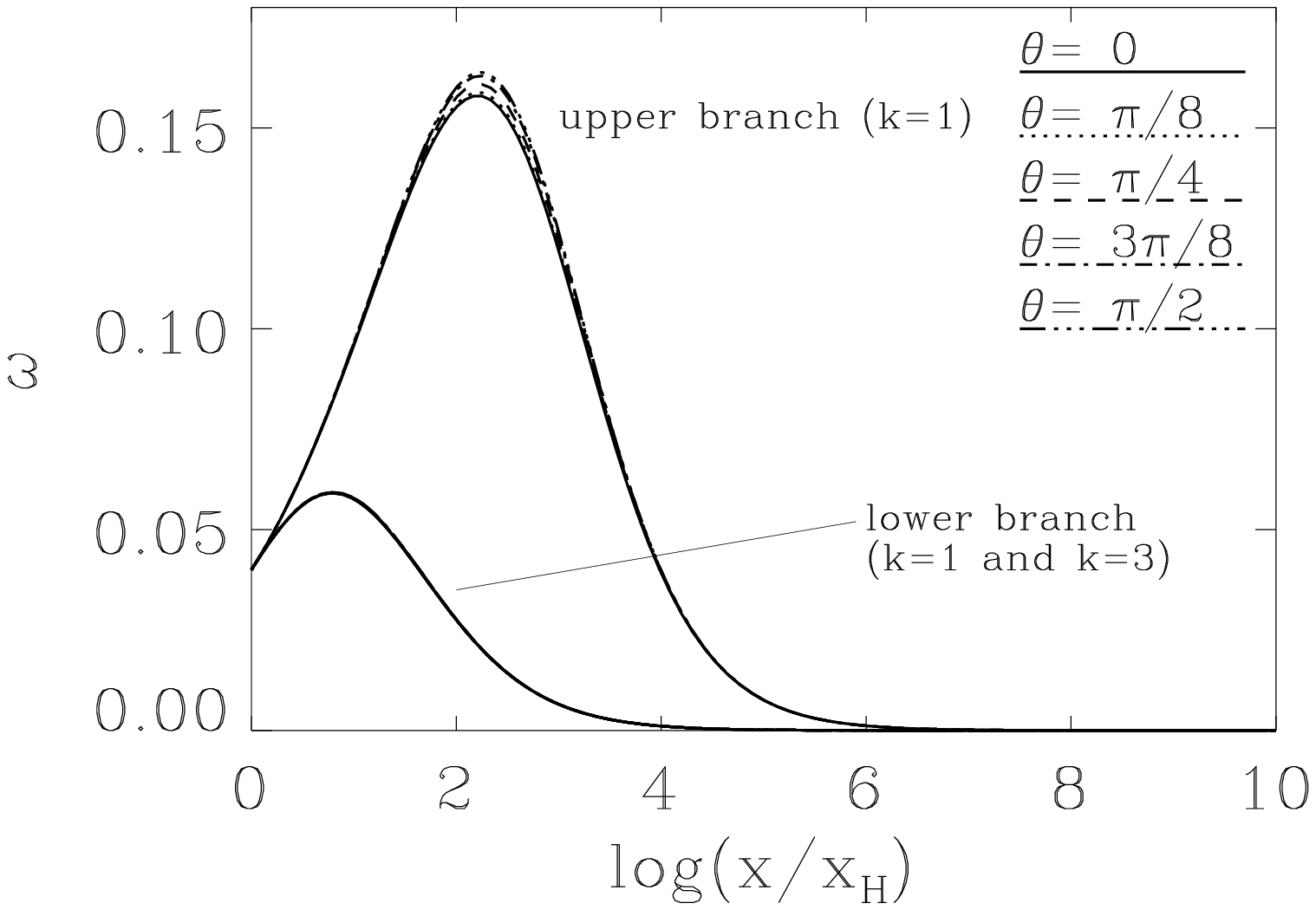}}
\caption{
Same as Fig.~4a for the function $\omega$.
The curves for $k=1$ and $k=3$ on the lower branch coincide.
}
\end{figure}
       \end{fixy}

 \clearpage

       \begin{fixy}{0}
\begin{figure}\centering
{\large Fig. 5a} \vspace{0.0cm}\\
\epsfysize=8cm
\mbox{\epsffile{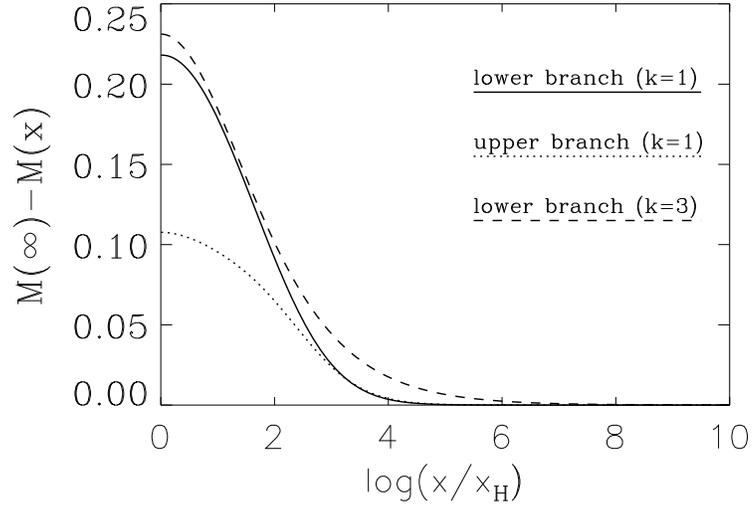}}
\caption{
The local mass  $M(x)$  is shown
for $x_{\rm H}=1.0$, $\omega_{\rm H}=0.04$ and $k=1$
on the lower and upper branch,
and for  $k=3$ on the lower  branch.
}
\end{figure}

\begin{figure}\centering
{\large Fig. 5b} \vspace{0.0cm}\\
\epsfysize=8cm
\mbox{\epsffile{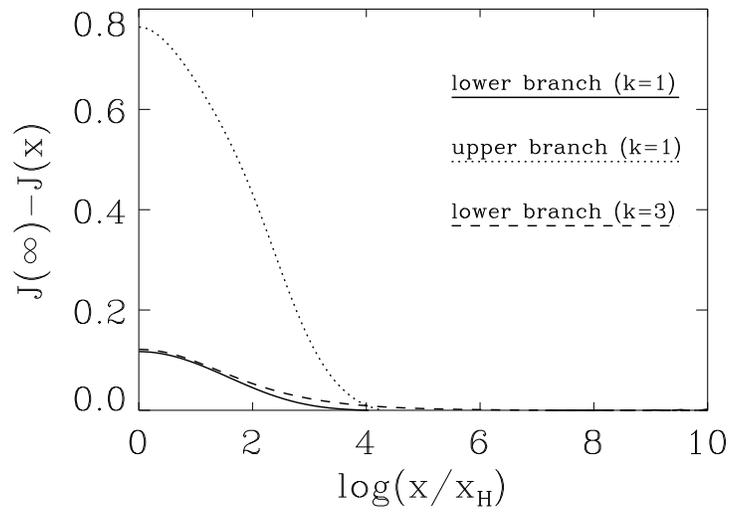}}
\caption{
The same as Fig.~5a for the local angular monetum $J(x)$.
}
\end{figure}

\begin{figure}\centering
{\large Fig. 5c} \vspace{0.0cm}\\
\epsfysize=8cm
\mbox{\epsffile{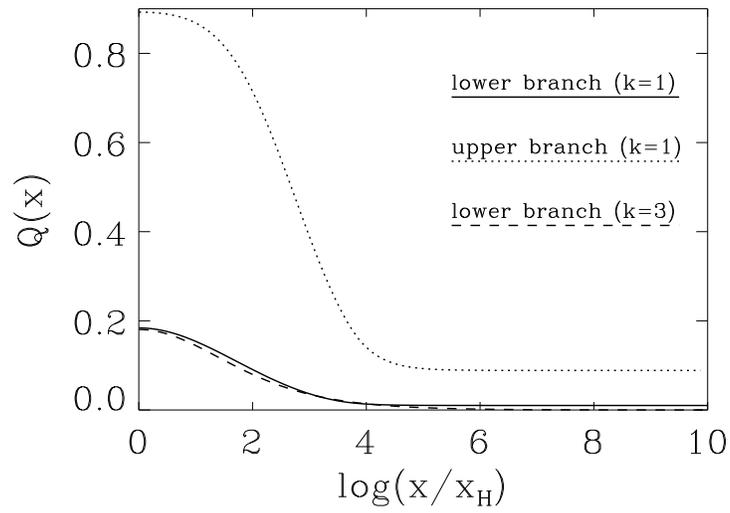}}
\caption{
The same as Fig.~5a for the local electric charge $Q(x)$.
}
\end{figure}

\begin{figure}\centering
{\large Fig. 5d} \vspace{0.0cm}\\
\epsfysize=8cm
\mbox{\epsffile{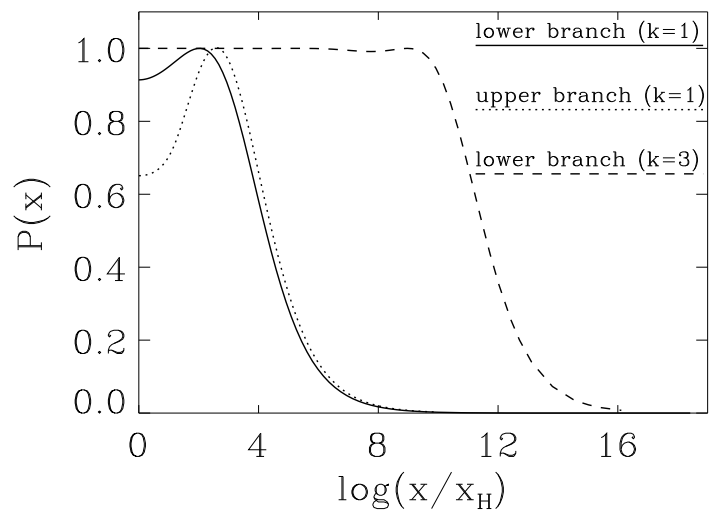}}
\caption{
The same as Fig.~5a for the local magnetic charge $P(x)$.
}
\end{figure}
     \end{fixy}

     \begin{fixy}{0}
\begin{figure}\centering
{\large Fig. 6a} \vspace{0.0cm}\\
\epsfysize=8cm
\mbox{\epsffile{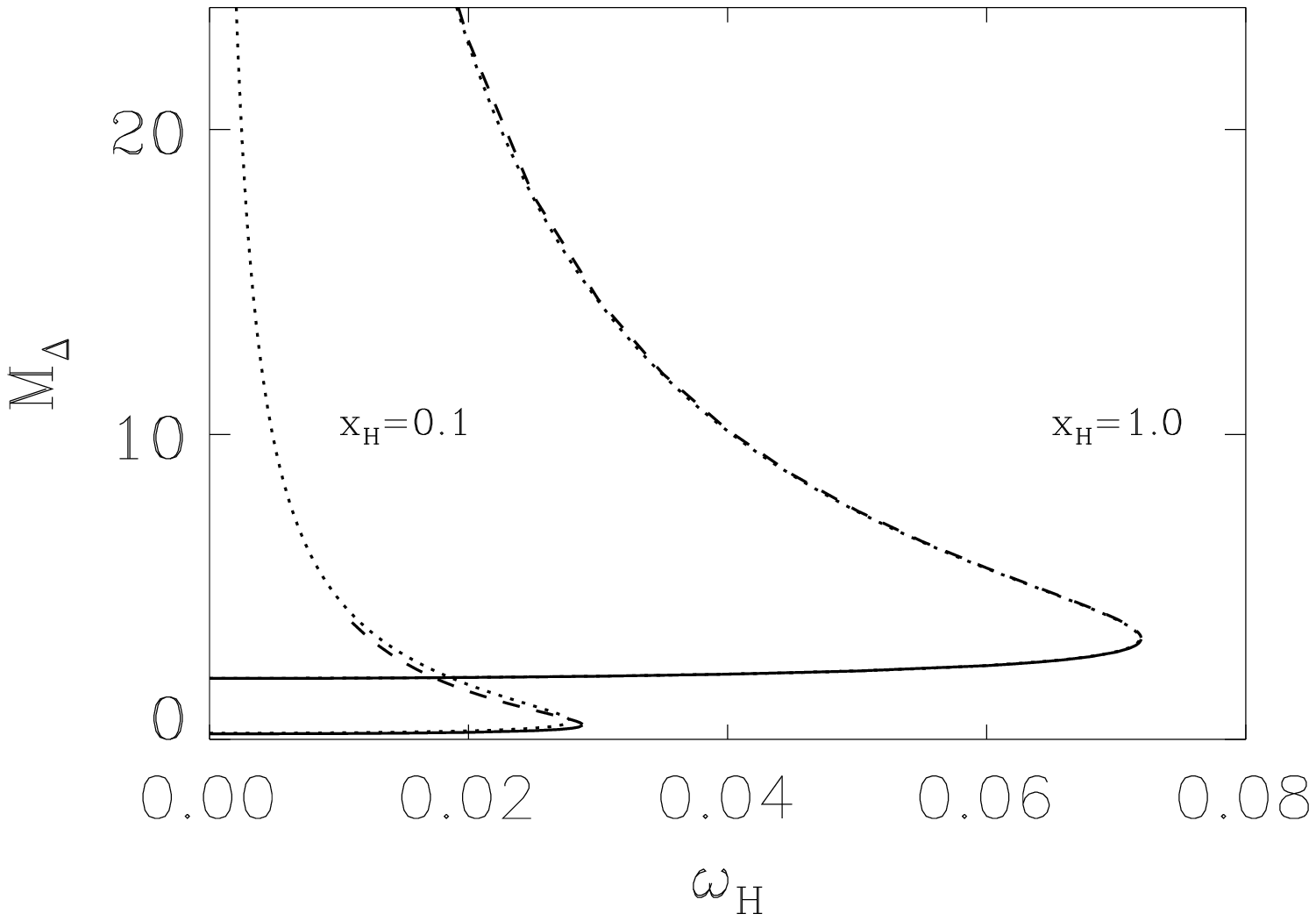}}
\caption{
The horizon mass $M_\Delta$ is shown as a function of $\omega_{\rm H}$
for $k=1$, $x_{\rm H}=1$ and $x_{\rm H}=0.1$ on the lower branch (solid)
and on the upper branch (dashed).
For the same values of parameters
the corresponding functions of the
Kerr solution (thin solid) and the
Kerr-Newman solution (dotted) for $Q=0$ and $|P|=1$ are also shown.
}
\end{figure}

\begin{figure}\centering
{\large Fig. 6b} \vspace{0.0cm}\\
\epsfysize=8cm
\mbox{\epsffile{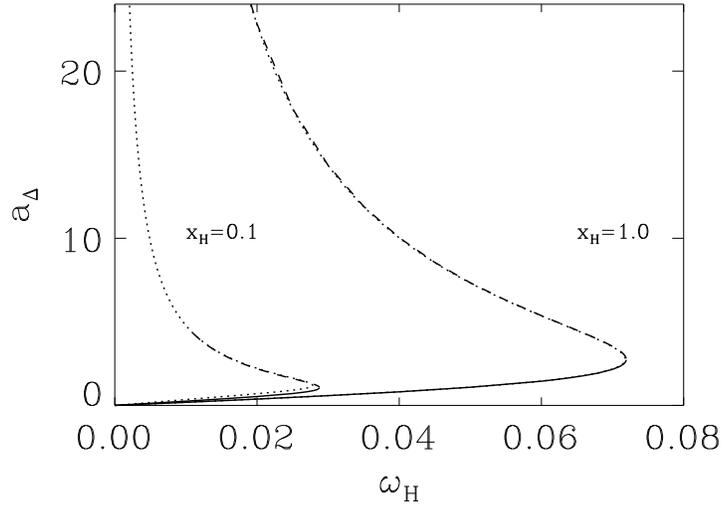}}
\caption{
The same as Fig.~6a for
the angular momentum per mass at the horizon
$a_\Delta=J_\Delta/M_\Delta$.
}
\end{figure}

\begin{figure}\centering
{\large Fig. 6c} \vspace{0.0cm}\\
\epsfysize=8cm
\mbox{\epsffile{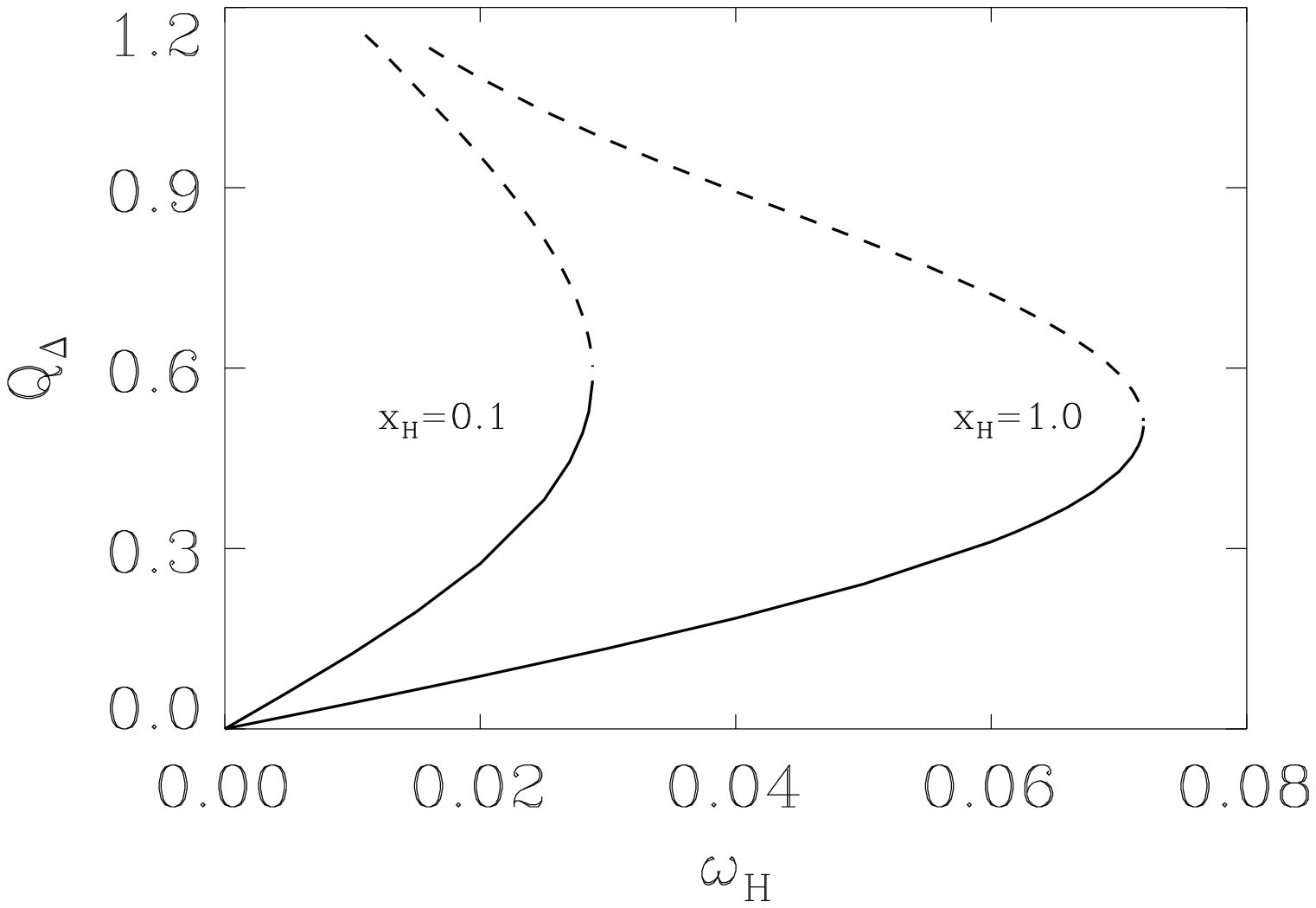}}
\caption{
The same as Fig.~6a for
the horizon electric charge
$Q_\Delta$.
}
\end{figure}

\begin{figure}\centering
{\large Fig. 6d} \vspace{0.0cm}\\
\epsfysize=8cm
\mbox{\epsffile{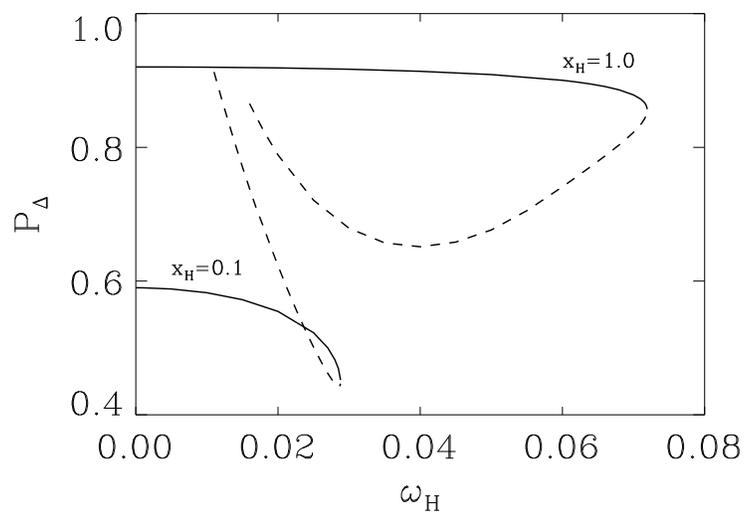}}
\caption{
The same as Fig.~6a for
the horizon magnetic charge
$P_\Delta$.
}
\end{figure}
       \end{fixy}

 \clearpage

       \begin{fixy}{0}
\begin{figure}\centering
{\large Fig. 7a} \vspace{0.0cm}\\
\epsfysize=8cm
\mbox{\epsffile{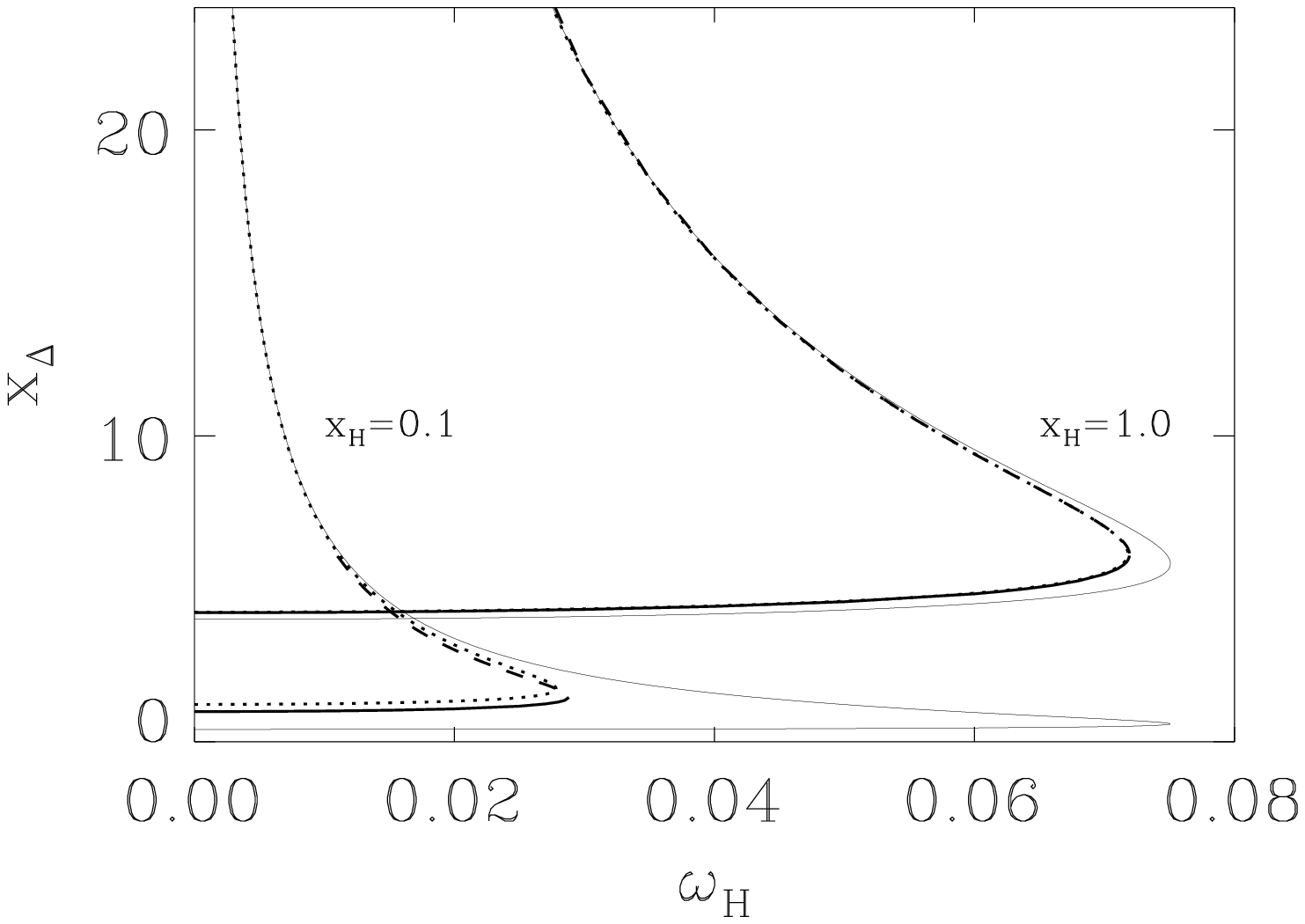}}
\caption{
The area parameter $x_\Delta$ is shown as a function of $\omega_{\rm H}$
for $k=1$, $x_{\rm H}=1$ and $x_{\rm H}=0.1$ on the lower branch (solid)
and on the upper branch (dashed).
For the same values of parameters
the corresponding functions of the
Kerr solution (thin solid) and the
Kerr-Newman solution (dotted) for $Q=0$ and $|P|=1$ are also shown.
}
\end{figure}

\begin{figure}\centering
{\large Fig. 7b} \vspace{0.0cm}\\
\epsfysize=8cm
\mbox{\epsffile{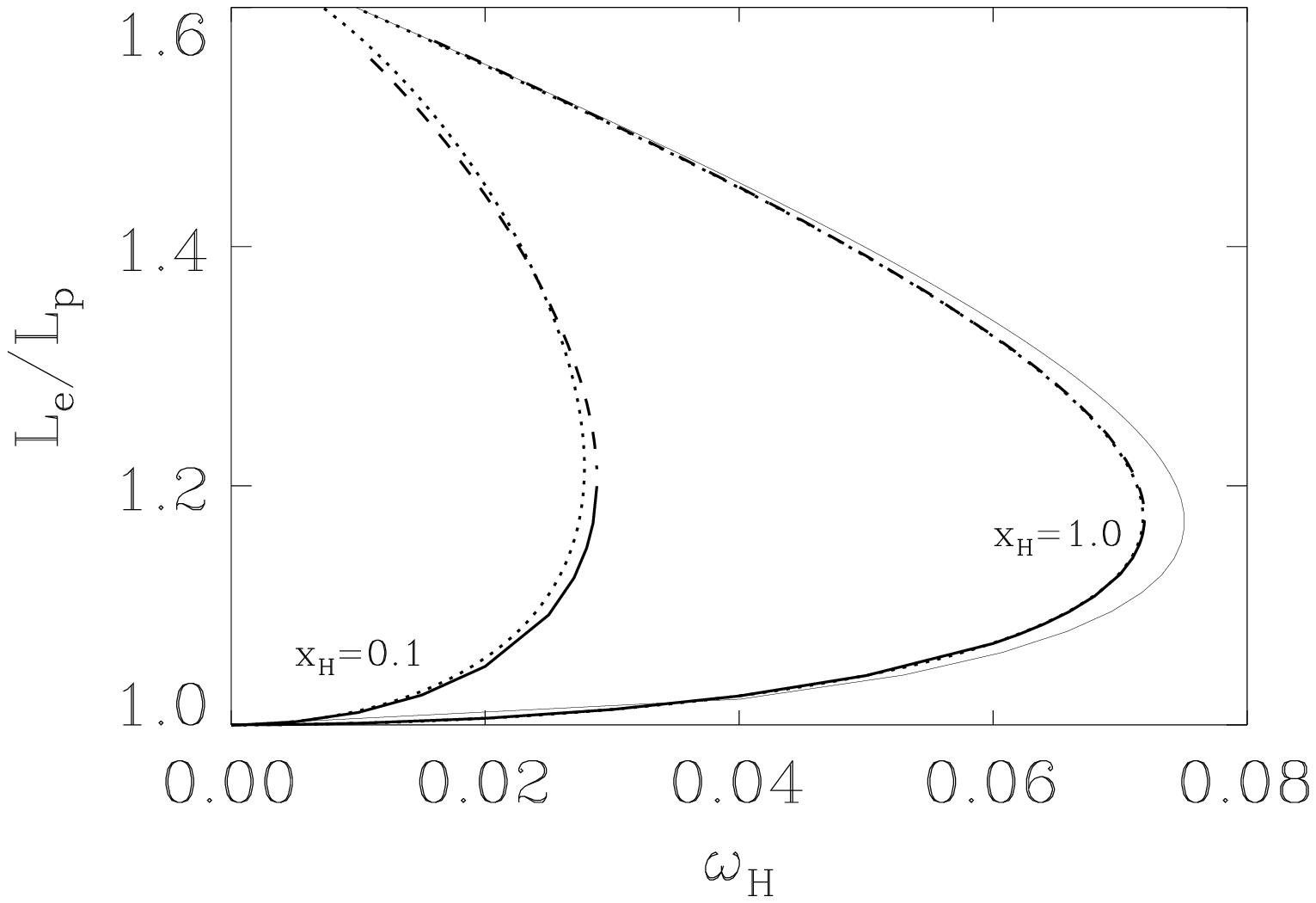}}
\caption{
The same as Fig.~7a for the horizon electric charge  $Q_\Delta$.
}
\end{figure}

\begin{figure}\centering
{\large Fig. 7c} \vspace{0.0cm}\\
\epsfysize=8cm
\mbox{\epsffile{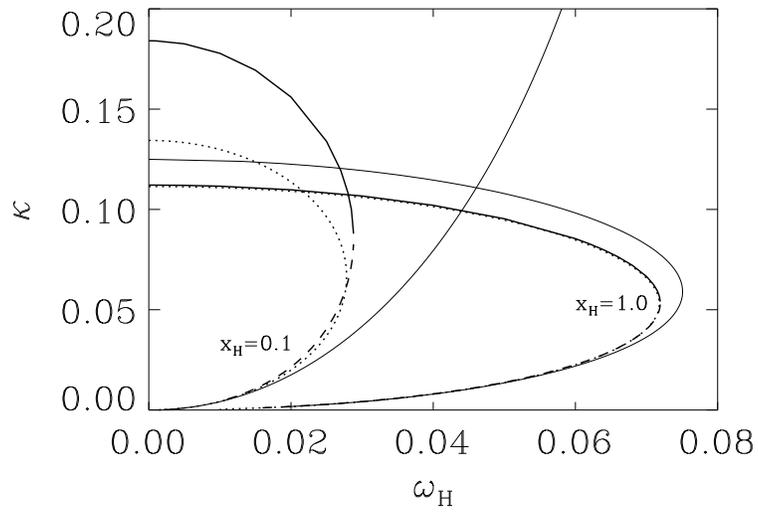}}
\caption{
The same as Fig.~7a for the surface gravity  $\kappa$.
}
\end{figure}
      \end{fixy}

      \begin{fixy}{-1}
\begin{figure}\centering
{\large Fig. 8} \vspace{0.0cm}\\
\epsfysize=8cm
\mbox{\epsffile{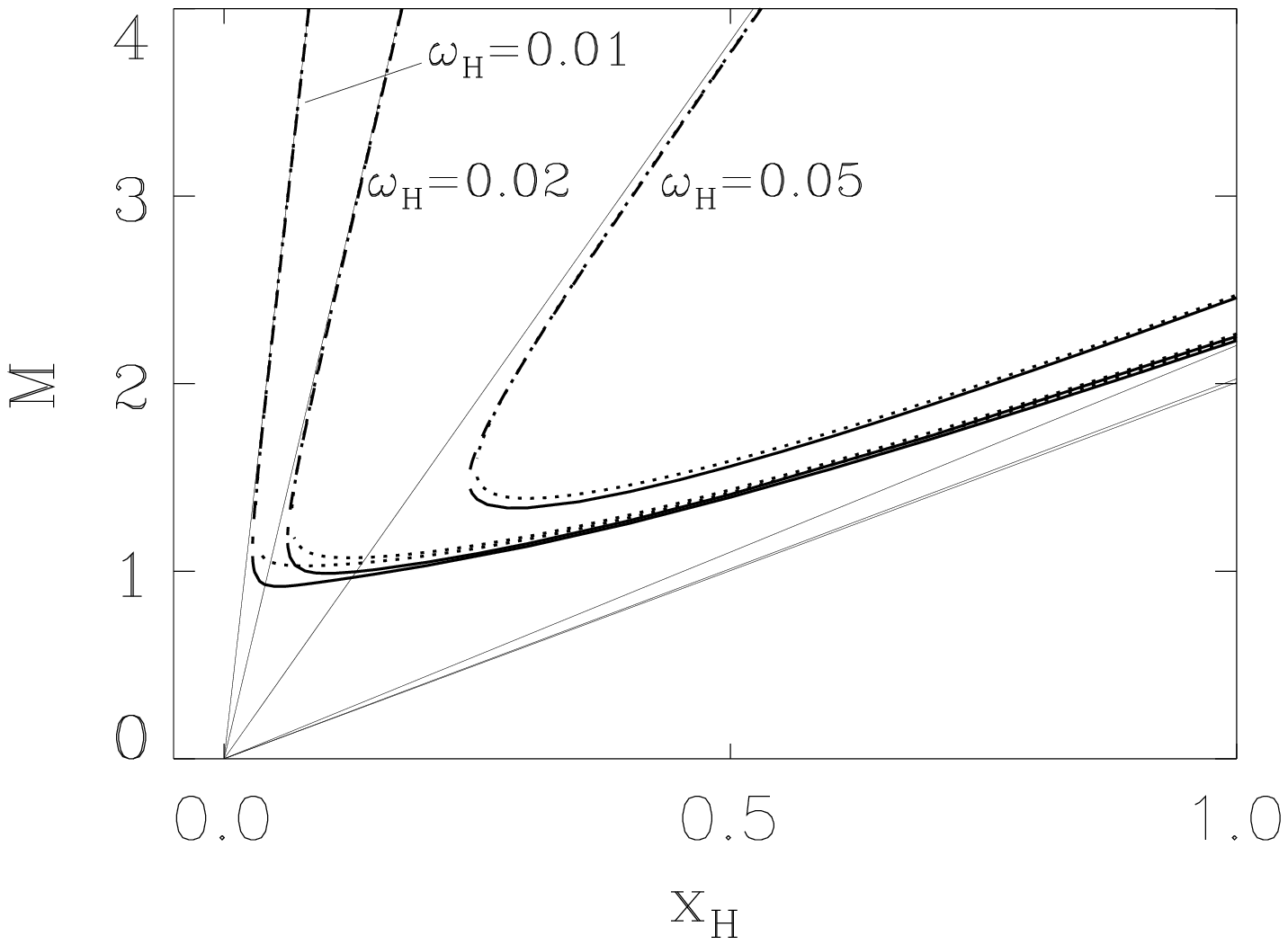}}
\caption{
The dimensionless mass $M$ is shown as a function of $x_{\rm H}$ for $k=1$ and
$\omega_{\rm H}=0.01$, $0.02$ and $0.05$ on the lower branch (solid)
and on the upper branch (dashed).
For the same values of parameters
the corresponding functions of the
Kerr solution (thin solid) the
Kerr-Newman solution (dotted) for $Q=0$ and $|P|=1$ are also shown.
}
\end{figure}
      \end{fixy}

       \begin{fixy}{-1}
\begin{figure}\centering
{\large Fig. 9} \vspace{0.0cm}\\
\epsfysize=8cm
\mbox{\epsffile{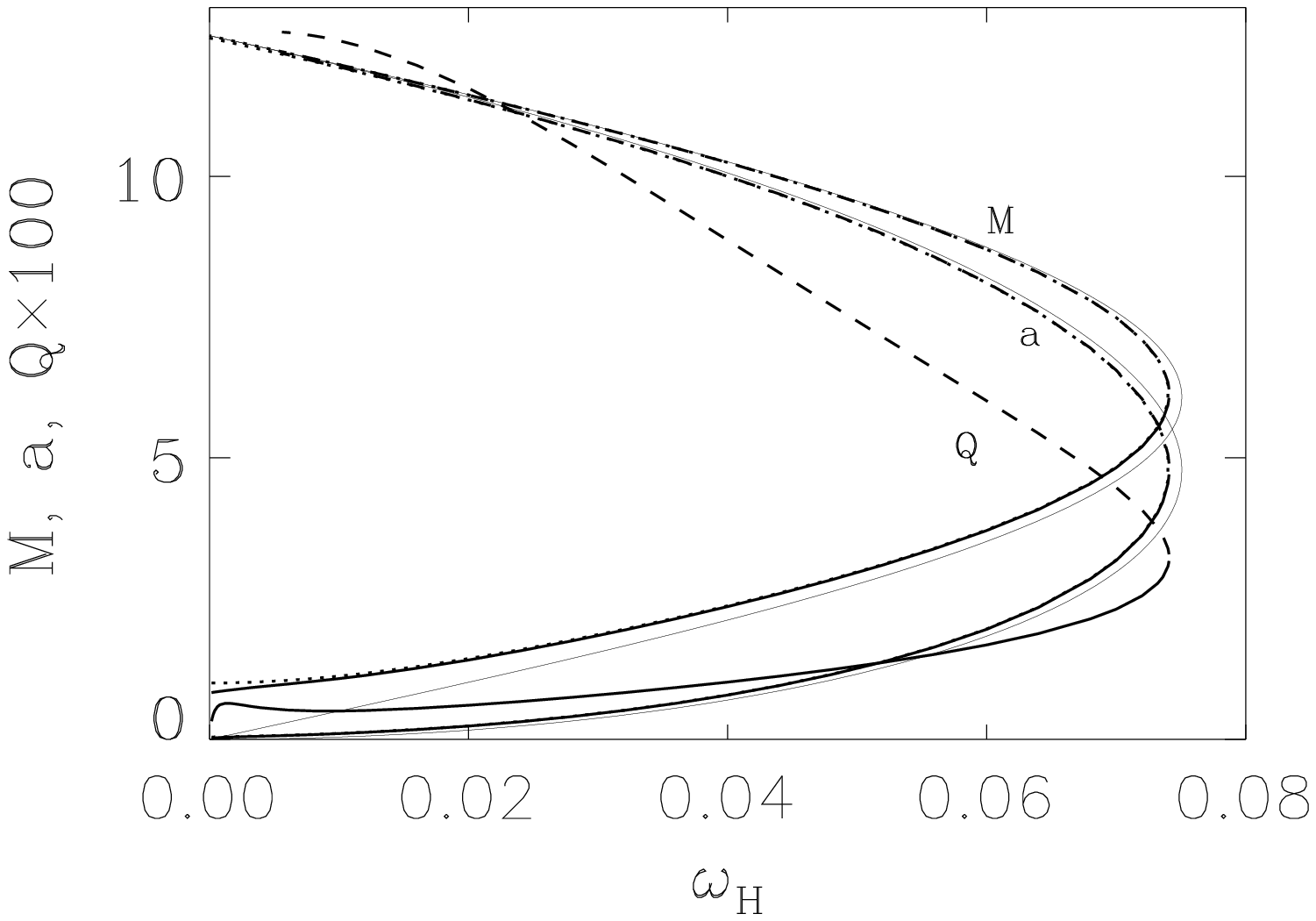}}
\caption{
The dimensionless mass $M$ , ratio $a=J/M$  and  electric charge $Q$
are shown as function of $x_{\rm H}$ for $k=1$ and
fixed $\omega_{\rm H}/x_{\rm H}= 0.04 $ on the lower branch (solid)
and on the upper branch (dashed).
For the same values of parameters
the corresponding functions of the
Kerr solution (thin solid) the
Kerr-Newman solution (dotted) for $Q=0$ and $|P|=1$ are also shown.
}
\end{figure}
        \end{fixy}

     \begin{fixy}{-1}
\begin{figure}\centering
{\large Fig. 10} \vspace{0.0cm}\\
\epsfysize=8cm
\mbox{\epsffile{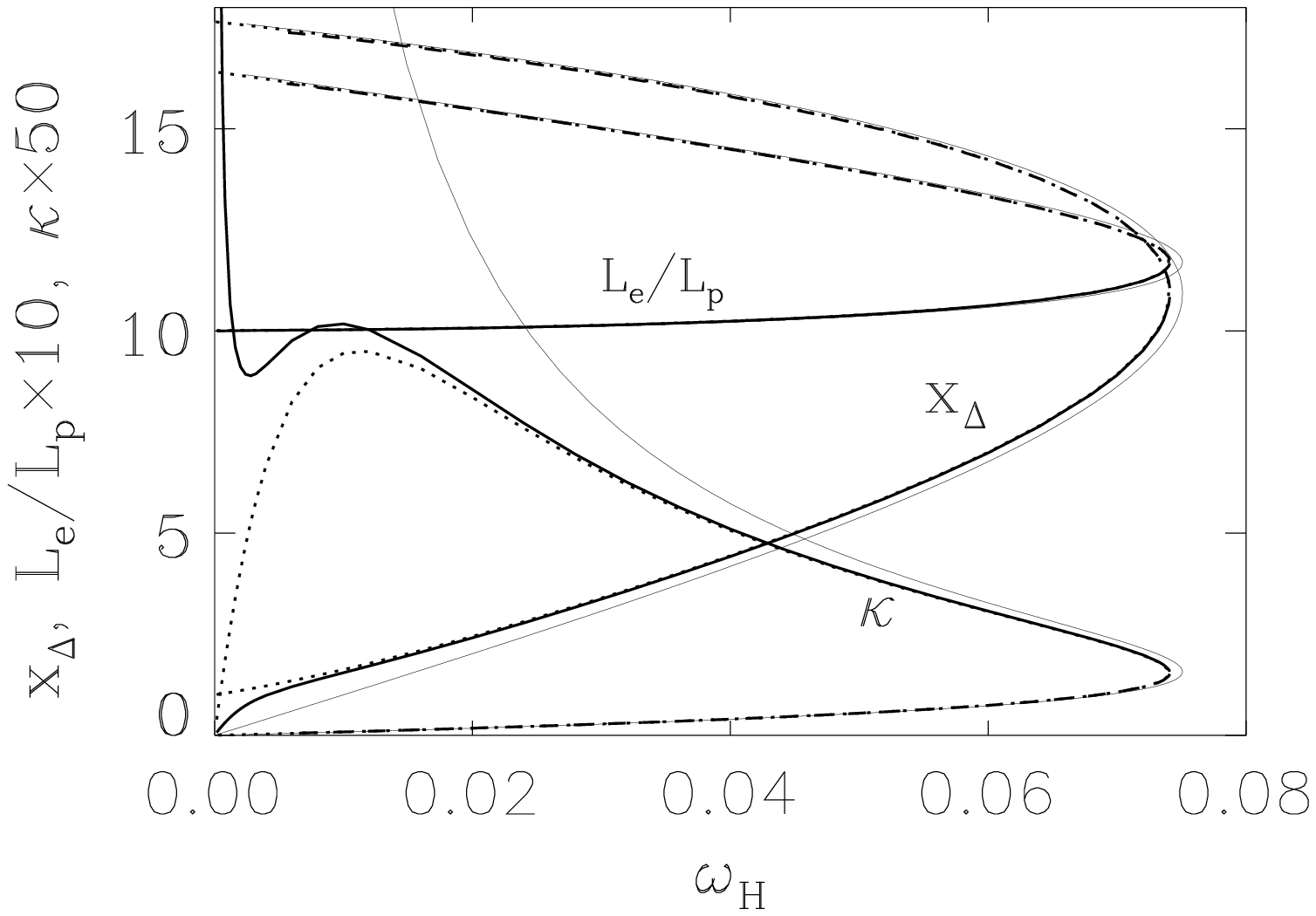}}
\caption{
The same as Fig.~9 for the horizon parameter $x_\Delta$,
the ratio $L_e/L_p$ and the surface curvature $\kappa$.
}
\end{figure}
       \end{fixy}


\newpage

\begin{thebibliography}{000}

\bibitem{nohair1}
 W. Israel,
%Event horizons in static electrovac space-times,
 Commun. Math. Phys. {\bf 8} (1968) 245;\\
 D.~C. Robinson,
%Uniqueness of the Kerr black hole,
 Phys. Rev. Lett. {\bf 34} (1975) 905;\\
 P. Mazur,
%Proof of uniqueness of the Kerr-Newman black hole solution,
 J. Phys. {\bf A15} (1982) 3173.

\bibitem{nohair2}
 M. Heusler,
 Black Hole Uniqueness Theorems,
%Cambridge Lecture Notes in Physics
 (Cambrigde University Press, 1996).

\bibitem{review}
 see e.g.~M.~S. Volkov and D.~V. Gal'tsov,
%Gravitating Non-Abelian Solitons and Black Holes with Yang-Mills Fields,
 Phys. Rept. {\bf 319} (1999) 1;\\
 D.~V. Gal'tsov,
%Gravitating lumps
 hep-th/0112038,
 {\sl Proceedings of the 16th International Conference
 on General Relativity and Gravitation,
 July 2001, Durban, South Africa}

\bibitem{su2bh}
 M.~S. Volkov and D.~V. Galt'sov,
%Black holes in Einstein-Yang-Mills theory,
 Sov. J. Nucl. Phys. {\bf 51} (1990) 747;\\
 P. Bizon,
%Colored black holes,
 Phys. Rev. Lett. {\bf 64} (1990) 2844;\\
 H.~P. K\"unzle and A.~K.~M. Masoud-ul-Alam,
%Spherically symmetric static SU(2) Einstein-Yang-Mills fields,
 J. Math. Phys. {\bf 31} (1990) 928.

\bibitem{kkbh}
 B. Kleihaus and J. Kunz, Phys. Rev. Lett. {\bf 79} (1997) 1595;\\
 B. Kleihaus and J. Kunz, Phys. Rev. {\bf D57} (1998) 6138.

\bibitem{map3}
 B. Kleihaus and J. Kunz,
 Phys. Rev. Lett. {\bf 85} (2000) 2430;\\
 B. Kleihaus and J. Kunz,
% Non-Abelian black holes with magnetic dipole hair,
 Phys. Lett. {\bf B494} (2000) 130.

\bibitem{hkk}
 B. Hartmann, B. Kleihaus, and J. Kunz,
 Phys. Rev. Lett. {\bf 86}, 1423 (2001);\\
 B. Hartmann, B. Kleihaus, and J. Kunz,
 Phys. Rev. {\bf D65} (2002) 0024027.

\bibitem{sud-wald}
 D. Sudarsky and R.~M. Wald,
 Phys. Rev. {\bf D46} (1992) 1453;\\
 D. Sudarsky and R.~M. Wald,
 Phys. Rev. {\bf D47} (1993) 5209.
                 
\bibitem{book}
 see e.~g.~D. Kramer, H. Stephani, E. Herlt, and M. MacCallum,
 Exact Solutions of Einstein's Field Equations, Ch.~17
 (Cambridge University Press, Cambridge, 1980)

\bibitem{circ}
 A. Papapetrou,
 Ann. Inst. H. Poincare {\bf A4} 1966 83;\\
 B. Carter,
 J. Math. Phys. {\bf 10} (1969) 70.

\bibitem{heus}
 M. Heusler and N. Straumann,
%{\sl The first law of black hole physics for a class of nonlinear
% matter models},
 Class. Quant. Grav. {\bf 10}, (1993) 21.

\bibitem{vs}
 M.~S. Volkov and N. Straumann,
%Slowly rotating nonAbelian black holes,
 Phys. Rev. Lett. {\bf 79} (1997) 1428.

\bibitem{galt}
 D.~V. Gal'tsov,
 Einstein-Yang-Mills solitons: towards new degrees of freedom,
 gr-qc/9808002.

\bibitem{ge}
 D.~V. Gal'tsov and A.~A. Ershov,
%{\sl Non-Abelian baldness of colored black holes},
 Phys. Lett. {\bf 138 A} (1989) 160.

\bibitem{bhsv}
 O. Brodbeck, M. Heusler, N. Straumann and M.~S. Volkov,
%Rotating solitons and non-rotating, non-static black holes,
 Phys. Rev. Lett. {\bf 79} (1997) 4310;\\
 O. Brodbeck and M. Heusler,
%Stationary perturbations and infinitesimal rotations
%of static Einstein-Yang-Mills configurations
%with bosonic matter,
 Phys. Rev. {\bf D56} (1997) 6278.

\bibitem{kkrot}
 B. Kleihaus and J. Kunz, 
 Phys. Rev. Lett. {\bf 86} (2001) 3704.

\bibitem{cs}
 A. Corichi, and D. Sudarsky,
%Mass of colored black holes,
 Phys. Rev. D61 (2000) 101501;\\
 A. Corichi, U. Nucamendi, and D. Sudarsky,
%Einstein-Yang-Mills isolated horizons:
%phase space, mechanics, hair and conjectures,
 Phys. Rev. D62 (2000) 044046.
%A. Corichi, U. Nucamendi, and D. Sudarsky,
%A mass formula for EYM solitons,
%Phys. Rev. D64 (2001) 107501.

\bibitem{kkreg}
 B. Kleihaus and J. Kunz, 
 Phys. Rev. Lett. {\bf 78} (1997) 2527;\\
 B. Kleihaus and J. Kunz, 
 Phys. Rev. {\bf D57} (1998) 834.

\bibitem{wald}
 R.~M. Wald,
 General Relativity
 (University of Chicago Press, Chicago, 1984)

\bibitem{islam}
 J.~N. Islam,
 Rotating fields in general relativity
 (Cambridge University Press, Cambridge, 1985)

\bibitem{isorot}
 A. Ashtekar, Ch. Beetle and J. Lewandowski,
%Mechanics of Rotating Isolated Horizons
 Phys. Rev. D64 (2001) 044016.

\bibitem{perry}
 M.~J. Perry,
%Black holes are coloured,
 Phys. Lett. {\bf B71} (1977) 234.

\bibitem{foot0}
 When both electric and magnetic non-Abelian charge is present,
 the vectors must be parallel to each other,
 i.e.~$Q^a = \lambda P^a$ with $\lambda$ constant \cite{perry}.

\bibitem{foot1}
Since the gauge fields wind $n$ times around,
while the azimuthal angle $\vphi$ covers the full trigonometric circle once,
we refer to the integer $n$ as the winding number of the solutions.

\bibitem{kksw}
 B. Kleihaus, J. Kunz, A. Sood and M. Wirschins, 
 Phys. Rev. {\bf D58} (1998) 084006.

\bibitem{foot2}
 The limiting solution corresponds to an embedded
 Reissner-Nordstr\o m black hole with magnetic charge 
 $|P|=n$ for horizon radius $\tilde x_{\rm H} >1$, 
 where $\tilde x$ represents a Schwarzschild-like radial coordinate.
 \cite{review,kkbh}.
 For $\tilde x_{\rm H} < 1$,
 the limiting solution also corresponds to an embedded
 Reissner-Nordstr\o m black hole with magnetic charge 
 $|P|=n$ in the region $1 < \tilde x$, but
 contains an oscillating non-Abelian part 
 in the region $\tilde x_{\rm H} < \tilde x < 1$
 \cite{review,kkbh}.

\bibitem{schoen}
 W. Sch\"onauer and R. Wei\ss ,
 J. Comput. Appl. Math. 27, 279 (1989) 279;
 M. Schauder, R. Wei\ss\ and W. Sch\"onauer,
 The CADSOL Program Package,
 Universit\"at Karlsruhe, Interner Bericht Nr. 46/92 (1992).

\bibitem{radu2}
 J.~J. van der Bij and E. Radu,
 On rotating regular nonabelian solutions,
 Int. J. Mod. Phys. {\bf A17} (2002) 1477.

\bibitem{next}
 B. Kleihaus, J. Kunz, F. Navarro, in preparation.

\end{thebibliography}
\end{document}